\documentclass[aps, prl, reprint, twocolumn,superscriptaddress,floatfix]{revtex4-2}
\usepackage[utf8]{inputenc}
\usepackage[english]{babel}
\usepackage[T1]{fontenc}
\usepackage{amsmath}
\usepackage{amssymb}
\usepackage{setspace}
\usepackage{blkarray}
\usepackage{tikz}
\usepackage{lipsum}
\usepackage{bbm, amsthm, bm,textcomp, nicefrac,geometry,ragged2e}
\usepackage{float}
\usepackage[bbgreekl]{mathbbol}
\usepackage{graphicx,epstopdf,color,verbatim,enumitem} 
\usepackage{svg}
\usepackage{pbox,array}
\usepackage{mathrsfs}
\usepackage{stackrel}
\usepackage{cancel,soul}
\makeatletter
\usepackage[caption=false]{subfig}
\addto\captionsenglish{}
\usepackage{dcolumn}    
\usepackage{ifpdf}
\usepackage{graphics}
\usepackage{rotating}
\usepackage{lineno,amsfonts}
\usepackage{graphicx}   
\usepackage{bm}         
\usepackage{bbm}
\usepackage{upgreek}
\usepackage{mathtools}
\usepackage{epstopdf}
\usepackage{setspace}
\usepackage{float}
\usepackage{natbib}
\usepackage{qcircuit}
\usepackage{dsfont}
\usepackage{xr}
\usepackage{cancel}
\usepackage{hyperref}
\usepackage{chngcntr}
\usepackage{pythonhighlight}
\usepackage{makecell}
\usepackage{multirow}
\usepackage{algorithm}
\usepackage{algpseudocode}
\usepackage{threeparttable, tablefootnote}
\usepackage{comment}
\makeatletter
\newcommand*{\addFileDependency}[1]{
  \typeout{(#1)}
  \@addtofilelist{#1}
  \IfFileExists{#1}{}{\typeout{No file #1.}}
}
\makeatother

\newcommand{\update}{\color{black}}

\begin{document}
\title{\textcolor{black}{Measurement-Based Infused Circuits for Variational Quantum Eigensolvers}}
\begin{abstract} 
Variational quantum eigensolvers (VQEs) are successful algorithms for studying physical systems on quantum computers. Recently, they were extended to the measurement-based model of quantum computing, bringing resource graph states and their advantages into the realm of quantum simulation. In this work, we incorporate such ideas into traditional VQE circuits. This enables novel problem-informed designs and versatile implementations of many-body Hamiltonians. We showcase our approach on real superconducting quantum computers by performing VQE simulations of testbed systems including the perturbed planar code, $\mathbb{Z}_2$ lattice gauge theory, 1D quantum chromodynamics, and the LiH molecule.
\end{abstract}
\author{Albie Chan}
    \email{a233chan@uwaterloo.ca}
    \thanks{this author contributed equally}
    \affiliation{Institute for Quantum Computing and Department of Physics \& Astronomy, University of Waterloo, Waterloo, Ontario, N2L 3G1, Canada}
\author{Zheng Shi}
    \thanks{This author contributed equally}
    \affiliation{Institute for Quantum Computing and Department of Physics \& Astronomy, University of Waterloo, Waterloo, Ontario, N2L 3G1, Canada}
\author{Luca Dellantonio}
    \affiliation{Institute for Quantum Computing and Department of Physics \& Astronomy, University of Waterloo, Waterloo, Ontario, N2L 3G1, Canada}
    \affiliation{Department of Physics and Astronomy, University of Exeter, Stocker Road, Exeter EX4 4QL, United Kingdom}
\author{Wolfgang D\"ur}
    \affiliation{Universität Innsbruck, Institut für Theoretische Physik, Technikerstraße 21a, 6020 Innsbruck, Austria}
\author{Christine A. Muschik}
    \affiliation{Institute for Quantum Computing and Department of Physics \& Astronomy, University of Waterloo, Waterloo, Ontario, N2L 3G1, Canada}    
    \affiliation{Perimeter Institute for Theoretical Physics, Waterloo, Ontario, N2L 2Y5, Canada}
    \maketitle  
Hybrid classical-quantum algorithms are a transformative approach to simulating physical systems~\cite{callison2022}. One such algorithm is the \textit{variational quantum eigensolver} (VQE), which optimizes an ansatz on classical hardware to determine a system's ground state (GS) energy~\cite{peruzzo2014variational, mcclean2016theory, tilly2022vqe, callison2022}. This delegation allows VQEs to mitigate the limitations of current noisy intermediate-scale quantum (NISQ) devices, including coherence times, number of qubits, and circuit depths~\cite{preskill2018quantum, callison2022}. As simulation tools, VQEs have successfully modelled systems in chemistry~\cite{motta2022chemistry, callison2022}, materials science~\cite{bauer2016correlated,bauer2020chemmater} and high-energy physics~\cite{banuls2020simulating, daley2022practical, kokail2019self, paulson2020qed, haase2021qed, atas2021su2}.
\par
Traditional VQEs represent ansatzes as parameterized circuits comprising gates and measurements~\cite{tilly2022vqe}. However, as proposed in Ref.~\cite{ferg2021_mbvqe}, an ansatz can exploit the measurement-based (MB) model of quantum computing (MB-QC). In MB-QC, computations are performed via measurements and local operations on an entangled resource graph state~\cite{owqc_old_ref, raussendorf_owqc, browne2006one, briegel2009measurement}. By parameterizing the measurements, one obtains an ansatz in the MB model, yielding the MB-VQE~\cite{ferg2021_mbvqe}. As a gate-free approach, MB-VQEs enable VQEs on platforms for which the implementation of entangling gates is challenging.
\par
Here, we infuse the MB-VQE ideas of graph states and MB-QC into traditional VQE circuits. This creates new design- and cost-effective ways of performing VQEs, while broadening the range of platforms compatible with MB methods. 
In particular, we perform the first demonstrations of circuit VQEs exploiting MB-VQE ideas on quantum computers (IBM Quantum). Practically, we address two important challenges: the design of an effective variational ansatz, and their cost-effective implementation on NISQ devices.
\par
\begin{figure*}
\subfloat{\includegraphics[width=1\textwidth]{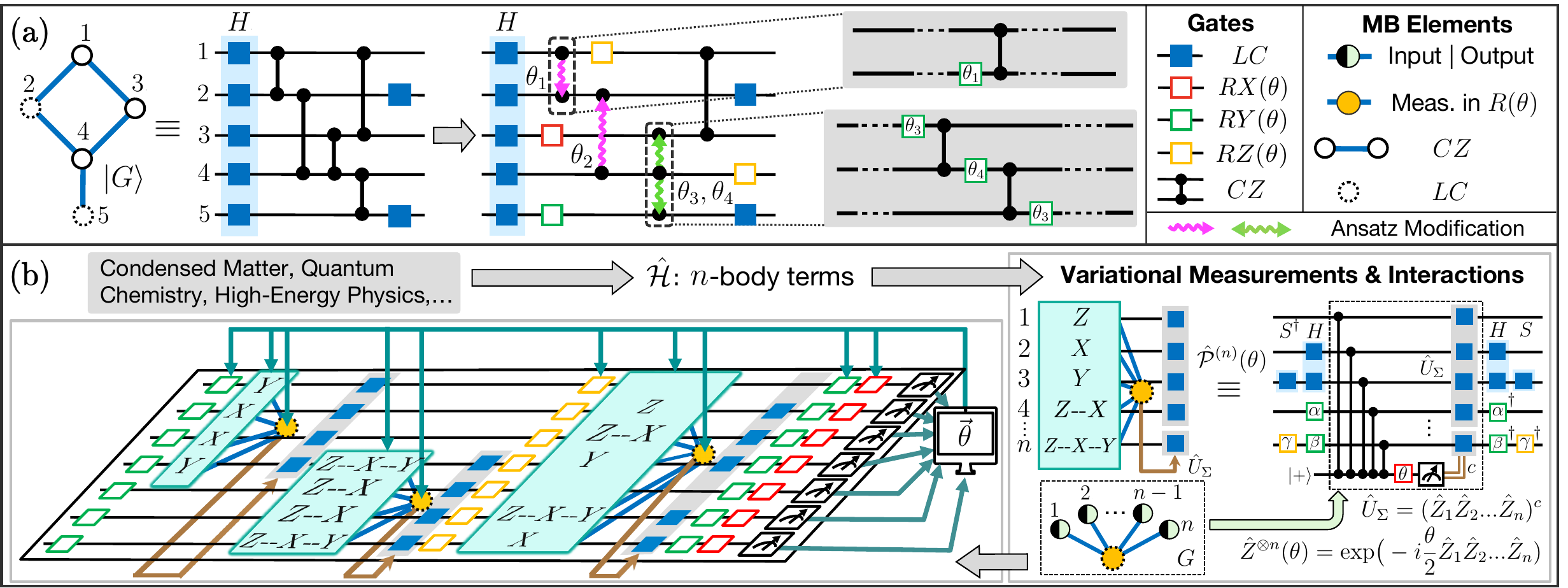}}
 \caption{
 \textbf{Infusing MB-VQE ideas into variational circuits.}
    (a) Ansatz modification (wavy arrows) allows for designing circuit ansatzes that access graph-inspired variational state families. From a circuit representation of a graph state $|G\rangle$, we dress each CZ with parameterized gates to tune the entanglement structure of $|G\rangle$. (b) The Pauli gadgets $\mathcal{\hat{P}}^{(n)}(\theta)$ combine resource graphs ($G$) with gates to yield efficient circuit ansatzes for NISQ platforms. $\mathcal{\hat{P}}^{(n)}(\theta)$ realizes a tunable $n$-body Pauli term in a target Hamiltonian. The parameterized ancilla (in orange) variationally tunes the measurement basis used to perform $G$, while $RY$ and $RZ$ gates permit variational tunability of the interaction basis. These gadgets provide potential resource advantages in the number of entangling operations and circuit depth. Grey, teal, and brown arrows denote the workflow, classical--quantum feedback loop, and feedforward involving byproduct Pauli operators $\hat{U}_{\Sigma}$ respectively. On circuit platforms (e.g. IBM Quantum), Pauli gadgets are implemented \textit{dynamically} via mid-circuit measurements (see SM~\cite{supp_mat}).
    }
    \label{fig: conceptual}
  \end{figure*}

For the first challenge, we exploit the entanglement structures of $n$-qubit graph states to design an effective circuit ansatz. By applying suitable gate modifications to a graph-equivalent circuit [Fig.~\ref{fig: conceptual}(a)], we can tune its entanglement structure variationally. This allows access to graph-inspired variational ansatz families without adaptively updating the ansatz~\cite{grimsley2019adaptive, ryabinkin2020iterative}. Our ansatz design is systematic, scalable ($\mathcal{O}(n^{2})$ parameters at most), and problem-informed when simulating models with graph state representations. Unlike tensor-network methods~\cite{watanabe2023, rudolph2022, huang2022}, it derives an initial circuit design from the graph state connectivity of the model. We apply this technique to the planar code, as an example of a problem-informed VQE.
\par
For the second challenge, we exploit resource graph states containing ancillas to implement variational measurements in the VQE circuit, i.e. their parameters determine the variable measurement basis. We demonstrate this idea with “Pauli gadgets”, which integrate resource graphs with local rotations to realize tunable $n$-body Pauli interactions [Fig.~\ref{fig: conceptual}(b)]. Such terms appear in Hamiltonians in condensed matter, high energy physics, and quantum chemistry~\cite{yamamoto2020, atas2022su3, xiao2012, kandala2017hardware}. By realizing Pauli interactions through measurements, our gadgets can be implemented efficiently in the circuit depth and number of entangling gates, both bottlenecks on NISQ devices. We leverage these circuits in VQE simulations of $\mathbb{Z}_{2}$ lattice gauge theory, 1D quantum chromodynamics, and the LiH molecule.
\par
\par
\textbf{\textit{Background.}}--- We begin by summarizing relevant concepts underlying our VQE techniques.
\par
\textit{VQE:} VQEs commonly approximate the GS $|\psi_{0}\rangle$ of a target Hamiltonian $\hat{\mathcal{H}}$ with an ansatz state $|\psi_{a}(\vec{\theta})\rangle$. From the ansatz, a quantum processor computes the energy expectation $E = \langle\psi_{a}(\vec{\theta})|\hat{\mathcal{H}}|\psi_{a}(\vec{\theta})\rangle$, which a classical optimizer minimizes towards the GS energy $E_{0} = \langle\psi_{0}|\hat{\mathcal{H}}|\psi_{0}\rangle$ by varying $\vec{\theta}$~\cite{mcclean2016theory}. Traditional VQEs represent the ansatz with a circuit whose gates are parameterized by $\vec{\theta}$.
\par
\textit{Graph states:} The MB model employs graph states $|G\rangle$~\cite{briegel2001cluster, duer2003graph, hein2004multiparty, hein2006entanglement}, see Fig.~\ref{fig: conceptual}(a). Vertices represent qubits in $|+\rangle$, and edges realize CZ operations defining the graph entanglement structure. $|G\rangle$ may be transformed into MB-QC resource graph states by adding input and output qubits, and designating the remainder as ancillas \cite{ferg2021_mbvqe}.
\par
\textit{MB-QC:} From a resource graph state, MB-QC proceeds as a sequence of projective measurements in an eigenbasis of the Pauli operators ${\hat{X}, \hat{Y}, \hat{Z}}$, or a rotated basis $R(\theta) = (|0\rangle \pm e^{i\theta}|1\rangle)/\sqrt{2}$. The angle $\theta$ parameterizes the desired operation.
Since measurement outcomes are random, feedforward is applied to ensure determinism via byproduct operators $\hat{U}_{\Sigma}$ and angle adaptations for qubits measured in $R(\theta)$.
Together, the entangling, measurement, and feedforward steps comprise an MB pattern realizing a specific unitary (see Ref.~\cite{raussendorf_owqc} and Supplemental Material (SM)~\cite{supp_mat}).
\par
\textbf{\textit{Ansatz modification.}}---
First, we describe an MB-inspired technique to generate the ansatz from an initial graph, by adding parameters that variationally tune its entanglement structure. While such ``modifications'' were used to design graph ansatzes (i.e. edge decoration) for MB-VQEs~\cite{ferg2021_mbvqe}, we focus on infusing this technique into the design of circuit VQEs.
\par
\begin{figure*}
    \centering
    \includegraphics[width=1\textwidth]{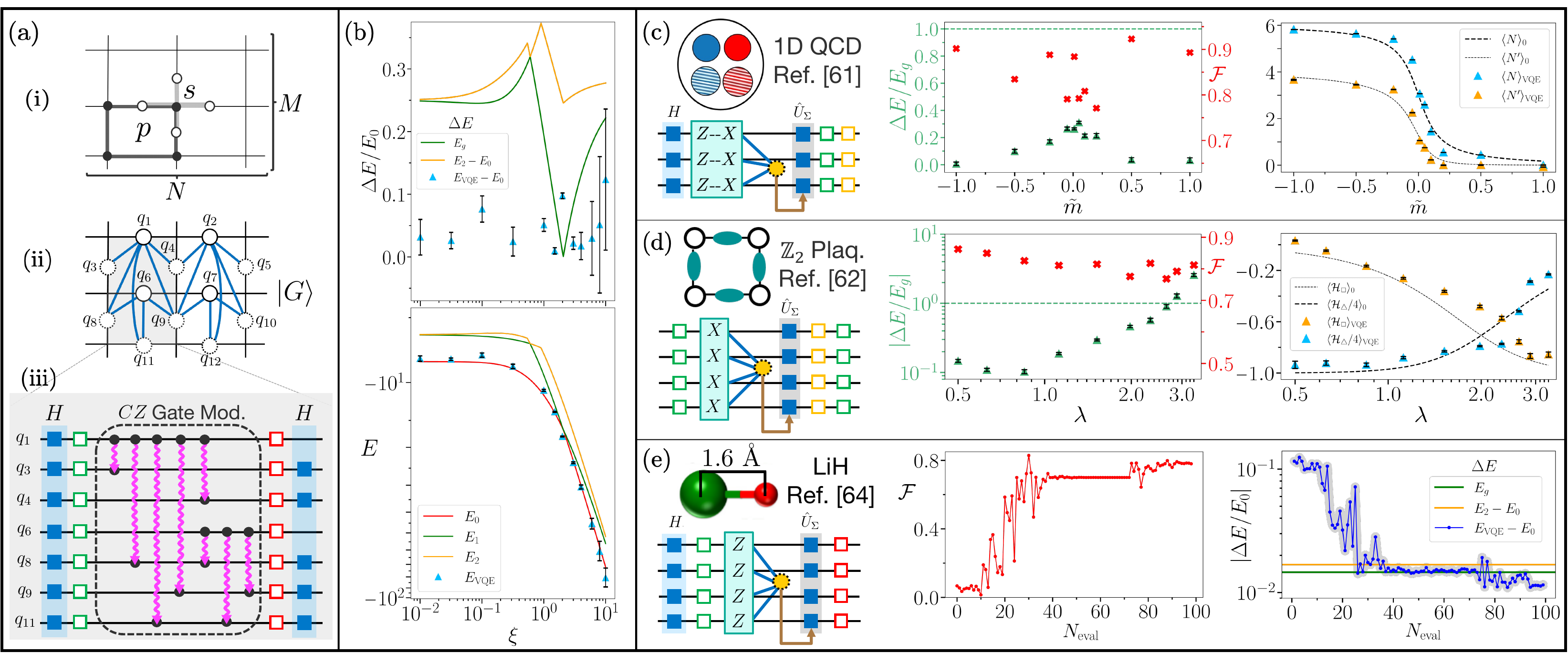}
    \caption{
    \textbf{MB-infused circuit VQE demonstrations on IBM Quantum}. Energy differences $\Delta E = |E_{\text{VQE}}-E_{0}|$ between the VQE ($E_{\text{VQE}}$) and exact GS ($E_{0}$) are expressed in units of the energy gap $E_{g} = E_{1}-E_{0}$ or $E_{0}$. (a; i) 2D planar code (PC) as an $M \times N$ lattice, with star ($s$) and plaquette ($p$) stabilizer operators. (a; ii) GS of the $M\times N$  PC ($|G\rangle$), showing qubits (circles) residing on the edges. (a; iii) Circuit ansatz for the $2 \times 1$ perturbed PC based on (a; ii), where each CZ gate is modified as $CZ_{1, 2} \mapsto CZ_{1, 2}RY(\theta_{i})_{2}$ [Fig.~\ref{fig: conceptual}(a)].
    (b) VQE demonstration for the $2 \times 1$ perturbed PC (on \texttt{ibm\_perth} and \texttt{ibm\_lagos}). $E_{0}$, $E_{\text{VQE}}$, relative error ($\Delta E/E_{0}$) and state fidelity ($\mathcal{F}$) are shown as functions of perturbation strength $\xi$. The two (exact) lowest excited state energies $E_{1}$ and $E_{2}$ are shown for comparison. 
    (c)--(e) VQE demonstrations with a single Pauli gadget. All runs were performed dynamically (see SM~\cite{supp_mat}). (c) 1D QCD model (on \texttt{ibm\_lagos} and \texttt{ibm\_peekskill}): $\langle\hat{N}\rangle$ and $\langle\hat{N'}\rangle$ vs. reduced bare quark mass $\tilde{m}$ ($x = 0.8$), with plots of $\Delta E/E_g$ and $\mathcal{F}$. (d) $\mathbb{Z}_2$ lattice gauge theory model (on \texttt{ibm\_peekskill}): $\langle \hat{\mathcal{H}}_{\square} \rangle$ and $\langle \hat{\mathcal{H}}_{\triangle}/4\rangle$ vs. coupling strength $\lambda$, with plots of $|\Delta E/E_g|$ and $\mathcal{F}$. (e) LiH molecule (at $R$ = 1.6 \AA, on \texttt{ibm\_perth}): Plots of |$\Delta E/E_{0}$| and $\mathcal{F}$ vs. number of function evaluations $N_{\text{eval}}$. For comparison, the errors for higher energy states ($\Delta E = E_{g}$ and $\Delta E = E_{2}-E_{0}$) are plotted as horizontal lines. Grey regions indicate statistical uncertainty.}
    \label{fig: vqe_results}
\end{figure*}

We demonstrate ansatz modification with the planar code (PC), an error-correcting surface code prominent in experimental efforts toward fault-tolerant quantum computation. The PC is defined on an $M \times N$ lattice with open boundary conditions and qubits residing on the edges. Its GS is the $+1$ eigenstate of the star ($\hat{S}_{s}$) and plaquette ($\hat{P}_{p}$) operators which act on neighbouring qubits of each vertex $s$ and plaquette $p$, respectively [Fig.~\ref{fig: vqe_results}(a; i)]~\cite{kitaev2003toric}. Consequently, the PC Hamiltonian $\mathcal{\hat{H}}^{(M, N)}_{\text{PC}} = \hat{\mathcal{H}}_+ + \hat{\mathcal{H}}_{\square}$, where $\hat{\mathcal{H}}_+$ and $\hat{\mathcal{H}}_{\square}$ contain all star and plaquette terms, respectively. We also add a perturbation term $\hat{\mathcal{H}}_{\triangle}$, mimicking a stray field in experimental settings. We express
\begin{equation}
\hat{\mathcal{H}}_+ = -\sum_{s=1}^{n_{\mathrm{s}}}\hat{S}_{s}; \hspace{0.2cm}
\hat{\mathcal{H}}_\square = - \sum_{p=1}^{n_{\mathrm{p}}}\hat{P}_{p}; \hspace{0.2cm} 
\hat{\mathcal{H}}_\triangle = \xi\sum_{q=1}^{n_{\mathrm{q}}}\hat{Z}_{q},
\label{eq: pcham}
\end{equation}
where $\hat{S}_{s} = \prod_{i \in +_{s}} \hat{Z}_{i}$ and $\hat{P}_{p} = \prod_{i \in \square_{p}} \hat{X}_{i}$. Here, $i \in \square_p \hspace{0.1cm} (i \in +_{s}$) indicates the set of qubits belonging to a plaquette $p$ (vertex $s$), $n_{\mathrm{s}} (n_{\mathrm{p}})$ is the number of vertices (plaquettes) in the lattice, and $\xi$ is the strength of the perturbation $\hat{\mathcal{H}}_{\triangle}$ acting uniformly on each of the $n_{\mathrm{q}}$ qubits. $\hat{\mathcal{H}}_{\triangle}$ changes the degree of entanglement in the graph, yielding a continuum of ground states ranging from maximally entangled $|G\rangle (\xi = 0)$ to the product $|1\rangle^{\otimes{n_{\mathrm{q}}}}$ ($\xi \rightarrow \infty)$. States in the intermediate regime characterize imperfect realizations of the PC~\cite{wolfgang_tensor_network} and are challenging to determine analytically.
\par
We construct our ansatz from $|G\rangle$, and translate it into a circuit of CZ and $H$ gates~\cite{raussendorf_owqc} [Fig.~\ref{fig: conceptual}(a)]. While $|G\rangle$ provides accurate GS energy approximations for $\xi \ll 1$, we obtain good approximations for larger $\xi$ via $L$-layer modifications of each CZ. Here, we consider $CZ_{1, 2} \mapsto CZ_{1, 2}RY(\theta_{i})_{2}$ (Mod.~1, magenta arrow) and $CZ_{2, 3}CZ_{1, 2} \mapsto RY(\theta_{i})_{3}CZ_{2,3}RY(\theta_{j})_{2}CZ_{1,2}RY(\theta_{i})_{1}$ (Mod.~2, green arrow) [Fig.~\ref{fig: conceptual}(a)]. Both modifications incur modest overheads of $\mathcal{O}(n_{\mathrm{q}}^{1.5}L)$ parameters and provide sufficient entanglement tunability for all $\xi$ when $(M, N)$ is small (see SM~\cite{supp_mat}).
\par 
For the main VQE demonstration on IBM Quantum, we consider two vertical plaquettes [$(M, N) =(2, 1)$; grey region in Fig.~\ref{fig: vqe_results}(a; ii)]. From the initial circuit, we apply Mod.~1 to each CZ once ($L=1$) and exploit model symmetries to obtain the 7-qubit, 11-parameter ($2\times4+3$) variational circuit in Fig.~\ref{fig: vqe_results}(a; iii), with $RY$ and $RX$ rotation layers appended (see SM~\cite{supp_mat} for Mod.~2 demonstrations). We ran the circuit on \texttt{ibm\_perth} and \texttt{ibm\_lagos} for $10^{-2} \leq \xi \leq 10^1$. The approximated energies $E_{\rm VQE}$ [Fig.~\ref{fig: vqe_results}(b)] agree closely with the exact GS energies for $\hat{\mathcal{H}}^{(2, 1)}_{\text{PC}}$, as confirmed by the small differences $\Delta E = E_{\rm VQE} - E_0$ attained, $0.091 \leq |\Delta E/E_{g}| \leq 0.556$ where $E_{g}=E_{1}-E_{0}$ is the exact first energy gap.
Importantly, the expected physical behavior is recovered as we identify a region $\xi \sim 1$ where the GS crosses over from describing a randomly oriented to fully polarized spin lattice (see SM~\cite{supp_mat}).
\par
\textbf{\textit{Variational measurements and Pauli gadgets.}}---
Next, we address the efficient implementation of VQE circuits on NISQ devices. We achieve this using MB-QC inspired~\cite{raussendorf_owqc, browne2006one} variational measurements on ancillas. Here, ancillas are parameterized with an angle $\theta$ and then measured mid-circuit (i.e. prior to readout). While there exists a variety of MB techniques~\cite{owqc_old_ref, raussendorf_owqc, browne2006one, briegel2009measurement, briegel2001cluster, gottesman1999universal, terhal2004adaptive, hoyer2005fanout, bravyi2005universal,browne2011mbdepth, jozsa2013simcomplexity, paetznick2014repeat, takahashi2016depth, zhu2020depth, piroli2021locc, verresen2022lre, tantivasadakarn2022lre, liu2022depth, bravyi2022adaptive, miguelramiro2023sqem, miguelramiro2023enhancing} capable of performing variational measurements, we demonstrate them with “Pauli gadgets” built from resource graphs $G$~\footnote{
We use $G$ rather than $|G\rangle$ to convey that the graph is not a stabilizer graph state, since the $n$ input and output qubits are not restricted to eigenstates of $\hat{X}$ (i.e. \{$|+\rangle$, $|-\rangle$\}
}
realizing $n$-body $Z$-interaction operators  $\hat{Z}^{\otimes n}(\theta) = e^{-i\theta \hat{Z}_{1}\hat{Z}_{2}...\hat{Z}_{n}/2}$.
$G$ consists of $n$ vertices (input and output) connected to a single ancilla initialized in $|+\rangle$ and measured in the $R(\theta)$-basis. Expressing $G$ as a stabilizer circuit and acting local Clifford (LC) gates $H$ and $S$~\cite{browne2006one, raussendorf_owqc}, we transform the basis of qubit $j \in \{1,2,...,n\}$ from $\hat{Z}_{j}$ to any $\mathcal{\hat{P}}_{j};\hspace{0.1cm} \mathcal{\hat{P}}_{j} \in \{\hat{X}_{j}, \hat{Y}_{j}, \hat{Z}_{j}\}$ [Fig.~\ref{fig: conceptual}(b)], thereby forming an $n$-qubit Pauli gadget $\mathcal{\hat{P}}^{(n)}(\theta)$. Alternatively, we may apply $RY$ and $RZ$ gates which allow continuous tunability between interaction bases: $\hat{\mathcal{P}}_{j} \in$ $\{\hat{Z}_{j}$-{}-$\hat{X}_{j}$, $\hat{Z}_{j}$-{}-$\hat{X}_{j}$-{}-$\hat{Y}_{j}$\}, where e.g. $\{\hat{X}_{j}$-{}-$\hat{Y}_{j}\}$ indicates qubit $j$ rotates around a tunable axis on the Bloch sphere's XY-plane. Both approaches may be performed with external rotation layers to widen the optimization search space [Fig.~\ref{fig: conceptual}(b)]. Such versatility in constructing gadgets allows one to readily implement a circuit ansatz capturing multi-qubit interactions. 
\par 
Our construction of $\mathcal{\hat{P}}^{(n)}(\theta)$ from $G$ has two motivations. First, $G$ acts as a unitary “blackbox” that is conveniently incorporated in circuits, since it contains input and output qubits generally absent in graphs~\cite{browne2006one}. Second, $G$ incurs fewer entangling operations by employing a single ancilla that is reusable after each measurement: $n$ CZs instead of the $2(n-1)$ cascaded CXs in the ancilla-free $\hat{Z}^{\otimes n}$ implementation~\cite{nielsen_chuang_2010, duncan2020zx, cowtan2020gadget, clinton2021hamiltonian, sriluckshmy2023multiqubitpauli} [Fig.~\ref{fig: conceptual}(b)]. This factor-of-two reduction in entangling gates with constant overhead is relevant for NISQ implementations when increasing the number of qubits, variational layers, or many-body interactions. Furthermore, the gadgets' dependence on $\mathcal{\hat{H}}$ implies they scale only in depth and maintain constant $n$ with increasing system size. These features highlight Pauli gadgets as a cost-efficient element for VQE circuits; for trade-offs between depth-reduction and noise from mid-circuit measurements, see SM~\cite{supp_mat}. Below, we showcase their broad applicability via three VQE demonstrations on IBM Quantum.
\par
\textit{1D lattice QCD:} Particle physics is a promising candidate for achieving quantum advantage~\cite{funcke2022nisq, daley2022practical, klco2022standard, zohar2022quantum}, motivating the study of benchmark models on quantum computers~\cite{martinez2016real, muschik2017u, kokail2019self, paulson2020qed, haase2021qed, atas2021su2, mildenberger2022probing}. Here, we examine quantum chromodynamics (QCD) on a one-dimensional lattice. Following Ref.~\cite{atas2022su3}, we express the Hamiltonian $\mathcal{\hat{H}}_{\text{QCD}}$ for a unit cell [Fig.~\ref{fig: vqe_results}(c)] possessing three quarks and three antiquarks in the zero-baryon number ($B=0$) sector. It is the sum of kinetic, mass, and electric field terms
\begin{equation}
\begin{split}
\hat{\mathcal{H}}_{k}^{(B=0)} &= -\frac{1}{2}(\hat{X}_{1}\hat{Z}_{2}\hat{Z}_{3} + \hat{Z}_{1}\hat{X}_{2}\hat{Z}_{3} + \hat{Z}_{1}\hat{Z}_{2}\hat{X}_{3}), \\
\hat{\mathcal{H}}_{m}^{(B=0)} &= \tilde{m}(3 - \hat{Z}_{1} - \hat{Z}_{2} - \hat{Z}_{3}), \\
\hat{\mathcal{H}}_{e}^{(B=0)} &= \frac{1}{6x}(3 - \hat{Z}_{1}\hat{Z}_{2} - \hat{Z}_{1}\hat{Z}_{3} - \hat{Z}_{2}\hat{Z}_{3}),
\end{split}
\label{eq: qcdham}
\end{equation}
where $\tilde{m}$ and $x$ are the dimensionless bare quark mass and coupling constant, respectively.
\par
We conducted a VQE on \texttt{ibm\_lagos} and \texttt{ibm\_peekskill} to determine the GS of $\mathcal{\hat{H}_{\text{QCD}}}$ for $x = 0.8$ and $\tilde{m} \in [-1, 1]$.
Since Eq.~\eqref{eq: qcdham} contains only Pauli strings with $\hat{X}$ and $\hat{Z}$, our circuit ansatz comprises a tunable $\hat{\mathcal{P}}^{(3)}$ gadget, $\hat{\mathcal{P}} = (\hat{Z}$-{}-$\hat{X})^{\otimes 3}$, with an external $RY$-$RZ$ layer placed after, see Fig.~\ref{fig: vqe_results}(c). 
Using the model's symmetries we reduce the number of free parameters from 10 ($3\times3+1$) to 4 (see SM~\cite{supp_mat}). We then calculate the quantities $\langle\hat{N}\rangle = \langle\hat{\mathcal{H}}_{m}\rangle/\tilde{m}$ and $\langle\hat{N'}\rangle = (1/3)\sum_{i < j}(1-\langle\hat{Z}\rangle_{i})(1-\langle\hat{Z}\rangle_{j})$ [$ i,j = 1,2,3$] for different values of $\tilde{m}$ to analyze transitions in the mean occupation number~\cite{atas2022su3}, see Fig.~\ref{fig: vqe_results}(c). 
\par
In the negative (positive) mass limits $(|x/\tilde{m}| < 1)$, $\langle\hat{N}\rangle$ and $\langle\hat{N'}\rangle$ attain their maximal (minimal) values, indicating the presence (absence) of quarks and antiquarks on all lattice sites. These correspond to the baryonium (vacuum) state. The state transitions (at intermediate occupation numbers) between the two extremes occur when $\tilde{m} \in [-0.2, 0.2]$, in accordance with Ref.~\cite{atas2022su3}.
Fig.~\ref{fig: vqe_results}(c) also exhibits good agreement with exact values in both large mass limits $|\tilde{m}| > 0.2$ $(\mathcal{F} \geq 0.834$; $|\Delta E/E_{g}| \leq 9.92 \times 10^{-2})$, and in the transition region $|\tilde{m}| \leq 0.2$ $(\mathcal{F} \geq 0.77$; $|\Delta E/E_{g}| \leq 3.11 \times 10^{-1})$. The lower fidelity in the latter is attributed to smaller energy gaps between the ground and first excited states. We note that a single gadget with local rotations suffices to resolve the gaps while maintaining $\mathcal{F} > 0.75$. These observations highlight the suitability of Pauli gadgets in VQE problems involving many-body interactions.
\par
\textit{2D gauge theory:} Next, we analyze a pure (matterless) $\mathbb{Z}_2$ lattice gauge theory. The lattice is analogous to that of the PC, with two-level gauge fields on the edges [Fig.~\ref{fig: vqe_results}(a; i)]. The Hamiltonian has electric and magnetic contributions, $\mathcal{\hat{H}}_{\mathbb{Z}_{2}} = \lambda \hat{\mathcal{H}}_{\square} + \hat{\mathcal{H}}_{\triangle}$, 
with coupling strength $\lambda$. $\hat{\mathcal{H}}_{\square}$ and $\hat{\mathcal{H}}_{\triangle}$ are defined in Eq.~\eqref{eq: pcham} with $\xi = \lambda^{-1}$  ~\cite{horn1979zn}.
We performed a VQE simulation on \texttt{ibm\_peekskill} for a single plaquette  [Fig.~\ref{fig: vqe_results}(d)], and plotted the normalized plaquette expectation value 
$\langle\hat{\mathcal{H}}_{\square}\rangle/n_{\mathrm{p}}$ and $E$-field operator $\langle\hat{\mathcal{H}}_{\triangle}\rangle/n_{\mathrm{q}}$ for $0.5 \leq \lambda \leq 3.3$, where ($n_{\mathrm{p}}$, $n_{\mathrm{q}}$) = $(4, 1)$. Since Eq.~\eqref{eq: pcham} is the sum of a plaquette $\hat{X}\hat{X}\hat{X}\hat{X}$ and local $\hat{Z}$ terms, our circuit ansatz comprises a ${\update\hat{\mathcal{P}}^{(4)}}$ gadget; $\mathcal{\hat{P}} = \hat{X}^{\otimes 4}$, with external $RY$ and $RZ$-$RY$ layers placed before and after, respectively. The model symmetries allow reduction of the number of circuit parameters from 13 ($3\times4+1$) to 4 (see SM~\cite{supp_mat}). \par                  
The target plot in Fig.~\ref{fig: vqe_results}(d) exhibits a competition between $\langle\hat{\mathcal{H}}_{\triangle}\rangle/4$ and $\langle\hat{\mathcal{H}}_{\square}\rangle$, with the former (latter) dominating in the small (large) $\lambda$ regime. The crossover regime around $\lambda = 2$ is well captured by the VQE, as evidenced by $0.768 \leq \mathcal{F} \leq 0.862$ and consistent resolution of the energy gap ($|\Delta E/E_{g}| \leq 0.573$)~\footnote{
The VQE runs for $\lambda \geq 2.65$ yielded $|\Delta E/E_{g}| > 0.9$, indicating measured energies close to or above $E_{1}$. However, this is expected given the near-degeneracy ($E_{g} \leq 0.2349$) at these points and the limited resource budget.
}.
\par
\textit{Quantum chemistry:} Lastly, we showcase the suitability of Pauli gadgets in simulating molecular systems. We determine the GS of the LiH molecule, following Ref.~\cite{kandala2017hardware} to obtain a reduced $4$-qubit Hamiltonian consisting of $99$ Pauli terms (see SM~\cite{supp_mat}). 
We executed a VQE on \texttt{ibm\_perth} at an interatomic distance $R = 1.6$ \AA. With $\hat{Z}\hat{Z}\hat{Z}\hat{Z}$ the largest amplitude 4-body term, our ansatz consists of a $\mathcal{\hat{P}}^{(4)}$ gadget; $\mathcal{\hat{P}} = \hat{Z}^{\otimes 4}$, with external $RY$ and $RX$ layers placed before and after, respectively [Fig.~\ref{fig: vqe_results}(e)]. The prior elimination of redundancies in Ref.~\cite{kandala2017hardware} imposes that all 9 ($2\times4+1$) optimization parameters are necessary.
\par
The progress plots in Fig.~\ref{fig: vqe_results}(e) verify the effective performance of the ansatz, as the VQE converges within $1.1\%$ of $E_0 \approx -7.88$ Ha after $99$ function evaluations, with $\mathcal{F}$ improving from $0.067$ to $0.78$. The final error~\cite{adaptive_estimation} and fidelity contain contributions from not only hardware noise, but also the limitation of a single gadget in capturing all other interaction terms in the Hamiltonian. We study a use-case for multiple gadgets in the SM~\cite{supp_mat} involving numerical VQE simulations of the strongly correlated rectangular $\mathrm{H}_{4}$ molecule.
\par
\textit{\textbf{Conclusions and Outlook.}}---
We developed a new approach to VQEs, exploiting MB-VQE ideas to design and implement circuit ansatzes. While purely MB approaches possess high error thresholds~\cite{raussendorf2007fault, zwerger2013mbthreshold, zwerger2014hybrid, zwerger2015mb} and permit more efficient implementations of certain operations~\cite{gottesman1999universal, raussendorf_owqc, owqc_old_ref, browne2006one, briegel2009measurement, briegel2001cluster, terhal2004adaptive, hoyer2005fanout, bravyi2005universal, browne2011mbdepth, jozsa2013simcomplexity, paetznick2014repeat, takahashi2016depth, zhu2020depth, ferg2021_mbvqe, piroli2021locc, verresen2022lre, tantivasadakarn2022lre, liu2022depth, bravyi2022adaptive, miguelramiro2023sqem, miguelramiro2023enhancing}, circuit approaches are more flexible, with high-fidelity implementations on a variety of physical platforms.

Our approach is showcased via two techniques: an ansatz modification scheme to design circuits with tunable entanglement structures, and variational mid-circuit measurements using Pauli gadgets that efficiently implement multi-qubit operations. By successfully capturing GS properties with high fidelities, our VQE simulations on IBM Quantum confirm the viability of MB-infused circuits. Our use of mid-circuit measurements is timely given their availability on multiple platforms~\cite{stricker2020midcircti, corcoles2021midcircsc, pino2021midcircti, gaebler2021midcircti, deist2022midcircneutral, graham2023midcircneutral, sainath2023midcircti}, with their classical control of operations fundamentally redefining how one delegates computational resources for hybrid algorithms.

Our VQE techniques can be extended to the broader class of variational quantum algorithms~\cite{cerezo2021vqa}, e.g. in quantum neural networks and quantum machine learning~\cite{kwak2021neural, abbas2021neural, kyriienko2021generalized, garcia2022qml, kyriienko2022protocols, liu2022noise, liu2022representation, zheng2023sncqa, Paine2023}. In addition, the Pauli gadget may be used to efficiently simulate real-time evolutions~\cite{lloyd1996universal, lanyon2011universal, salathe2015digital, martinez2016real, atas2022su3}. Finally, it will be interesting to develop novel readout schemes using such gadgets to improve the measurement efficiency for many-body observables~\cite{mcclean2016theory, huang2022qml, adaptive_estimation}.
\par   

\section*{Acknowledgments} We thank PINQ$^2$ and CMC Microsystems for providing access to the IBM Quantum platform, Alexandre Choquette, Sarah Sheldon, Sieglinde M.-L. Pfaendler, Winona Murphy, Matthew Stypulkoski, and Udson Mendes for their support, and three anonymous Referees for their invaluable comments that greatly improved this manuscript. LD acknowledges the EPSRC Quantum Career Development grant EP/W028301/1. This work was supported by the Austrian Science Fund (FWF) through projects No. P36009-N and P36010-N. Finanziert von der Europäischen Union - NextGenerationEU. This work has been supported by the Natural Sciences and Engineering Research
Council of Canada (NSERC), New Frontiers in Research Fund (NFRF), Ontario Early Researcher Award, and the Canadian Institute for Advanced Research (CIFAR). This research was undertaken thanks in part to funding from the Canada First Research Excellence Fund. We also thank Mike and Ophelia Lazaridis and Innovation, Science and Economic Development Canada (ISED).

\bibliography{Bibliography}
\end{document}


\title{\textcolor{black}{Supplemental Material for: Measurement-Based Infused Circuits for Variational Quantum Eigensolvers}}
\begin{abstract} 
 In this Supplementary Material, we provide further details of how the VQE demonstrations on IBM Quantum systems were conducted on the hardware (quantum) and numerical (classical) levels. We also present concepts relevant to our work, and additional simulation data to highlight the scalability of our techniques to larger systems. 
\end{abstract}
\author{Albie Chan}
    \email{a233chan@uwaterloo.ca}
    \thanks{this author contributed equally}
    \affiliation{Institute for Quantum Computing and Department of Physics \& Astronomy, University of Waterloo, Waterloo, Ontario, N2L 3G1, Canada}
\author{Zheng Shi}
    \thanks{This author contributed equally}
    \affiliation{Institute for Quantum Computing and Department of Physics \& Astronomy, University of Waterloo, Waterloo, Ontario, N2L 3G1, Canada}
\author{Luca Dellantonio}
    \affiliation{Institute for Quantum Computing and Department of Physics \& Astronomy, University of Waterloo, Waterloo, Ontario, N2L 3G1, Canada}
    \affiliation{Department of Physics and Astronomy, University of Exeter, Stocker Road, Exeter EX4 4QL, United Kingdom}
\author{Wolfgang D\"ur}
    \affiliation{Universität Innsbruck, Institut für Theoretische Physik, Technikerstraße 21a, 6020 Innsbruck, Austria}
\author{Christine A. Muschik}
    \affiliation{Institute for Quantum Computing and Department of Physics \& Astronomy, University of Waterloo, Waterloo, Ontario, N2L 3G1, Canada}
    \affiliation{Perimeter Institute for Theoretical Physics, Waterloo, Ontario, N2L 2Y5, Canada}
    \maketitle  
\newpage
\section{Overview of measurement-based quantum computation}
\label{app: mbqc}
In this section, we give a more detailed description of the MB-QC protocol. At the outset, one prepares an entangled resource graph or cluster state containing input ($I$), body ($B$) and output ($O$) qubits connected via $\text{CZ}$-edges. The body and output qubits are initialized in $|+\rangle$, while the input qubits encode the input state $|\psi_{\mathrm{in}}\rangle$. The initial state of the resource $|G\rangle$ is therefore
\begin{equation}
|G\rangle = \prod_{\{i, j\} \in E} CZ_{i,j}\bigg(|\psi_{\mathrm{in}}\rangle_{I} \bigotimes_{m \in \{B, O\}} |+\rangle_{m}\bigg),
\end{equation}
where $E$ is the set of prescribed edges. This is a $+1$ eigenstate of the operators $\hat{S}_{n}$ (i.e. $\hat{S}_{n}|G\rangle = |G\rangle$), where
\begin{equation}
\hat{S}_{n} = \hat{X}_{n}\prod_{k \in \mathcal{N}(n)}\hat{Z}_{k},
\label{EA1}
\end{equation}
and $\mathcal{N}(n)$ are the neighbours of qubit $n$ (i.e. those connected to qubit $n$ via a $\text{CZ}$-edge). \par
The main computation is driven by a series of projective measurements on the input and body qubits, typically in the eigenbasis of the Pauli operators $\hat{X}$, $\hat{Y}$, $\hat{Z}$, or the rotated basis
$R(\theta) = \Big\{\frac{|0\rangle + e^{i\theta}|1\rangle}{\sqrt{2}}, \frac{|0\rangle - e^{i\theta}|1\rangle}{\sqrt{2}}\Big\}$.
Conventionally, all input qubits $I$ are chosen to be measured in the $X$-basis~\cite{raussendorf_owqc}. Each measurement applied to a qubit removes it from the resource graph and modifies the remaining connections depending on the prescribed basis. Concurrently, $|\psi_{\mathrm{in}}\rangle$ is propagated through each body qubit (known as feedforward), and upon reaching the output qubits it will have undergone a linear transformation 
\begin{equation}
V: |\psi_{\mathrm{in}}\rangle \rightarrow |\psi_{\mathrm{out}}\rangle: \mathcal{H}_{I} \rightarrow \mathcal{H}_{O}, 
\label{EA2}
\end{equation}
where $\mathcal{H}_{I}$ ($\mathcal{H}_{O}$) denotes the Hilbert space spanned by the input (output) qubits, and  $|I| = |O|$ for a unitary transformation. To ensure the result $|\psi_{\mathrm{out}}\rangle$ is deterministic, i.e. the same for each possible set of random measurement outcomes, corrective operations are applied. These comprise byproduct Pauli operators $\hat{U}_{\Sigma} \in \{\hat{I}, \hat{X}, \hat{Y}, \hat{Z}\}^{\otimes |O|}$ acting on $|\psi_{\mathrm{out}}\rangle$ and angle adaptations $M_{i}(R(\theta)) \rightarrow M_{i}(R(\pm\theta))$ for qubits $i$ measured in non-Clifford bases (i.e. $\theta \not\equiv 0 \hspace{0.1cm} \text{mod} \hspace{0.1cm} \pi/2$).
The angle adaptations require knowledge of prior outcomes, implying that a temporal (causal) order for measuring the adaptive qubits must exist for a pattern to yield deterministic results: an adaptive qubit may be measured only after all measurements that it depends on have taken place. The projective measurements ($\hat{P}$), the byproduct operators and the angle adaptations form an MB \textit{pattern} realizing $V$, with
\begin{equation}
 |\psi_{\mathrm{out}}\rangle = V|\psi_{\mathrm{in}}\rangle = \hat{U}_{\Sigma}\hat{P}|G\rangle.
 \label{EA3}
\end{equation}
\par
\begin{figure}
    \centering \includegraphics[width=1\textwidth]{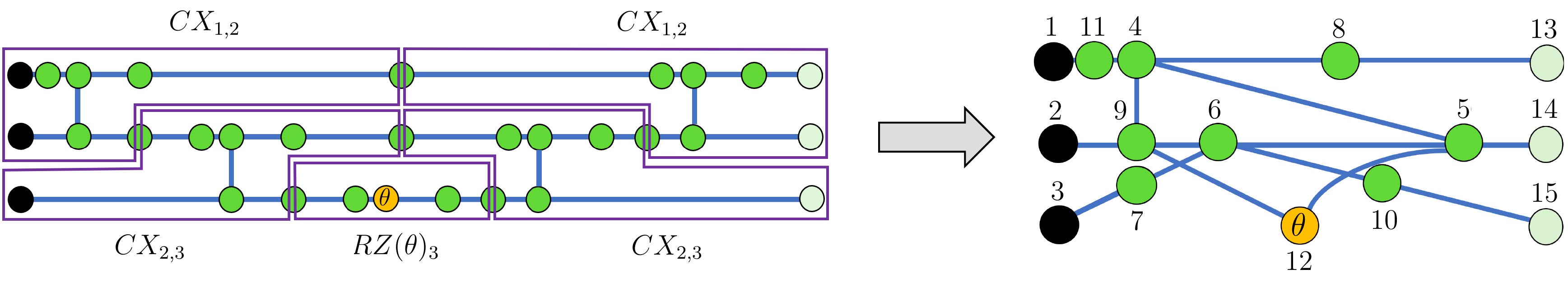}
    \caption[MB pattern for $\hat{Z}^{\otimes 3}(\theta)$]{\textbf{MB pattern for $Z^{\otimes 3}(\theta)$, obtained by concatenating smaller patterns corresponding to each operation in Eq.~\eqref{EA4}.} To reduce the number of initial qubits, the pattern is simplified via the techniques of Ref.~\cite{duncan2020zx}. Black and light green circles denote input and output qubits respectively, blue lines denote $\text{CZ}$-edges, green circles denote nonadaptive qubits measured in the $X$-basis, and orange circles denote adaptive qubits measured in the $R(\theta)$-basis. The numbers label the qubits and their measurement order (excluding outputs).}
    \label{fig:mbqc_pattern}
 \end{figure}

Elementary one- and two-logical qubit gates correspond to well-known patterns~\cite{raussendorf_owqc}, which are sufficient for universal computation. An arbitrary gate sequence may be achieved by concatenating each output of a pattern to the corresponding input of the next~\cite{raussendorf_owqc}.
\\ \\
\textit{\ul{Example}}. To illustrate the MB-QC procedure, we will derive and verify the MB pattern corresponding to the $\hat{Z}^{\otimes 3}(\theta)$ operation. In the circuit model, it realizes the unitary
\begin{align}
\begin{split}
 \hat{U}_{\text{circ}}(\theta) &= CX_{1,2}CX_{2,3}RZ_{3}(\theta)CX_{2,3}CX_{1,2} \\
&= \mathrm{diag}(e^{-i\theta/2}, e^{i\theta/2}, e^{i\theta/2}, e^{-i\theta/2}, e^{i\theta/2}, e^{-i\theta/2}, e^{-i\theta/2}, e^{i\theta/2}). 
\label{EA4}
\end{split}
\end{align}
From the gate sequence in Eq.~\eqref{EA4}, we may construct an equivalent MB pattern via concatenation, which yields a 31-qubit pattern [Fig.~\ref{fig:mbqc_pattern}]. (Note CX can be expressed in terms of CZ and Hadamards.)
To reduce the number of initial qubits, we apply graph-theoretic simplifications to the pattern (Ref.~\cite{duncan2020zx}), which enables a reduction from 31 to 15 qubits. The simplified pattern is shown in Fig.~\ref{fig:mbqc_pattern}, where the edges are
\begin{align}
\begin{split}
E = \hspace{0.05cm} &\{(4, 5), (6, 7), (13, 8), (9, 4), (9, 12), (12, 5), (5, 14), (15, 10), (4, 11), (1, 11), (6, 5), (3, 7), (9, 6), (6, 10), (4, 8), (2, 9)\}.
\label{EA6}
\end{split}
\end{align}
The relation between input and output states is therefore
\begin{equation}
|\psi_{\mathrm{out}, {\text{MB}}}\rangle_{13,14,15} = \hat{P}\hat{U}_{\Sigma}|G\rangle_{1,2,...,15} = \hat{P}\hat{U}_{\Sigma}\bigg[\prod_{\{i, j\} \in E} CZ_{i,j}\bigg(|G_{\psi_{\text{in}, {\text{MB}}}}\rangle_{1,2,3} \otimes |+\rangle^{\otimes 12}_{4,5,...,15}\bigg)\bigg]
 \label{EA7},
\end{equation}
where $|G\rangle$ is the term in square brackets. By convention, we measure the input qubits first, followed by the non-adaptive (Clifford) and adaptive (non-Clifford) qubits. Since the inputs and non-adaptive qubits are not affected by the temporal ordering, they may be measured simultaneously. To calculate $\hat{U}_{\Sigma}$, we solve eigenvalue equations (Eq.~(82) of Ref.~\cite{raussendorf_owqc}) for each qubit $i$. These equations are a direct consequence of Eq.~\eqref{EA1}, and determine whether to apply $\hat{X_{i}}, \hat{Z_{i}}$ or both on qubit $i$ if it measures 1 (denoted by $s_{i}$). A similar step is performed to solve for the adaptive corrections on qubit 12, which indicate which outcomes of qubit 1--11 require the measurement angle of qubit 12 to be flipped (i.e. $\theta \rightarrow -\theta$). Here, we must obey the temporal ordering by considering \textit{only} qubits that have been measured up to that point (i.e. qubits 1--11). For the pattern in Fig.~\ref{fig:mbqc_pattern}, we find the projective measurements and byproducts to be
\begin{equation}
\hat{P} = M_{12}\big({R({\theta(-1)^{s_{5}+s_{8}+s_{10}})\big)\prod_{k=1}^{11}}}M_{k}(X);
\label{EA8}
\end{equation}
\begin{equation}
\hat{U}_{\Sigma} = \hat{Z}_{15}^{s_{3}+s_{6}+s_{12}}\hat{X}_{15}^{s_{7}+s_{10}}\hat{Z}_{14}^{s_{2}+s_{12}}\hat{X}_{14}^{s_{5}+s_{9}}\hat{Z}_{13}^{s_{1}+s_{4}+s_{12}}\hat{X}_{13}^{s_{8}+s_{11}},
\label{EA9}
\end{equation}
where $M_{k}$ is the measurement operator on qubit $k$ and $s_{k} \in \{0,1\}$ its corresponding outcome. Because of the outcome dependence of $\hat{U}_{\Sigma}$, they must be applied \textit{after} the measurements and therefore require commutation over the operations in Eq.~\eqref{EA4}, which $V$ implements. Here, we note that the diagonal form of the unitary implies that $\hat{U}_{\Sigma}$ remains unchanged after commutation~\cite{browne2006one}; however, the adaptive corrections on qubit 12 are affected. Eqs.~\eqref{EA7} and \eqref{EA8} become
\begin{equation}
|\psi_{\mathrm{out}, {\text{MB}}}\rangle_{13,14,15} = \hat{U}_{\Sigma}\hat{P'}|G\rangle_{1,2,...,15};
\end{equation}
\begin{equation}
\hat{P}' = M_{12}\big(R{(\theta(-1)^{s_{7}+s_{9}+s_{11}})\big)\prod_{k=1}^{11}}M_{k}(X),
\label{EA10}
\end{equation}
where we have used the commutation relations in Eqs.~(72) and (73) of Ref.~\cite{raussendorf_owqc}. 
For simplicity, we consider a scenario where each qubit of the input state is a Pauli eigenvector $|+\rangle$; however, any valid measurement pattern can process an arbitrary input. \par
One may now proceed and perform Eq.~\eqref{EA10} directly. However, since the majority of qubits (1--11) are non-adaptive, they can be efficiently simulated classically via the stabilizer formalism of Ref.~\cite{improved_sim}. To do this, we employ the tableau representation of stabilizer states and track the input state as it undergoes Clifford operations and measurements. For all qubits initialized in $|+\rangle$, the stabilizer state is given by
\begin{equation}
\{(+\hat{X}_{1})...(+\hat{X}_{15})\}.
\end{equation}
After performing all $\text{CZ}$s and non-adaptive ($X$-basis) measurements in Eq.~\eqref{EA10}, the canonical form stabilizers over the remaining adaptive and output qubits (assuming an outcome of 0 for qubits 1 to 11) are
\begin{equation}
\{(+\hat{Z}_{1}), (+\hat{Z}_{2}),..., (+\hat{Z}_{11}), (+\hat{X}_{12}\hat{X}_{15}), (+\hat{Z}_{12}\hat{Z}_{13}\hat{Z}_{14}\hat{Z}_{15}), (+\hat{X}_{13}\hat{X}_{15})\},
\label{EA12}
\end{equation}
where elementary row operations were used to reduce each of qubits 1--11 to a single $+Z$. These stabilizers, corresponding to the post-measured state $|0\rangle$ may be subsequently traced out from the overall state.
Using the Gottesman-Knill theorem~\cite{improved_sim}, we then work out a unitary $\hat{U}_{\mathrm{Cliff}}$ that produces a \textit{reduced} graph state stabilized by the operators above. To achieve the form of Eq.~\eqref{EA1} (which describes the connectivity of graphs), we express Eq.~\eqref{EA12} as a generator matrix over qubits 12-15 and reduce the stabilizer ($\hat{X}$) portion to row-echelon form, where each manipulation incurs a local Clifford (LC) operation $H$ or $S$ (see Ref.~\cite{improved_sim}). The resulting stabilizers and circuit are
\begin{flalign}
\begin{blockarray}{cccccc}
\begin{block}{cc[cccc]}
\hat{Z}_{1} & & 0 & 1 & 0 & 0 \\
\hat{Z}_{2} & & 0 & 1 & 0 & 0 \\
\hat{Z}_{3} & & 0 & 1 & 0 & 0 \\ 
\hat{Z}_{4} & & 0 & 1 & 0 & 0 \\ 
\hat{X}_{1} & & 1 & 0 & 0 & 0 \\
\hat{X}_{2} & & 0 & 0 & 1 & 0 \\
\hat{X}_{3} & & 0 & 0 & 0 & 1 \\
\hat{X}_{4} & & 1 & 0 & 1 & 1 \\
\end{block} 
\end{blockarray}
\hspace{0.1cm} &\xrightarrow{H_{4}} \hspace{0.1cm} 
\begin{blockarray}{cccccc}
\begin{block}{cc[cccc]}
\hat{Z}_{1} & & 0 & 0 & 0 & 1 \\
\hat{Z}_{2} & & 0 & 0 & 0 & 1 \\
\hat{Z}_{3} & & 0 & 0 & 0 & 1 \\ 
\hat{Z}_{4} & & 1 & 1 & 1 & 0 \\
\hat{X}_{1} & & 1 & 0 & 0 & 0 \\
\hat{X}_{2} & & 0 & 1 & 0 & 0 \\
\hat{X}_{3} & & 0 & 0 & 1 & 0 \\
\hat{X}_{4} & & 0 & 0 & 0 & 1 \\
\end{block} 
\end{blockarray} \\
\begin{split}
\hspace{0.1cm} &\rightarrow
\{(+\hat{X}_{1}\hat{Z}_{4}), (+\hat{X}_{2}\hat{Z}_{4}), (+\hat{X}_{3}\hat{Z}_{4}), (+\hat{X}_{4}\hat{Z}_{1}\hat{Z}_{2}\hat{Z}_{3})\}  
\label{EA13} 
\end{split} \\
\begin{split}
&\implies \hat{U}_{\mathrm{Cliff}} = H_{4}CZ_{1,4}CZ_{1,3}CZ_{1,2},
\label{EA14}
\end{split}
\end{flalign}
where $H_{4}$ is a LC-operation and we have relabelled qubits 13, 14, 15, 12 as 1, 2, 3, 4 respectively. Consequently, $V$ reduces to a 4-qubit (3 + 1 ancilla) pattern $V_{\text{red}}$, and we have
\begin{flalign}
\begin{split}
 |\psi_{\mathrm{out}, {\text{MB}}}\rangle_{1, 2, 3} &= V_{\text{red}}\hat{U}_{\mathrm{Cliff}}|+\rangle^{\otimes 4}_{1,2,3,4} \\
 &= (\hat{Z}_{3}\hat{Z}_{2}\hat{Z}_{1})^{s_4}M_{4}{\big(R( \theta(-1)^{s_{4}})\big)}\hat{U}_{\mathrm{Cliff}}|+\rangle^{\otimes 4}_{1,2,3,4}.
 \end{split}
\end{flalign}
\label{EA15}One observes that the graph corresponding to Eq.~\eqref{EA13} is a three-pointed star, which is the $n=3$ case of the $\hat{Z}^{\otimes n}$ graph shown in Fig.~1(b).
If qubit 4 is measured with outcome 1 ($s_{4} = 1$), we obtain the final output
\begin{flalign}
\begin{split}
 |\psi_{\mathrm{out}, {\text{MB}}} \rangle_{1, 2, 3} &= {\frac{1}{\sqrt{8}}\hat{Z}_{3}\hat{Z}_{2}\hat{Z}_{1}\Big[e^{-i\theta/2}\mathlarger{\sum}_{x \in \{b_{1}...b_{3} \in \vec{b} |b_{1}+...+b_{3}=0\}}|x\rangle -e^{i\theta/2}\mathlarger{\sum}_{x' \in \{b_{1}...b_{3} \in \vec{b}  |b_{1}+...+b_{3}=1\}}|x'\rangle\Big]} \vspace{1.5cm}
\\ &= {\frac{1}{\sqrt{8}}\Big[e^{-i\theta/2}\mathlarger{\sum}_{x \in \{b_{1}...b_{3} \in \vec{b} |b_{1}+...+b_{3}=0\}}|x\rangle +e^{i\theta/2}\mathlarger{\sum}_{x' \in \{b_{1}...b_{3} \in \vec{b}|b_{1}+...+b_{3}=1\}}|x'\rangle\Big]},
\label{EA16}
\end{split}
\end{flalign}
where $\vec{b} = \{0,1\}^{\otimes 3}$. It is straightforward to verify that this state is equivalent (up to a global phase) to  $\hat{U}_{\text{circ}}$ acting on 
$|+\rangle^{\otimes 3}$.  \par
In summary, {\text{MB}}-QC constitutes a natural method of computation for platforms that implement highly entangled graph or cluster states, including photonic and spin qubits. Despite the larger number of ancillary qubits employed in comparison to the  circuit model, the method is beneficial when the resource graph contains a large Clifford portion, as shown in the example above. By performing the majority of the pattern classically, one drastically reduces the size of the resource graph to be implemented. On a practical level, this can yield fewer computational steps, and in noisy cases, opportunities for the overall resource to decohere.
\newpage
\section{VQE demonstrations --- methods and techniques} \label{app: methods}
\begin{figure}
    \centering \subfloat{\includegraphics[width=1\textwidth]{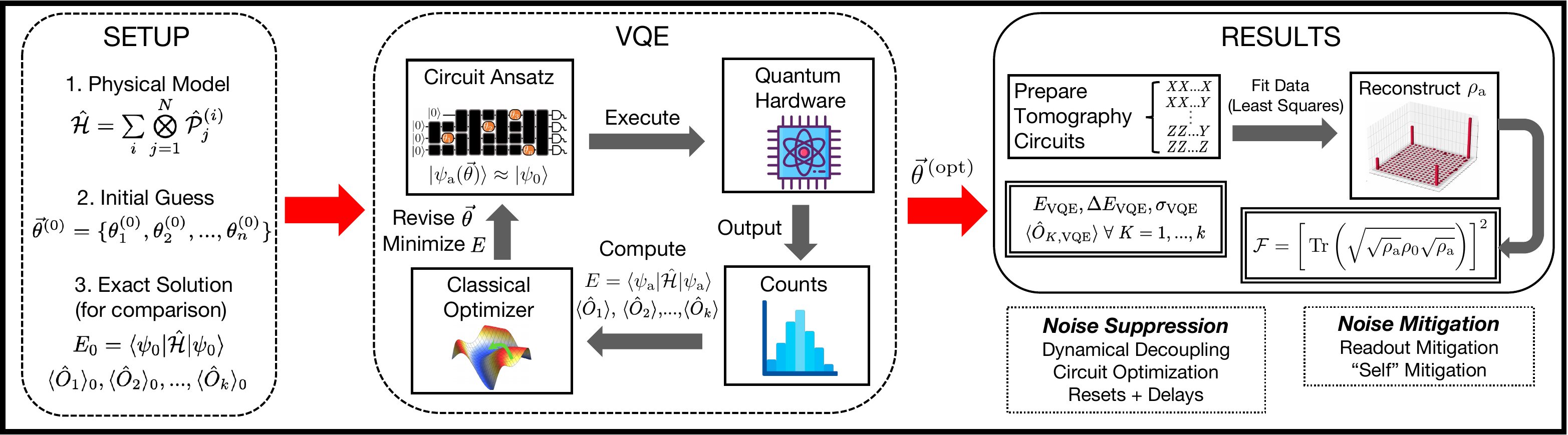}}
    \caption[Workflow for the numerical VQE program]{\textbf{Workflow for the numerical VQE program.} Readout mitigation and all noise suppression techniques are employed during execution of each variational circuit, while self-mitigation occurs immediately after $\vec{\theta}\hspace{0.05cm}^{(\text{opt})}$ is determined.}
    \label{fig:vqe_program}
 \end{figure}

In this section, we discuss the numerical program and associated methods used to perform VQE demonstrations on IBM Quantum systems. We also detail the error mitigation and suppression techniques implemented to combat noise and decoherence effects prevalent on real hardware. A schematic of the entire program is depicted in Fig.~\ref{fig:vqe_program}. \par  
\textit{\ul{Setup.}} The program takes in three main inputs, which are specified by the user:
\begin{enumerate}
\item{A target Hamiltonian $\hat{\mathcal{H}}$ describing the physical model, expressed as a linear combination of Pauli terms
\begin{equation}
\bigotimes_{j=1}^{n}\mathcal{\hat{P}}_{j}; \hspace{0.3cm}\mathcal{\hat{P}}_{j} \in \{\hat{I},\hat{X},\hat{Y},\hat{Z}\} \ \forall j.
\label{eq: pauliham}
\end{equation}
}
\item{A parameterized circuit ansatz that outputs $|\psi_{\mathrm{a}}{(\theta)}\rangle$ to compute the mean energy $E = \langle\psi_{\mathrm{a}}(\theta)|\hat{\mathcal{H}}|\psi_{\mathrm{a}}(\theta)\rangle$.}
\item{An initial guess for the variational parameters ($\vec\theta^{(0)}$).}
\end{enumerate}
 To gauge VQE performance, we use the above to calculate the exact GS ($|\psi_{0}\rangle$) and GS energy $E_{0} = \langle\psi_{0}|\mathcal{\hat{H}}|\psi_{0}\rangle$ via exact diagonalization. We note that this is feasible only for small to intermediate system sizes. \par
 \textit{\ul{Measurement protocol}}. To calculate the total energy, we group all observables in $\hat{\mathcal{H}}$ into $n$ \textit{bitwise} commuting sets $g_{n}$, where all bitwise commuting observables within a set are measured \textit{simultaneously} by a single circuit (for a total of $n$; see Sec.~\ref{app: err} for details). For example, in the $\mathbb{Z}_{2}$ demonstration, we require two circuits --- one measuring $X_{1}X_{2}X_{3}X_{4}$ and the other measuring $Z_{1}Z_{2}Z_{3}Z_{4}$, since they commute bitwise with all terms in $\hat{\mathcal{H}}_{\square}$ and $\hat{\mathcal{H}}_{\triangle}$ respectively. \par
Given the iterative nature of VQE, a practical approach involves minimizing the allocated budget, which encompasses the number of optimizer iterations or function evaluations, and the total number of measurement shots over all circuit executions. While there are standard choices for the former, selecting an appropriate number for the latter while balancing performance is more complicated. To address this, we employ a tailored version of the approach developed in Ref.~\cite{adaptive_estimation} based on the size of the first energy gap $E_{g} = E_{1}-E_{0}$. We regard $E_{g}$ as an indicator of the optimization difficulty in that the smaller it is, the more statistical counts are needed to resolve the gap. In this sense, we can relate $E_{g}$ to the standard deviation over all $i$ commuting sets $g_{i} \in \hat{\mathcal{H}}$, weighted by their respective number of shots $N_{g_{i}}$, i.e.
\begin{equation}
E_{g} \simeq \sqrt{\frac{\langle\hat{\mathcal{H}}_{g_{1}}\rangle^{2}}{N_{g_{1}}} + \frac{\langle\hat{\mathcal{H}}_{g_{2}}\rangle^{2}}{N_{g_{2}}} +...+\frac{\langle\hat{\mathcal{H}}_{g_{i}}\rangle^{2}}{N_{g_{i}}}}.
\label{EB1}
\end{equation}
Here, we assume that the variation due to $g_{i}$ is weighted by the sum of the $n$ coefficients corresponding to the $n$ Pauli terms in the group
\begin{equation}
\langle\hat{\mathcal{H}}_{g_{i}}\rangle^{2} = \sum_{n}{(g_{i}^{(n)})^{2}}.
\end{equation}
To minimize the LHS, we require that all terms in Eq.~\eqref{EB1} are equal: 
\begin{equation}
\frac{\sum{g_{1}}^{2}}{N_{g_{1}}} = 
\frac{\sum{g_{2}}^{2}}{N_{g_{2}}} =... 
=\frac{\sum{g_{i}}^{2}}{N_{g_{i}}}.
\end{equation}  
If we execute each circuit with the same number of shots, the problem reduces to optimizing the total number $N_{\mathrm{s}} = N_{g_{1}} + N_{g_{2}} +...+ N_{g_{i}}$. 
One can employ a classical optimizer to solve for each $N_{g_{i}}$; however, in the case of two groups, the problem may be solved directly and we find that
\begin{align}
 N_{g_{1}} &= R{N_{g_{2}}}; \hspace{0.3cm} R = {\frac{\sum{g_{1}}^{2}}{\sum{g_{2}}^{2}}} \\
\implies N_{g_{1}} &= \frac{R}{1+R}N_{\mathrm{s}} 
\implies E_{g} \simeq \sqrt{2\bigg(\frac{1+R}{RN_{\mathrm{s}} }\bigg)\sum{g_{1}^{2}}} \\
\implies N_{\mathrm{s}} \hspace{0.15cm} &\simeq \hspace{0.1cm} 2\bigg(\frac{1+R}{R(E_{g})^{2}}\bigg)\sum{g_{1}^2}
\label{EB7}.
\end{align}
We then extend Eq.~\eqref{EB7} directly to the general case by replacing the prefactor with a generic scaling factor $s$ that can be chosen depending on the range of shots desired. We also drop the weighting factor $(1+R)/R$ and match $N_{\mathrm{s}}$ to the commuting group with the largest coefficient sum. Thus, we have
\begin{equation}
N_{\mathrm{s}} = \bigg\lceil\frac{s}{(E_{g})^{2}}\max_{\substack{g \in G}}\bigg(\sum_{i}(c_{g}^{(i)})^{2}\bigg)\bigg\rceil, 
\label{EB8}
\end{equation}
where $G$ is the set of all commuting groups belonging to $\hat{\mathcal{H}}$ and $c_{g}^{(i)}$ is the coefficient of the $i$th Pauli in group $g$. Here, we impose that $2.5 \times 10^{2} \leq N_{\mathrm{s}} \leq 5 \times 10^{4}$, which ensures a sufficiently representative sample while avoiding large processing demands on the hardware. Any calculated $N_{\mathrm{s}}$ falling outside these ranges are set to the closer of the two limits. We remark that in view of constraints on IBM Quantum systems, we cannot implement the adaptive allocation of $N_{\mathrm{s}}$ as in Ref.~\cite{adaptive_estimation}, and must divide them equally between all groups of observables. However, we retain other advantages such as potential overlap between groups (i.e. the same observable is measured by multiple circuits), and the possibility of estimating not only the average value of the system energy, but also its variance.
\par
\textit{\ul{Readout mitigation}}. This is a standard technique to reduce errors from readout measurements~\cite{neeley2010generation,bialczak2010quantum}, which yield the counts data for calculating observable expectation values. The main principle involves capturing the extent of the readout errors via a \textit{calibration matrix} $M$. Each element in $M$ is the probability of initializing a computational basis state $|x\rangle$ and subsequently measuring $|x'_{\mathrm{meas}}\rangle$ (i.e. $|\langle x'_{\mathrm{meas}}|x\rangle|^2$). This requires the execution of $2^{n}$ basis circuits, where $n$ is the number of qubits in $\hat{\mathcal{H}}$. For example, $M$ has the following structure for $n=3$ (acquired from a sample run on \texttt{ibm\_perth}):
\begin{equation}
M = 
\begin{blockarray}{cccccccccc}
 & & |000\rangle  & |001\rangle & |010\rangle & |011\rangle & |100\rangle & |101\rangle & |110\rangle & |111\rangle \\
\begin{block}{cc(cccccccc)} 
\langle 000| &  & 0.95446 &  0.02772 &  0.01029 &  0.00000 &  0.00980 &  0.00000 &  0.00000 & 0.00000 \\ 
\langle 001| &  & 0.02178 &  0.95248 &  0.00000 &  0.01437 &  0.00000 &  0.00596 &  0.00000 &  0.00000 \\    
\langle 010| &  & 0.01386 &  0.00000 &  0.97942 &  0.01848 &  0.00000 &  0.00000 &  0.00648 & 0.00000\\ 
\langle 011| &  & 0.00000 &  0.01386 &  0.00617 &  0.95277 &  0.00000 &  0.00000 &  0.00000 &  0.00386\\ 
\langle 100| &  & 0.00594 &  0.00000 &  0.00000 &  0.00000 &  0.95686 &  0.02187 &  0.01296 &  0.00000\\
\langle 101| &  & 0.00000 &  0.00396 &  0.00000 &  0.00000 &  0.02157 &  0.94433 &  0.00000 &  0.01544\\
\langle 110| &  & 0.00000 &  0.00000 &  0.00000 &  0.00000 &  0.00784 &  0.00199 &  0.96544 &  0.00579\\
\langle 111| &  & 0.00000 &  0.00000 &  0.00000 &  0.01027 &  0.00000 &  0.02187 &  0.01296 &  0.96525\\
\end{block}
\end{blockarray} \hspace{0.3cm}.
\end{equation}
 For low noise levels, $M$ is expected to be predominantly diagonal and close to the identity (as shown above). Furthermore, if ancillary qubits are present (i.e. in the Pauli gadget demonstrations), we determine a separate calibration matrix over $n$ qubits for each possible set of ancillary outcomes. \par
If $C_{\text{ideal}}$ is the counts distribution corresponding to the exact $E_{0}$, then $MC_{\text{ideal}}$ leads to an approximation of the noisy distribution $C_\text{noisy}$. Naturally, we wish to perform the reverse. N\"aively, this is achieved by applying $M^{-1}C_\text{noisy}$, where $M^{-1}$ is the inverse of $M$; however, such an operation can yield negative counts (quasiprobabilities). We remedy this problem by performing a least-square fit to the closest valid distribution. After the calibration matrix is determined, it is stored and applied to the counts distribution from every subsequent circuit evaluation. This is most ideal when the hardware-specific parameters (i.e. error rates, decoherence times) are constant; however, as drifts in the parameters over time are expected, we determine $M$ more than once (i.e. after each data point) to ensure the errors are accurately captured throughout the course of the VQE run. In all demonstrations, we employ $10^{4}$ measurement shots for each basis circuit used to determine $M$. \par 
\textit{\ul{VQE algorithm}}. As shown in Fig.~\ref{fig:vqe_program}, the VQE proceeds as a closed feedback loop between the quantum processor and classical optimizer. 
All circuits are performed on IBM Quantum systems, with readout measurements chosen depending on the desired basis.
Prior to circuit execution, we apply various techniques to suppress noise:
\begin{enumerate}
\item{Performing the circuits requires one to specify a mapping between virtual and physical hardware qubits. Since the latter is restricted by the qubit topology, SWAP operations (each incurring three CXs) are added to match the connectivity of the considered circuit. With CXs being the most expensive operation on IBM Quantum systems, we choose mappings that lead to the fewest number of added SWAPs. At the time of the VQE run, we also ensure that the chosen qubits have reasonably small CX errors and long decoherence times.}
\item{Resets and delays are added to the beginning of each circuit. This improves the likelihood of each qubit being properly initialized in the ground state $|0\rangle$.}
\item{On the pulse level, we schedule circuits to incorporate dynamical decoupling (DD)~\cite{viola1998dd,viola1999dd,viola1999dd2}. This mitigation technique inserts extra pulses (equivalent to the identity) whenever a particular qubit is idle. Although it introduces further noise, the periodicity of the applied pulses enable undesired state rotations from system-environmental interactions to be reversed. Here, we utilize $X$-$X$ pulses which are commonly employed in spin-refocusing techniques to increase the coherence time of qubits~\cite{carr1954refocusing}.} \par
\end{enumerate}
 During the VQE run, the circuit ansatz is sent to and executed on the IBM Quantum system, which outputs results in the form of counts data. From the counts, we compute the total energy $E$ and contributions $\langle\hat{O}_{1}\rangle, \langle\hat{O}_{2}\rangle...\langle\hat{O}_{N}\rangle$ from specific observables of interest. The value of $E$ is then sent to a classical optimizer, which modifies $\vec{\theta}$ with the goal of minimizing $E$. The above process repeats until a stopping condition for convergence is met on the optimizer, at which point the optimized energy $E_{\text{VQE}}$ and parameters $\vec\theta^{({\text{opt}})}$ are returned. \par 
We employ the COBYLA and DIRECT methods for the classical optimization. COBYLA~\cite{COBYLA} enables a rapid and expansive search of the energy landscape with few function evaluations, while DIRECT~\cite{DIRECT} enables a more systematic search, identifying promising areas that may contain the global minimum. When using the latter, we follow the recommendation in Ref.~\cite{NLopt} and perform a second, local minimization (i.e. COBYLA) afterwards for a more refined search in the most promising region.  While SPSA is often recommended for noisy simulations~\cite{kandala2017hardware,liu2022noise}, we avoid its use as it generally requires large numbers of function evaluations to converge. Indeed, our circuits possess reasonably small depths and few parameters such that they achieve a low-enough level of noise for other optimizers to be effective. \par
\begin{figure}
       \centering \subfloat{\includegraphics[width=1\textwidth]{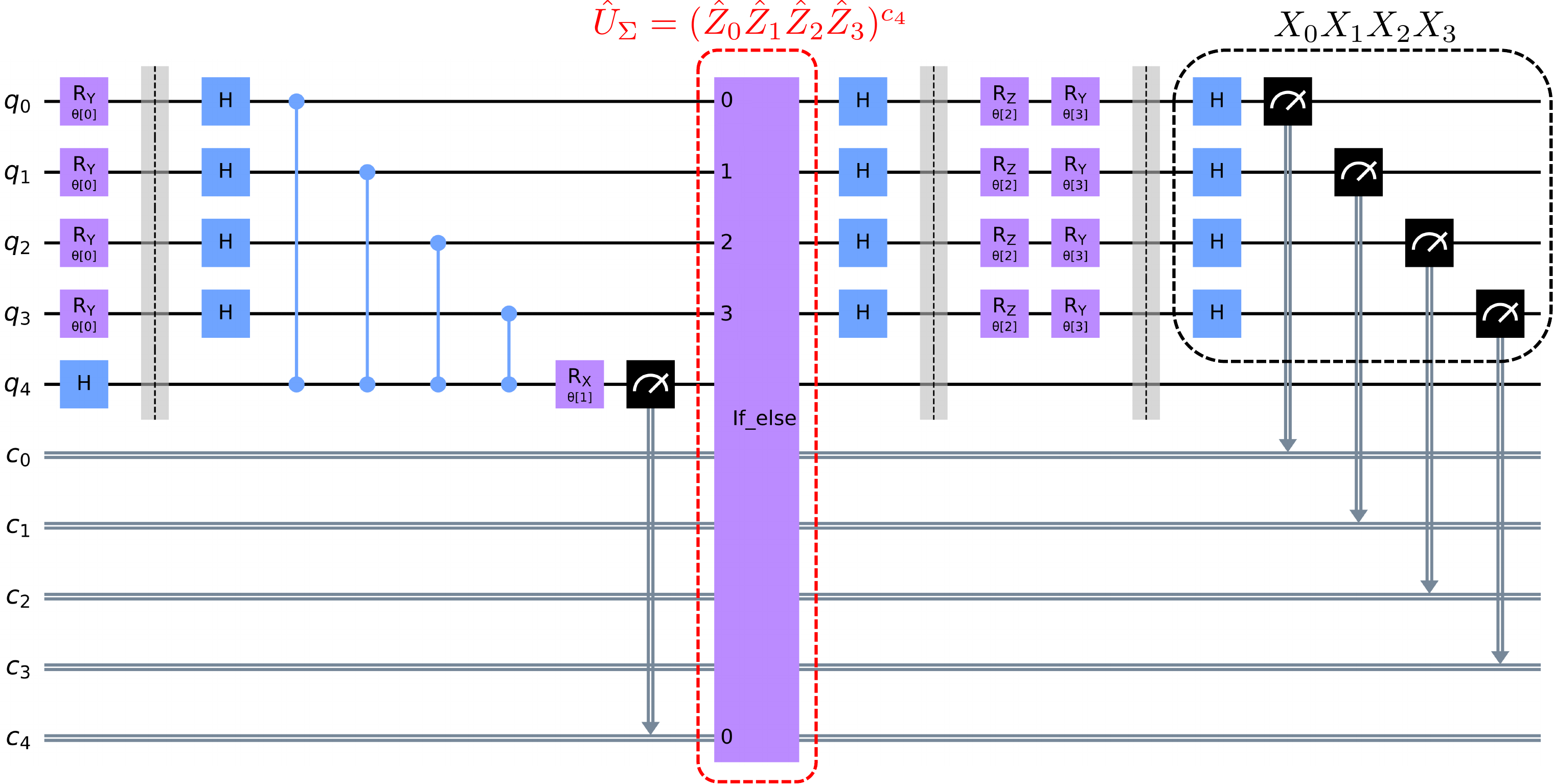}}
    \caption[Circuit ansatz used in the VQE demonstration of pure $\mathbb{Z}_{2}$ lattice gauge theory, single plaquette]{\textbf{Circuit ansatz used in the VQE demonstration of pure $\mathbb{Z}_{2}$ lattice gauge theory, single plaquette.} The qubits and corresponding classical registers for storing measurement outcomes are labelled by $q$ and $c$ respectively. After the mid-circuit measurement of the ancilla ($q_4$), the dynamic \texttt{if\_else} instruction (highlighted in red) implements the byproduct operators $\hat{U}_{\Sigma} =(\hat{Z}_{0}\hat{Z}_{1}\hat{Z}_{2}\hat{Z}_{3})^{c_{4}}$ corresponding to the $\hat{Z}^{\otimes 4}(\theta)$ operation (see Sec.~\ref{app: mbqc}). The measurement basis of the circuit is $X_{0}X_{1}X_{2}X_{3}$.}
    \label{z2_ansatz}
 \end{figure}

\textit{\ul{Dynamical circuits}}. All circuits employing Pauli gadgets are performed \textit{dynamically}. These circuits incorporate added \texttt{if\_test} instructions in the circuit, which apply corrective byproduct operations $\hat{U}_{\Sigma}$ on the main qubits based on the outcomes of the preceding measurements on the ancillary qubits~\footnote{The adaptive bases discussed in Sec.~\ref{app: mbqc} are not needed for the Pauli gadget, provided the byproduct operators are applied immediately after each gadget~\cite{browne2006one}.}. This enables one in practice to perform feedforward in a circuit (i.e. as opposed to clusters or graphs). By convention, the byproduct operator for a Pauli gadget only acts nontrivially when the ancilla measurement outcome is `1'. Fig.~\ref{z2_ansatz} depicts the dynamic circuit used in the $\mathbb{Z}_{2}$ demonstration. \par
As discussed in the main text, the use of Pauli gadgets in the form of mid-circuit measurements reduces the circuit depth on NISQ devices compared to cascaded CX gates. However, it is also important to note that this approach does not unambiguously improve the implementation of Pauli gadget unitaries on NISQ devices. This is because measuring the ancillas mid-circuit can introduce new sources of errors different from gate errors. For instance, they can incur delay times in which \textit{all} qubits are subject to noise and decoherence. This can hamper performance as the unmeasured qubits are intended to remain coherent throughout the computation. Furthermore, imperfections in performing these measurements can lead to incorrect outcomes being obtained, consequently affecting the counts statistics and proper application of byproduct operators. Therefore, the variational measurement approach introduces a trade-off between the reduced circuit depth on the one hand, and the additional sources of noise from mid-circuit measurements. Despite this, mid-circuit measurements have become increasingly high-quality~\cite{deist2022midcircneutral,pino2021midcircti,sainath2023midcircti} such that these points become less of a limiting factor towards performance. The net effect of this trade-off ultimately depends on the platform and the concrete form of the protocol. For example, reusing the ancillary qubits may increase the delay time, but one can combat this by using various DD schemes to improve the coherence of unmeasured qubits.\par
\textit{\ul{Self-mitigation}}. At the conclusion of the VQE run, we employ the self-mitigation technique~\cite{rahman2022self,atas2022su3}. It specifically addresses CX errors by evaluating the circuit ansatz in two ways. First, there is a {\update physics} run using the circuit parameterized with $\vec\theta^{(\text{opt})}$, then there is a mitigation run which involves a modified version of the {\update physics run} as a reference. In both runs, each CX is dressed with randomly chosen single-qubit {\update Pauli} operations (see Ref.~\cite{atas2022su3} for details). These do not alter the action of the CX (they only change the basis), but instead transform the CX noise from  being coherent to incoherent in nature. The correction acts as a scale factor on any given Pauli-string observable $\langle\hat{O}\rangle$, based on the results of evaluating the {\update physics} and mitigation runs. It is given by
\begin{equation}
    \langle \hat{O} \rangle_{\text{phys, true}} = \langle\hat{O}\rangle_{\text{phys, meas}} \times \Bigg(\frac{\langle \hat{O} \rangle_{\text{mitig, true}}}{\langle \hat{O} \rangle_\text{mitig, meas}}\Bigg)^{\kappa},
\label{EB9}
\end{equation}
where $\kappa$ is the ratio of $\text{CX}$ gate numbers between the {\update physics} and mitigation circuits, and ``true'' (``meas'') refers to the exact (noisy) expectation values. To minimize added resources, all observables that commute with the measurement basis of the circuit ansatz are corrected with the same factor. For simplicity, we also formulate our mitigation circuits so that $\kappa = 1$ and $\langle\hat{O}\rangle_\text{mitig, true}$ =  1. We achieve $\kappa = 1$ by including only the CXs present in the ansatz circuit~\cite{urbanek2021}. In both cases, a suitable initial state is chosen to ensure $\langle\hat{O}\rangle_\text{mitig, true}$ = 1, with LC-operations added where necessary.  \\ \par
\textit{\ul{State fidelity}}.
In the final step, we perform full state tomography to determine the ansatz state corresponding to $\vec\theta^{(\text{opt})}$. This requires the evaluation of $3^{n}$ circuits (i.e. all $n$-qubit combinations of $X$, $Y$, and $Z$ Paulis). In all demonstrations, we employ $10^{4}$  measurement shots for each circuit evaluation.
If ancillas are present, their counts are marginalized over the counts of the readout qubits.
The $3^{n}$ counts distributions are then fitted to a valid density matrix $\rho_{\mathrm{a}}$ using maximum likelihood estimation, which corresponds to the reconstructed ansatz state. This enables us to compute the state fidelity $\mathcal{F}$ between $\rho_{\mathrm{a}}$ and the exact GS ($\rho_{0}$) as
\begin{equation}
\mathcal{F}(\rho_{\mathrm{a}}, \rho_{0}) = \bigg[\operatorname{Tr}\bigg(\sqrt{\sqrt{\rho_{\mathrm{a}}}\rho_{0}\sqrt{\rho_{\mathrm{a}}}}\bigg)\bigg]^{2},
\end{equation}
where $\operatorname{Tr}$ is the trace operation. As with exact diagonalization, calculating $\mathcal{F}$ via tomography is practical only for small system sizes owing to the exponential scaling. 

\newpage
\section{VQE demonstrations --- supplementary data} 
\label{app: suppdat}
This section presents target Hamiltonians, ground state properties, {\update variational circuits, and raw data from the VQE demonstrations on IBM Quantum systems. All demonstrations employed single-layer ansatz modifications or single Pauli gadgets. \\ \\
\textit{\ul{Hamiltonians, GS properties, and variational circuits}} \\ \\
\textbf{PC}: {\update $\hat{\mathcal{H}}_{\text{PC}}^{(M=2, N=1)}$ defined in Eq.~\ref{eq:2x1PC}; variational circuit shown in Fig.~\ref{pc_ansatz}.}
\begin{flalign}
\begin{split}
\xi = 0&: |\psi_{0}\rangle = \frac{1}{2}(|0000000\rangle + |1111000\rangle + |0001111\rangle + |1110111\rangle); \\
\xi \rightarrow \infty&: |\psi_{0}\rangle = |1111111\rangle \hspace{0.2cm} 
\end{split}
\end{flalign}
For an approximation of $|\psi_{0}\rangle$ at intermediate $\xi$, refer to the perturbative analysis in Sec.~\ref{app: pc_perturb}. \\ 
\\
\textbf{1D QCD}: {\update $\hat{\mathcal{H}}_{\text{QCD}}$ defined in Eq.~(2); $|\psi_{0}\rangle$ difficult to determine analytically. For more details regarding model properties, refer to Ref.~\cite{atas2022su3}. Variational circuit shown in Fig.~\ref{qcd_ansatz}.} \\ \\
\textbf{Pure $\mathbb{Z}_{2}$ LGT, single plaquette}: Variational circuit shown in Fig.~\ref{z2_ansatz}. 
\begin{gather}
{\update \hat{\mathcal{H}}_{\mathbb{Z}_{2}} = \lambda\hat{\mathcal{H}}_{\square} + {\hat{\mathcal{H}}}_{\triangle}; \hspace{0.2cm} \hat{\mathcal{H}}_{\square} = \hat{X}_{1}\hat{X}_{2}\hat{X}_{3}\hat{X}_{4}; \hspace{0.2cm} \hat{\mathcal{H}}_{\triangle} = \frac{1}{\lambda}\sum_{i=1}^{4}\hat{Z}_{i}}; \\
|\psi_{0}\rangle = \sqrt{\frac{1}{2} + \frac{2}{\sqrt{16+\lambda^{4}}}}\bigg(\frac{4-\sqrt{16+\lambda^{4}}}{\lambda^{2}}|0\rangle^{\otimes 4} + |1\rangle^{\otimes 4}\bigg);\ \hspace{0.5cm}
E_{0} = -\sqrt{\frac{16}{\lambda^{2}} + \lambda^{2}} 
\end{gather} \\ 
\textbf{LiH}: Hamiltonian derived from Qiskit VQE tutorial~\cite{Qiskit,QiskitNature} (shown in Table~\ref{table:LiHham}), with parity transformation to map fermionic operators to qubits~\cite{bravyi2002parity}. State simplifications include freezing core orbitals~\cite{kandala2017hardware} to reduce the number of qubits from 12 to 4. Variational circuit shown in Fig.~\ref{lih_ansatz}.
\begin{equation}
\update{|\psi_{0}\rangle \approx 0.9877 \hspace{0.1 cm} |1100\rangle - 0.1154 \hspace{0.1 cm} |0011\rangle;} \hspace{0.3cm} E_{0} = -7.881072044030926
\end{equation} 
\begin{figure}[htb]
\centering\subfloat{\includegraphics[width=18.5cm]{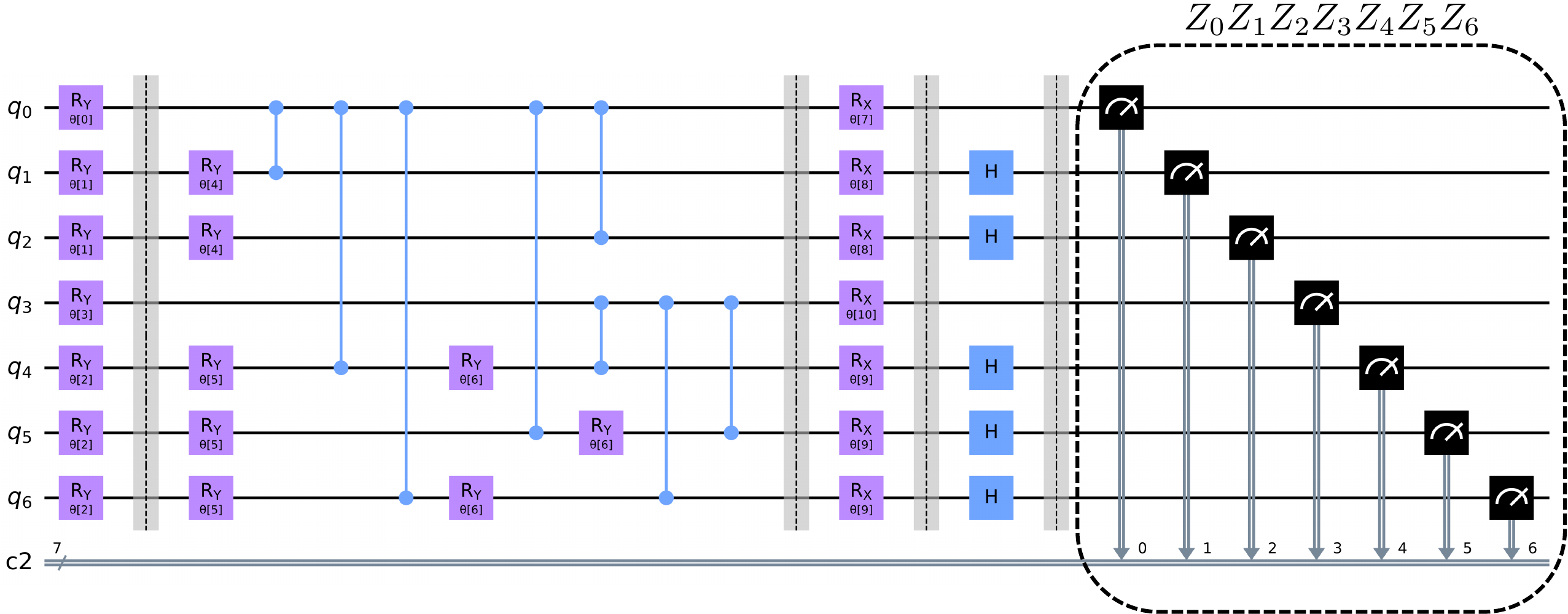}}
\caption[Circuit ansatz used in the VQE demonstration of the $2 \times 1$ planar code.]{\textbf{Circuit ansatz used in the VQE demonstration of the $2 \times 1$ planar code.} The qubits and corresponding classical registers for storing measurement outcomes are labelled by $q$ and $c$ respectively. The measurement basis of the circuit is $Z_{0}Z_{1}Z_{2}Z_{3}Z_{4}Z_{5}Z_{6}$.}
\label{pc_ansatz}
\end{figure}

\begin{figure}[htb]
\centering\subfloat{\includegraphics[width=18.5cm]{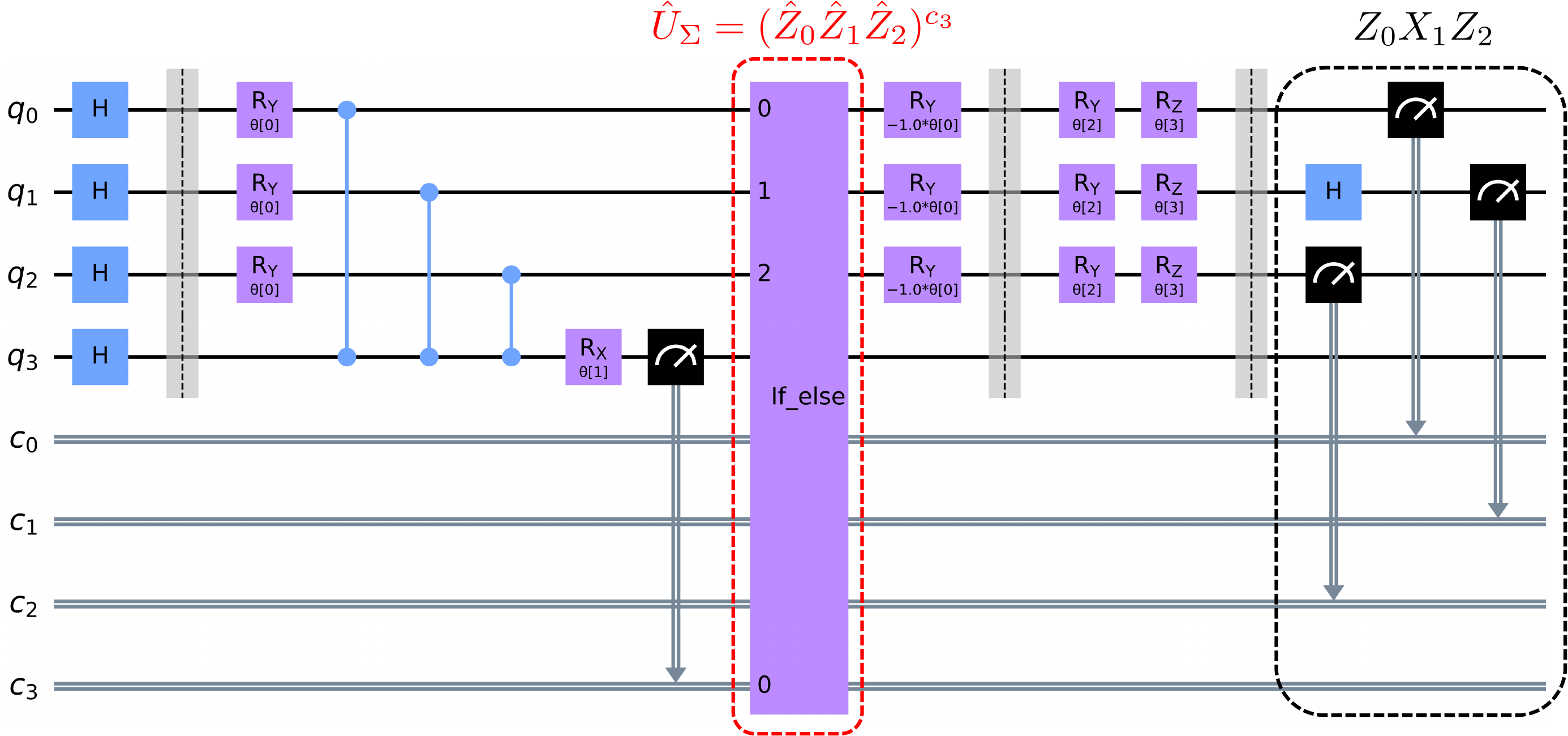}}
\caption[Circuit ansatz used in the VQE demonstration of 1D quantum chromodynamics.]{\textbf{Circuit ansatz used in the VQE demonstration of 1D quantum chromodynamics.} The qubits and corresponding classical registers for storing measurement outcomes are labelled by $q$ and $c$ respectively. After the mid-circuit measurement of the ancilla ($q_3$), the dynamic \texttt{if\_else} instruction (highlighted in red) implements the byproduct operators $\hat{U}_{\Sigma} =(\hat{Z}_{0}\hat{Z}_{1}\hat{Z}_{2})^{c_{3}}$ corresponding to the $\hat{Z}^{\otimes 3}(\theta)$ operation (see Sec.~\ref{app: mbqc}). The measurement basis of the circuit is $Z_{0}X_{1}Z_{2}$.}
\label{qcd_ansatz}
\end{figure}

\begin{figure}[htb]
\centering\subfloat{\includegraphics[width=18.5cm]{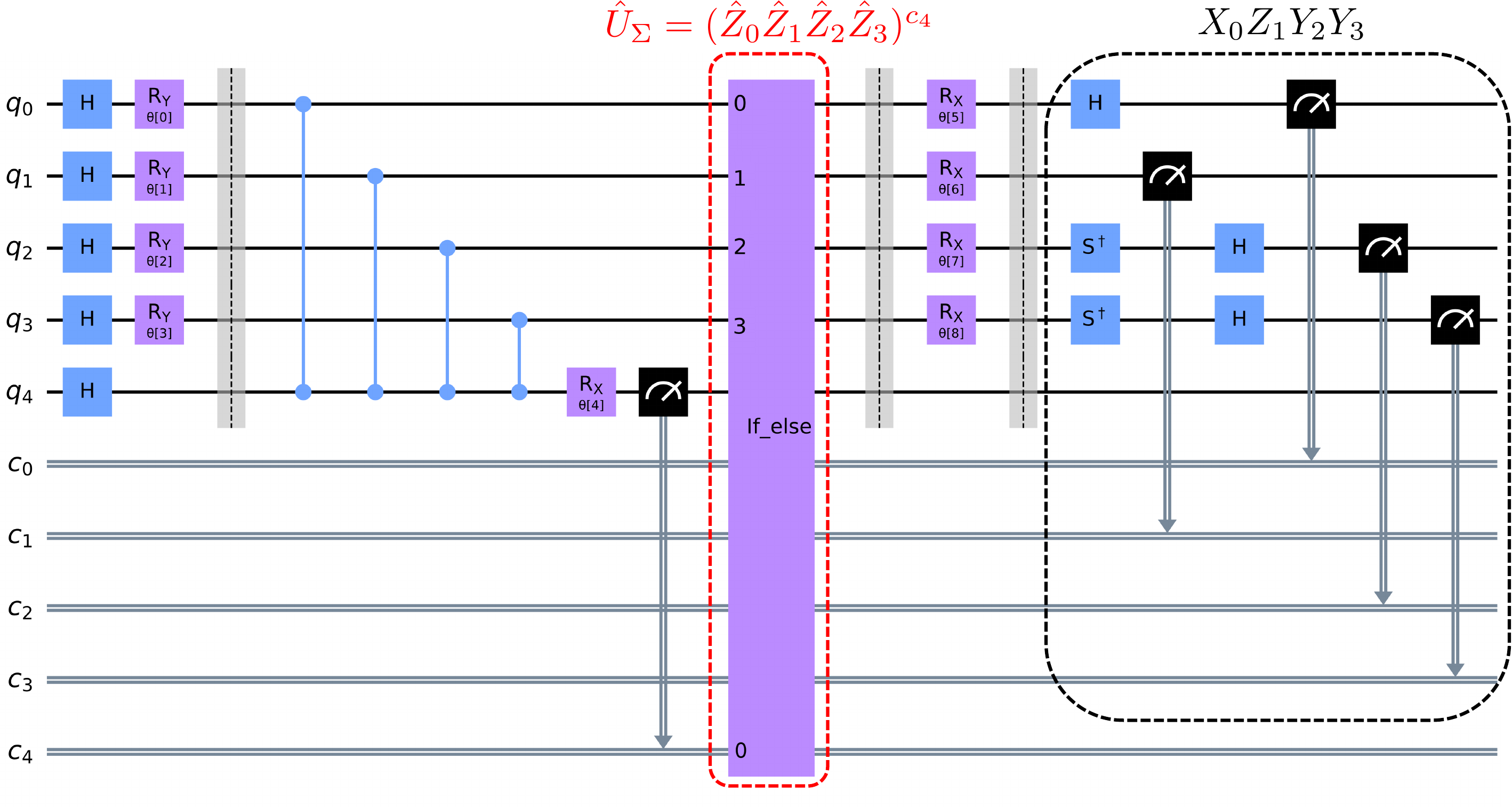}}
\caption[Circuit ansatz used in the VQE demonstration of LiH.]{\textbf{Circuit ansatz used in the VQE demonstration of LiH.} The qubits and corresponding classical registers for storing measurement outcomes are labelled by $q$ and $c$ respectively. After the mid-circuit measurement of the ancilla ($q_4$), the dynamic \texttt{if\_else} instruction (highlighted in red) implements the byproduct operators $\hat{U}_{\Sigma} =(\hat{Z}_{0}\hat{Z}_{1}\hat{Z}_{2}\hat{Z}_{3})^{c_{4}}$ corresponding to the $\hat{Z}^{\otimes 4}(\theta)$ operation (see Sec.~\ref{app: mbqc}). The measurement basis of the circuit is $X_{0}Z_{1}Y_{2}Y_{3}$.}
\label{lih_ansatz}
\end{figure} 
\clearpage
\begin{table}[htb]
\caption{4-qubit Hamiltonian for the LiH molecule. All Pauli terms are weighted by the coefficients shown and summed. The 4-body terms are highlighted in bold.}
\begin{center}
\begin{tabular}{|c|c|c|c|c|c|c|c|c|c|c|}
 \hline
 \makecell{$IIIZ$ \\ $-0.0938$}  & \makecell{$IIZX$ \\ $-0.00318$} & \makecell{$IIIX$ \\ $0.00318$} & \makecell{$IIXX$ \\ $-0.00125$} & \makecell{$IIYY$ \\ $0.00125$} & 
 \makecell{$IIZZ$ \\ $-0.212$} & \makecell{$IIXZ$ \\ $0.0192$} & \makecell{$IIXI$ \\ $0.0192$} & \makecell{$IIZI$ \\ $0.358$} & \makecell{$IZII$ \\ $0.0938$} & \makecell{$ZXII$ \\ $0.00318$} \\ \hline \makecell{$IXII$ \\ $0.00318$} &
\makecell{$XXII$ \\ $-0.00125$} & \makecell{$YYII$ \\ $0.00125$} & \makecell{$ZZII$ \\ $-0.212$} & \makecell{$XZII$ \\ $-0.0192$} & \makecell{$XIII$ \\ $0.0192$} & \makecell{$ZIII$ \\ $-0.358$} & \makecell{$IZIZ$ \\ $-0.122$} & \makecell{$IZZX$ \\ $0.0121$} & \makecell{$IZIX$ \\ $-0.0121$} & \makecell{$IZXX$ \\ $0.0317$} \\ \hline \makecell{$IZYY$ \\ $-0.0317$} & \makecell{$IXIZ$ \\ $0.0121$} &
\makecell{$ZXIZ$ \\ $0.0121$} & \makecell{$IXZX$ \\ $-0.00327$} & \makecell{\bm{$ZXZX$} \\ \bm{$0.00327$}} & \makecell{$IXIX$ \\ $0.00327$} & \makecell{$ZXIX$ \\ $0.00327$} & \makecell{$IXXX$ \\ $-0.00865$} & \makecell{\bm{$ZXXX$} \\ \bm{$-0.00865$}} & \makecell{$IXYY$ \\ $0.00865$} & \makecell{\bm{$ZXYY$} \\ \bm{$0.00865$}} \\ \hline \makecell{$YYIZ$ \\ $0.0317$} &
\makecell{$XXIZ$ \\ $-0.0317$} & \makecell{\bm{$YYZX$} \\ \bm{$-0.00865$}} &
\makecell{\bm{$XXZX$} \\ \bm{ $0.00865$}} & \makecell{$YYIX$ \\ $0.00865$} & \makecell{$XXIX$ \\ $-0.00865$} & \makecell{\bm{$YYXX$} \\ \bm{$-0.031$}} & \makecell{\bm{$XXXX$} \\ \bm{$0.031$}} & \makecell{\bm{$YYYY$} \\ \bm{$0.031$}} & \makecell{\bm{$XXYY$} \\ \bm{$-0.031$}} & \makecell{$ZZIZ$ \\ $0.0559$} \\ \hline \makecell{\bm{$ZZZX$} \\ \bm{$0.00187$}} & \makecell{$ZZIX$ \\ $-0.00187$} & \makecell{\bm{$ZZXX$} \\ \bm{$0.0031$}} & \makecell{\bm{$ZZYY$} \\ \bm{$-0.0031$}} &\makecell{$XIIZ$ \\ $0.0128$} & \makecell{$XZIZ$ \\ $-0.0128$} & \makecell{$XIZX$ \\ $-0.00235$}  
 & \makecell{\bm{$XZZX$} \\ 
 \bm{$0.00235$}} & \makecell{$XIIX$ \\ $0.00235$} & \makecell{$XZIX$ \\ $-0.00235$} & \makecell{$XIXX$ \\ $-0.00798$} \\ \hline \makecell{\bm{$XZXX$} \\ \bm{$0.00797$}} & 
\makecell{$XIYY$ \\ $0.00797$} & \makecell{\bm{$XZYY$} \\ \bm{$-0.00797$}} & \makecell{$ZIIZ$ \\ $0.113$} & \makecell{$ZIZX$ \\ $-0.0108$} &
\makecell{$ZIIX$ \\$0.0108$} & \makecell{$ZIXX$ \\ $-0.0336$} & \makecell{$ZIYY$ \\ $0.0336$} & \makecell{$IZZZ$ \\ $-0.0559$} &
 \makecell{$IZXZ$ \\ $-0.0128$} & \makecell{$IZXI$ \\ $-0.0128$} \\ \hline \makecell{$IXZZ$ \\ $-0.00187$} & \makecell{\bm{$ZXZZ$} \\ \bm{$-0.00187$}} & \makecell{$IXXZ$ \\ $0.00235$} & \makecell{\bm{$ZXXZ$} \\ \bm{$0.00235$}} & \makecell{$IXXI$ \\ $0.00235$} & \makecell{$ZXXI$ \\ $0.00235$} &
 \makecell{\bm{$YYZZ$} \\ \bm{$-0.0031$}} & \makecell{\bm{$XXZZ$} \\ \bm{$0.0031$}} & \makecell{\bm{$YYXZ$} \\ \bm{$0.00798$}} &
 \makecell{\bm{$XXXZ$} \\ \bm{$-0.00798$}} & \makecell{$YYXI$ \\ $0.00798$} \\ \hline
 \makecell{$XXXI$ \\ $-0.00798$} & \makecell{\bm{$ZZZZ$} \\ \bm{$0.0845$}} & \makecell{\bm{$ZZXZ$} \\ \bm{$-0.00899$}} & \makecell{$ZZXI$ \\ $-0.00899$} & \makecell{$XIZZ$ \\ $-0.00899$} & \makecell{\bm{$XZZZ$} \\ \bm{$0.00899$}} & \makecell{$XIXZ$ \\ $0.00661$} &
 \makecell{\bm{$XZXZ$} \\ \bm{$-0.00661$}} & \makecell{$XIXI$ \\ $0.00661$} & \makecell{$XZXI$ \\ $-0.00661$} & \makecell{$ZIZZ$ \\ $0.0604$} \\ \hline \makecell{$ZIXZ$ \\ $0.011$} & \makecell{$ZIXI$ \\ $0.011$} & \makecell{$IZZI$ \\ $0.113$} & \makecell{$IXZI$ \\ $-0.0108$} & \makecell{$ZXZI$ \\ $-0.0108$} & \makecell{$YYZI$ \\ $-0.0336$} & \makecell{$XXZI$ \\ $0.0336$} & \makecell{$ZZZI$ \\ $-0.0604$} &
 \makecell{$XIZI$ \\ $-0.011$} & \makecell{$XZZI$ \\ $-0.011$} & \makecell{$ZIZI$ \\ $-0.113$} \\ \hline \makecell{$IIII$ \\ $-7.012$} & & & & & & & & & & \\
 \hline
\end{tabular}
\end{center}
\label{table:LiHham}
\end{table}

\clearpage

\noindent\textit{\ul{Calibration and VQE data}}. \\ \\
Notes: \\
1. For all data presented, we refer to Sec.~\ref{app: methods} for a discussion of how the numbers of measurement shots were selected. \\
2. $N_\text{CX}$ refers to the number of CX gates in a VQE circuit after its qubits were mapped to the respective IBM Quantum system. \\
3. Unless otherwise stated, all reported energies are unitless and may be expressed in absolute units (e.g. eV, J, Ha) using a scaling factor with appropriate dimensions.
\begin{table}[htb] 
\caption{Calibration data for the IBM Quantum systems and associated qubits used throughout the VQE demonstrations. Data was obtained from IBM Quantum Service and captured around the time of the VQE runs.}
\begin{center}
\begin{threeparttable}
\resizebox{\textwidth}{!}{\begin{tabular}{|c|c|c|c|c|c|c|c|c|} 
\hline
 System & Qubits used & Qubit \# & T1($\mu$s) & T2($\mu$s) & Frequency (GHz) & Anharmonicity (GHz) & Readout error \\
 \hline\hline
\multirowcell{7}{\texttt{ibm\_lagos}} & \multirowcell{7}{$0,2,3,5,1^{\text{a, b}}$ \\ $3,6,1,4,5^{\text{a, b}}$ \\ $0,1,2,3,4,5,6$} & $0$ & $129.66$ & $44.43$ & $5.235$ & $-0.33987$ & $1.07\times 10^{-2}$\\
\cline{3-8}
& & $1$ & $129.27$ & $82.46$ & $5.1$ & $-0.34325$ & $1.86\times 10^{-2}$\\
\cline{3-8}
& & $2$ & $163.41$ & $111.7$ & $5.188$ & $-0.34193$ & $2.09\times 10^{-2}$\\
\cline{3-8}
& & $3$ & $138.25$ & $89.69$ & $4.987$ & $-0.34529$ & $1.41\times 10^{-2}$\\
\cline{3-8}
& & $4$ & $91.73$ & $29.79$ & $5.285$ & $-0.33923$ & $2.33\times 10^{-2}$\\
\cline{3-8}
& & $5$ & $116.37$ & $73.95$ & $5.176$ & $-0.34079$ & $2.01\times 10^{-2}$\\
\cline{3-8}
& & $6$ & $160.91$ & $102.98$ & $5.064$ & $-0.34276$ & $1.58\times 10^{-2}$\\
\hline
\multirowcell{7}{\texttt{ibm\_perth}} & \multirowcell{7}{$0,2,3,5,1^{\text{a, b}}$ \\ $3,6,1,4,5^{\text{a, b}}$ \\ $0,1,2,3,4,5,6$} & $0$ & $178.99$ & $86.52$ & $5.158$ & $-0.34152$ & $2.79\times 10^{-2}$\\
\cline{3-8}
& & $1$ & $187.57$ & $47.78$ & $5.034$ & $-0.34437$ & $2.07\times 10^{-2}$\\
\cline{3-8}
& & $2$ & $281.46$ & $124.09$ & $4.863$ & $-0.34727$ & $1.94\times 10^{-2}$\\
\cline{3-8}
& & $3$ & $222.63$ & $166.89$ & $5.125$ & $-0.34044$ & $1.63\times 10^{-2}$\\
\cline{3-8}
& & $4$ & $109.04$ & $86.32$ & $5.159$ & $-0.33337$ & $1.68\times 10^{-2}$\\
\cline{3-8}
& & $5$ & $152.81$ & $188.18$ & $4.979$ & $-0.34602$ & $2.31\times 10^{-2}$\\
\cline{3-8}
& & $6$ & $212.75$ & $219.43$ & $5.157$ & $-0.34045$ & $5.3\times 10^{-3}$\\
\hline
\multirowcell{5}{\texttt{ibm\_peekskill}} & \multirowcell{5}{$22, 26, 24, 23, 25$} & $22$ & $137.10$ & $221.55$ & $4.922$ & $-0.34481$ & $1.17\times 10^{-2}$\\
\cline{3-8}
& & $23$ & $366.55$ & $468.08$ & $5.021$ & $-0.34171$ & $4.5\times 10^{-3}$\\
\cline{3-8}
& & $24$ & $353.76$ & $485.12$ & $5.133$ & $-0.34087$ & $5.3\times 10^{-3}$\\
\cline{3-8}
& & $25$ & $426.40$ & $286.41$ & $4.987$ & $-0.34407$ & $3.36\times 10^{-2}$\\
\cline{3-8}
& & $26$ & $169.19$ & $389.76$ & $5.109$ & $-0.33923$ & $5.9\times 10^{-3}$\\
\hline
\end{tabular}}
\begin{tablenotes}\footnotesize
\item [a] At the time of running a particular VQE demonstration, the set of qubits that possessed the smaller CX error and larger T1 and T2 values overall was chosen.
\item [b] These sets of physical qubits incur the fewest additional CXs from SWAP operations (used to fit the CX connectivity of a circuit to the CX connectivity of each system).  
\end{tablenotes}
\end{threeparttable}
\label{tab:IBM_devices_calib_1}
\end{center}
\end{table}

\begin{table}
 \caption{Calibration data for the IBM Quantum systems and associated qubits used throughout the VQE demonstrations, continued.}
\begin{center} 
\begin{threeparttable}
\resizebox{\textwidth}{!}{\begin{tabular}{|c|c|c|c|c|c|c|c|c|} 
\hline
 System & Qubit \# & P(meas 0, prep 1) & P(meas 1, prep 0) & Readout len. (ns) & $I/X/\sqrt{X}$ error & CX error & Gate time (ns) \\
 \hline\hline
\multirowcell{12}{\texttt{ibm\_lagos}} & $0$ & $0.0096$ & $0.0118$ & \multirowcell{12}{$789.333$} & $1.938\times 10^{-4}$ & \makecell{$0\_1: 0.01358$} & \makecell{$0\_1: 576$}\\
\cline{2-4}\cline{6-8}
& $1$ & $0.0206$ & $0.0166$ &  & $3.853\times 10^{-4}$ & \makecell{$1\_3: 0.00653$\\$1\_2: 0.00882$\\$1\_0: 0.01358$} & \makecell{$1\_3: 334.222$\\$1\_2: 327.111$\\$1\_0: 611.556$}\\
\cline{2-4}\cline{6-8}
& $2$ & $0.0134$ & $0.0284$ &  & $3.003\times 10^{-4}$ & \makecell{$2\_1: 0.00882$} & \makecell{$2\_1: 291.556$}\\
\cline{2-4}\cline{6-8}
& $3$ & $0.0146$ & $0.0136$ &  & $1.712\times 10^{-4}$ & \makecell{$3\_1: 0.00653$\\$3\_5: 0.01053$} & \makecell{$3\_1: 298.667$\\$3\_5: 334.222$}\\
\cline{2-4}\cline{6-8}
& $4$ & $0.0254$ & $0.0212$ &  & $1.937\times 10^{-4}$ & \makecell{$4\_5: 0.00671$} & \makecell{$4\_5: 362.667$}\\
\cline{2-4}\cline{6-8}
& $5$ & $0.0184$ & $0.0218$ &  & $2.451\times 10^{-4}$ & \makecell{$5\_4: 0.00671$\\$5\_6: 0.00837$\\$5\_3: 0.01053$} & \makecell{$5\_4: 327.111$\\$5\_6: 256$\\$5\_3: 298.667$}\\
\cline{2-4}\cline{6-8}
& $6$ & $0.0186$ & $0.013$ &  & $2.57\times 10^{-4}$ & \makecell{$6\_5: 0.00837$} & \makecell{$6\_5: 291.556$}\\
\hline
\multirowcell{12}{\texttt{ibm\_perth}} & $0$ & $0.0306$ & $0.0252$ & \multirowcell{12}{$675.556$} & $2.122\times 10^{-4}$ & \makecell{$0\_1: 0.00854$} & \makecell{$0\_1: 391.111$}\\
\cline{2-4}\cline{6-8}
& $1$ & $0.0234$ & $0.018$ &  & $3.607\times 10^{-4}$ & \makecell{$1\_3: 0.00698$\\$1\_2: 0.00777$\\$1\_0: 0.00854$} & \makecell{$1\_3: 405.333$\\$1\_2: 355.556$\\$1\_0: 426.667$}\\
\cline{2-4}\cline{6-8}
& $2$ & $0.0208$ & $0.018$ &  & $2.808\times 10^{-4}$ & \makecell{$2\_1: 0.00777$} & \makecell{$2\_1: 320$}\\
\cline{2-4}\cline{6-8}
& $3$ & $0.0174$ & $0.0152$ &  & $2.792\times 10^{-4}$ & \makecell{$3\_5: 0.01048$\\$3\_1: 0.00698$} & \makecell{$3\_5: 284.444$\\$3\_1: 369.778$}\\
\cline{2-4}\cline{6-8}
& $4$ & $0.0216$ & $0.012$ &  & $7.687\times 10^{-4}$ & \makecell{$4\_5: 0.01346$} & \makecell{$4\_5: 590.222$}\\
\cline{2-4}\cline{6-8}
& $5$ & $0.0244$ & $0.0218$ &  & $3.553\times 10^{-4}$ & \makecell{$5\_6: 0.01221$\\$5\_4: 0.01346$\\$5\_3: 0.01048$} & \makecell{$5\_6: 640$\\$5\_4: 625.778$\\$5\_3: 320$}\\
\cline{2-4}\cline{6-8}
& $6$ & $0.0054$ & $0.0052$ &  & $2.885\times 10^{-4}$ & \makecell{$6\_5: 0.01221$} & \makecell{$6\_5: 604.444$}\\
\hline
\multirowcell{9}{\texttt{ibm\_peekskill}} & $22$ & $0.012$ & $0.0114$ & \multirowcell{9}{$860.444$} & $2.012\times 10^{-4}$ & \makecell{$22\_25: 0.00529$} & \makecell{$22\_25: 405.333$}\\
\cline{2-4}\cline{6-8}
& $23$ & $0.0032$ & $0.0058$ &  & $7.613\times 10^{-5}$ & \makecell{$23\_21: 0.00771$\\$23\_24: 0.00243$} & \makecell{$23\_21: 640$\\$23\_24: 419.556$}\\
\cline{2-4}\cline{6-8}
& $24$ & $0.0068$ & $0.0038$ &  & $1.06\times 10^{-4}$ & \makecell{$24\_25: 0.0038$\\$24\_23: 0.00243$} & \makecell{$24\_25: 384$\\$24\_23: 384$}\\
\cline{2-4}\cline{6-8}
& $25$ & $0.0376$ & $0.0296$ &  & $1.112\times 10^{-4}$ & \makecell{$25\_24: 0.0038$\\$25\_26: 0.00414$\\$25\_22: 0.00529$} & \makecell{$25\_24: 419.556$\\$25\_26: 384$\\$25\_22: 440.889$}\\
\cline{2-4}\cline{6-8}
& $26$ & $0.0058$ & $0.006$ &  & $1.822\times 10^{-4}$ & \makecell{$26\_25: 0.00414$} & \makecell{$26\_25: 348.444$}\\
\cline{2-4}\cline{6-8}
\hline
\end{tabular}}
\end{threeparttable}
\label{tab:IBM_devices_calib_2}
\end{center}
\end{table}
\clearpage
\begin{table}[hbt!]
\centering
\caption{Optimized energies, operator expectation values, fidelities, and errors for the $\mathbb{Z}_2$ VQE demonstration ($s=100$).}
\begin{tabular}{|c|c|c|c|c|c|c|c|c|c|c|c|}
 \hline
 $\lambda$ & System & Optimizer & $N_{\mathrm{s}}$, $N_{\text{eval}}$ & $N_{\mathrm{CX}}$ & $\kappa$ &
 $E_{\mathrm{VQE}}^{(0)}$ & $E_{\mathrm{VQE}}$ & $E_{0}$ & $E_{g}$ \\
 \hline\hline
 \makecell{$0.5$ \\ $0.63$ \\ $0.85$ \\ $1.12$ \\ $1.52$ \\ $1.98$ \\ $2.33$ \\ $2.65$ \\ $2.88$ \\ $3.3$} 
 & \makecell{\texttt{ibm\_peekskill} \\ \texttt{ibm\_peekskill} \\ \texttt{ibm\_peekskill} \\ \texttt{ibm\_peekskill} \\ \texttt{ibm\_peekskill} \\ \texttt{ibm\_peekskill} \\ \texttt{ibm\_peekskill} \\ \texttt{ibm\_peekskill} \\ \texttt{ibm\_peekskill} \\ \texttt{ibm\_peekskill}} 
 & \makecell{COBYLA \\ COBYLA  \\ COBYLA \\ COBYLA \\ COBYLA \\ COBYLA \\ COBYLA \\ COBYLA \\ COBYLA \\ COBYLA} 
 & \makecell{$13857, 50$ \\ $16267, 50$ \\ $22187, 50$ \\ $34372, 50$ \\ $50000, 50$ \\ $50000, 50$ \\ $50000, 50$ \\ $50000, 50$ \\ $50000, 50$ \\ $50000, 50$}
 & \makecell{$7$ \\ $7$ \\ $7$ \\ $7$ \\ $7$ \\ $7$ \\ $7$ \\ $7$ \\ $7$ \\ $7$}
 & \makecell{$1$ \\ $1$ \\ $1$ \\ $1$ \\ $1$ \\ $1$ \\ $1$ \\ $1$ \\ $1$ \\ $1$}
 & \makecell{$-3.958$ \\ $-3.162$ \\ $-2.197$ \\ $-1.651$ \\ $-1.157$ \\ $-0.793$ \\ $-0.824$ \\ $-0.587$ \\ $-0.604$ \\ $-0.392$} 
 & \makecell{$-7.427$ \\ $-6.036$ \\ $-4.547$ \\ $-3.436$ \\ $-2.735$ \\ $-2.548$ \\ $-2.659$ \\ $-2.784$ \\ $-2.896$ \\ $-3.104$} 
 & \makecell{$-8.016$ \\ $-6.380$ \\ $-4.782$ \\ $-3.743$ \\ $-3.039$ \\ $-2.829$ \\ $-2.894$ \\ $-3.050$ \\ $-3.197$ \\ $-3.516$} 
 & \makecell{$3.984$ \\ $3.144$ \\ $2.280$ \\ $1.635$ \\ $1.029$ \\ $0.606$ \\ $0.411$ \\ $0.294$ \\ $0.235$ \\ $0.160$} \\
 \hline
\end{tabular}
\begin{threeparttable}
\vspace{0.25cm}
\begin{tabular}{|c|c|c|c|c|c|c|c|c|c|}
 \hline
 $\lambda$ & $\langle\hat{\mathcal{H}}_{\square}\rangle_{\mathrm{VQE}}$ & $\langle\hat{\mathcal{H}}_{\square}\rangle_{0}$ & $\langle\hat{\mathcal{H}}_{\triangle}/4\rangle_{\mathrm{VQE}}$ & $\langle\hat{\mathcal{H}}_{\triangle}/4\rangle_{0}$ & $\mathcal{F}$ & $\sigma, E$ & $\sigma, \langle \hat{\mathcal{H}}_{\square}\rangle$ & $\sigma, \langle \hat{\mathcal{H}}_{\triangle}/4\rangle$ & $\Delta E/E_{g}$ \\
 \hline\hline
 \makecell{$0.5$ \\ $0.63$ \\ $0.85$ \\ $1.12$ \\ $1.52$ \\ $1.98$ \\ $2.33$ \\ $2.65$ \\ $2.88$ \\ $3.3$} 
 & \makecell{$0.0321$ \\ $-0.0468$ \\ $-0.166$ \\ $-0.262$ \\ $-0.363$ \\ $-0.481$ \\ $-0.572$ \\ $-0.755$ \\ $-0.868$ \\ $-0.8559$} 
 & \makecell{$-0.0624$ \\ $-0.0987$ \\ $-0.178$ \\ $-0.299$ \\ $-0.500$ \\ $-0.700$ \\ $-0.805$ \\ $-0.869$ \\ $-0.901$ \\ $-0.939$}
 & \makecell{$-0.936$ \\ $-0.922$ \\ $-0.936$ \\ $-0.880$ \\ $-0.830$ \\ $-0.790$ \\ $-0.773$ \\ $-0.518$ \\ $-0.286$ \\ $-0.230$}
 & \makecell{$-0.998$ \\ $-0.995$ \\ $-0.984$ \\ $-0.954$ \\ $-0.866$ \\ $-0.714$ \\ $-0.593$ \\ $-0.495$ \\ $-0.434$ \\ $-0.345$}
 & \makecell{$0.862$ \\ $0.850$ \\ $0.826$ \\ $0.811$ \\ $0.815$ \\ $0.776$ \\ $0.818$ \\ $0.768$ \\ $0.792$ \\ $0.812$}
 & \makecell{$0.0289$ \\ $0.0178$ \\ $0.0153$ \\ $0.0109$ \\ $0.00976$ \\ $0.0112$ \\ $0.0143$ \\ $0.0160$ \\ $0.0197$ \\ $0.0206$}
 & \makecell{$0.00482$ \\ $0.00560$ \\ $0.00676$ \\ $0.00665$ \\ $0.00763$ \\ $0.0100$ \\ $0.0135$ \\ $0.0150$ \\ $0.0189$ \\ $0.0199$}
 & \makecell{$0.0285$ \\ $0.0169$ \\ $0.0137$ \\ $0.00870$ \\ $0.00608$ \\ $0.00514$ \\ $0.00493$ \\ $0.00552$ \\ $0.00579$ \\ $0.00510$} 
 & \makecell{$0.148$ \\ $0.109$ \\ $0.103$ \\ $0.188$ \\ $0.296$ \\ $0.464$ \\ $0.573$ \\ $0.904$ \\ $1.282$ \\ $2.569$} \\
 \hline
\end{tabular}
\end{threeparttable}
\label{tab:VQE_data_Z2}
\vspace{2cm}
\centering
    \caption{Optimized energies, operator expectation values, fidelities, and errors for the 1D QCD VQE demonstration ($s=10$).}
\begin{tabular}{|c|c|c|c|c|c|c|c|c|c|c|}
 \hline
 $\tilde{m}^{\text{a}}$ & System$^{\text{b}}$ & Optimizer & $N_{\mathrm{s}}$, $N_{\text{eval}}$ & $N_{\mathrm{CX}}$ & $\kappa$ &
 $E_{\mathrm{VQE}}^{(0)}$ & $E_{\mathrm{VQE}}$ & $E_{0}$ & $E_{g}$ \\
 \hline\hline
\makecell{$-1.0$ \\ $-0.5$ \\ $-0.2$ \\ $-0.05$ \\ $0.01$ \\ $0.05$ \\ $0.1$ \\ $0.2$ \\ $0.5$ \\ $1.0$}
& \makecell{\texttt{ibm\_lagos} \\ \texttt{ibm\_lagos} \\ \texttt{ibm\_lagos} \\ \texttt{ibm\_lagos} \\ \texttt{ibm\_lagos} \\ \texttt{ibm\_peekskill} \\ \texttt{ibm\_peekskill} \\ \texttt{ibm\_lagos} \\ \texttt{ibm\_lagos} \\ \texttt{ibm\_lagos}} & \makecell{COBYLA \\ COBYLA \\ COBYLA \\ COBYLA \\ COBYLA \\ COBYLA \\ COBYLA \\ COBYLA \\ COBYLA \\ COBYLA}
& \makecell{$750, 50$ \\ $1522, 50 $ \\ $3953, 50$ \\ $9680, 55$ \\ $10402, 56$ \\ $9380, 57$ \\ $7290, 50$ \\ $3953, 50$ \\ $1131, 50$ \\ $454, 50$}
& \makecell{$4$ \\ $4$ \\ $4$ \\ $4$ \\ $4$ \\ $4$ \\ $4$ \\ $4$ \\ $4$ \\ $4$}
& \makecell{$1$ \\ $1$ \\ $1$ \\ $1$ \\ $1$ \\ $1$ \\ $1$ \\ $1$ \\ $1$ \\ $1$}
& \makecell{$-3.574$ \\ $-2.179$ \\ $-1.322$ \\ $-0.831$ \\ $-0.704$ \\ $-0.518$ \\ $-0.299$ \\ $1.228$ \\ $2.073$ \\ $3.446$}
& \makecell{$-6.240$ \\ $-3.220$ \\ $-1.626$ \\ $-0.952$ \\ $-0.763$ \\ $-0.623$ \\ $-0.570$ \\ $-0.384$ \\ $-0.335$ \\ $-0.164$}
& \makecell{$-6.259$ \\ $-3.395$ \\ $-1.791$ \\ $-1.122$ \\ $-0.924$ \\ $-0.822$ \\ $-0.723$ \\ $-0.591$ \\ $-0.395$ \\ $-0.259$}
& \makecell{$2.878$ \\ $1.771$ \\ $0.965$ \\ $0.639$ \\ $0.610$ \\ $0.639$ \\ $0.719$ \\ $0.965$ \\ $1.771$ \\ $2.878$} \\
\hline
\end{tabular}
\begin{threeparttable}
\vspace{0.25cm}
\begin{tabular}{|c|c|c|c|c|c|c|c|c|c|c|}
\hline
$\tilde{m}^{\text{a}}$ &
 $\langle\hat{N}\rangle_{\mathrm{VQE}}$ & $\langle\hat{N}\rangle_{0}$ & $\langle\hat{N'}\rangle_{\mathrm{VQE}}$ & $\langle\hat{N'}\rangle_{0}$ & $\mathcal{F}$ & $\sigma, E$ & $\sigma, N$ &  $\sigma, N'$ & $\Delta E/E_{g}$ \\
 \hline\hline
\makecell{$-1.0$ \\ $-0.5$ \\ $-0.2$ \\ $-0.05$ \\ $0.01$ \\ $0.05$ \\ $0.1$ \\ $0.2$ \\ $0.5$ \\ $1.0$}
& \makecell{$5.817$ \\ $5.638$ \\ $5.414$ \\ $4.518$ \\ $3.067$ \\ $2.591$ \\ $1.441$ \\ $0.548$ \\ $0.443$ \\ $0.00737$}
& \makecell{$5.822$ \\ $5.580$ \\ $4.963$ \\ $3.725$ \\ $2.849$ \\ $2.275$ \\ $1.702$ \\ $1.037$ \\ $0.420$ \\ $0.178$}
& \makecell{$3.662$ \\ $3.454$ \\ $3.247$ \\ $2.250$ \\ $1.060$ \\ $0.751$ \\ $0.808$ \\ $0.00223$ \\ $0.00905$ \\ $-0.0801$}
& \makecell{$3.768$ \\ $3.469$ \\ $2.815$ \\ $1.780$ \\ $1.170$ \\ $0.814$ \\ $0.497$ \\ $0.197$ \\ $0.0298$ \\ $0.00468$}
& \makecell{$0.902$ \\ $0.834$ \\ $0.888$ \\ $0.791$ \\ $0.884$ \\ $0.792$ \\ $0.808$ \\ $0.771$ \\ $0.923$ \\ $0.893$}
& \makecell{$0.0368$ \\ $0.0245$ \\ $0.0120$ \\ $0.00741$ \\ $0.00461$ \\ $0.00598$ \\ $0.00845$ \\ $0.0152$ \\ $0.0273$ \\ $0.0499$}
& \makecell{$0.00258$ \\ $0.00692$ \\ $0.00147$ \\ $0.00164$ \\ $0.00150$ \\ $0.000951$ \\ $0.000730$ \\ $0.00320$ \\ $0.00352$ \\ $0.00716$}
& \makecell{$0.0242$ \\ $0.0170$ \\ $0.0103$ \\ $0.00817$ \\ $0.00404$ \\ $0.00194$ \\ $0.00716$ \\ $0.00902$ \\ $0.0140$ \\ $0.00648$}
& \makecell{$0.00673$ \\ $0.0992$ \\ $0.171$ \\ $0.266$ \\ $0.264$ \\ $0.311$ \\ $0.214$ \\ $0.214$ \\ $0.0363$ \\ $0.0332$} \\
\hline
\end{tabular}
\begin{tablenotes}\footnotesize
\vspace{0.05cm}
\item [\text{a}] At $x = 0.8$.
\item [\text{b}] Owing to logistical challenges with the IBM Quantum platform, data points were collected over multiple systems.
\end{tablenotes}
\end{threeparttable}
\label{tab:VQE_data_SU3}
\end{table}

\begin{table}
\caption{Optimized energies, operator expectation values, fidelities, and errors for the LiH VQE demonstration (interatomic distance $=$ 1.6 \AA). All energies are expressed in Hartrees (Ha).}
\begin{center}
\begin{threeparttable}
\begin{tabular}{|c|c|c|c|c|c|c|c|c|c|c|c|c|c|c|c|}
 \hline
 System & Optimizer & $N_{\mathrm{s}}$, $N_{\text{eval}}$ & $N_{\mathrm{CX}}$ & $\kappa$ &
  $E_{\mathrm{VQE}}^{(0)}$ & $E_{\mathrm{VQE}}$ & $E_{0}$ & $E_{1}$ & $E_{2}$ & $\mathcal{F}^{(0)}$ & $\mathcal{F}$ & $\sigma, E$ & $\Delta E/E_{0}$ \\
 \hline\hline
 \texttt{ibm\_perth} & COBYLA & $10000^{\text{a}}, 99$ & $7$ & $1$
 & $-6.967$ & $-7.794^{\text{b}}$ & $-7.881$ & $-7.766$ & $-7.748$ & $0.0673^{\text{c}}$ & $0.782^{\text{c}}$ & $0.000534$ & $0.0110$ \\
\hline
\end{tabular}
\begin{tablenotes}\footnotesize
\item [\text{a}] $N_{\mathrm{s}}$ chosen from trial-and-error runs on fake IBM backends, as a minimal value yielding acceptable convergence and statistical errors.
\item [\text{b}] Since convergence behavior is analyzed, we do not apply self-mitigation to the final energy.
\item [\text{c}] See Fig.~2(e).
\end{tablenotes}
\end{threeparttable}
\label{tab:VQE_data_LiH}
\end{center}

\caption{Optimized energies, operator expectation values, fidelities, and errors for the $2\times 1$ PC VQE demonstration ($s=25$).$^{\text{a}}$}
\begin{center}
\begin{threeparttable}
\begin{tabular}{|c|c|c|c|c|c|c|c|c|c|c|c|c|c|c|}
 \hline
 $\xi$ & System$^{\text{b}}$ & Optimizer$^{\text{c}}$ & $N_{\mathrm{s}}$, {\update $N_{\text{eval}}$}$^{\text{c}}$ & $N_{\mathrm{CX}}$$^{\text{d}}$ & $\kappa^{\text{d}}$ &
 $E_{\mathrm{VQE}}^{(0)}$ & $E_{\mathrm{VQE}}$ & $E_{0}$ & $E_{g}$ & $\sigma, E$ & ${\update \Delta E/E_{0}}$ \\[0.01ex]
 \hline\hline
  \makecell{$0.01$ \\ $0.0316$ \\ $0.1$ \\ $0.316$ \\ $1.0$ \\ $1.5$ \\ $2.0$ \\ $3.0$ \\ $4.0$ \\ $6.0$ \\ $8.0$ \\$10.0$} & 
  \makecell{\texttt{ibm\_perth} \\ \texttt{ibm\_lagos} \\ \texttt{ibm\_perth} \\ \texttt{ibm\_perth} \\ \texttt{ibm\_lagos} \\ \texttt{ibm\_lagos} \\ \texttt{ibm\_perth} \\ \texttt{ibm\_lagos}  \\ \texttt{ibm\_lagos} \\ \texttt{ibm\_perth}  \\ \texttt{ibm\_lagos} \\ \texttt{ibm\_lagos}} & \makecell{DIRECT+COBYLA \\ DIRECT+COBYLA \\ DIRECT+COBYLA \\ DIRECT+COBYLA \\ DIRECT+COBYLA \\ DIRECT+COBYLA \\ DIRECT+COBYLA \\ DIRECT+COBYLA \\ DIRECT+COBYLA \\ DIRECT+COBYLA \\ DIRECT+COBYLA \\ DIRECT+COBYLA}
  & \makecell{$946, 50+50$ \\ $960, 50+50$ \\ $965, 50+50$ \\ $781, 50+50$ \\ $1653, 50+50$ \\ $10172, 50+50$ \\ $50000, 50+50$ \\ $11722, 50+50$ \\ $4755, 50+50$ \\ $2546, 50+50$ \\ $1981, 50+50$ \\ $1729, 50+50$}
  & \makecell{$27$ \\ $27$ \\ $27$ \\ $27$ \\ $27$ \\ $27$ \\ $27$ \\ $27$ \\ $27$ \\ $27$ \\ $26$ \\ $27$}
  & \makecell{$43/27$ \\ $36/27$ \\ $40/27$ \\ $40/27$ \\ $36/27$ \\ $36/27$ \\ $40/27$ \\ $36/27$ \\ $36/27$ \\ $40/27$ \\ $39/26$ \\ $36/27$}
  & \makecell{$-1.525$ \\ $-1.516$ \\ $-1.284$ \\ $-1.216$ \\ $ -1.755$ \\ $-1.850$ \\ $-0.668$ \\ $-2.269$ \\ $-1.914$ \\ $-2.179$ \\ $-3.106$ \\ $ -2.168 $}
  & \makecell{$-7.750$ \\ $-7.800$ \\ $-7.465$ \\ $-8.422$ \\ $-10.880$ \\ $-13.688$ \\ $-17.828$ \\ $-23.588$ \\ $-30.585$ \\ $-45.345$ \\ $-61.001$ \\ $-80.911$}
  & \makecell{$-8.001$ \\ $-8.009$ \\ $-8.083$ \\ $-8.632$ \\ $-11.465$ \\ $-13.822$ \\ $-16.245$ \\ $-23.083$ \\ $-30.062$ \\ $-44.042$ \\ $-58.031$ \\ $-72.025$}
  & \makecell{$1.991$ \\ $1.977$ \\ $1.983$ \\ $2.316$ \\ $2.217$ \\ $1.156$ \\ $0.120$ \\ $1.918$ \\ $3.938$ \\ $7.959$ \\ $11.969$ \\ $15.975$}
  & \makecell{$0.227$ \\ $0.105$ \\ $0.169$ \\ $0.199$ \\ $0.117$ \\ $0.0696$ \\ $0.0743$ \\ $0.227$ \\ $0.654$ \\ $2.594$ \\ $6.316$ \\ $8.107$} 
  & \makecell{$0.0314$ \\ $0.0261$ \\ $0.0764$ \\ $0.0244$ \\ $0.051$ \\ $ 0.00966$ \\ $0.0975$ \\ $0.0219$ \\ $0.0174$ \\ $0.0296$ \\ $0.0512$ \\ $0.123$} \\
  \hline
\end{tabular}
\begin{tablenotes}\footnotesize
\item [$\text{a}$] Since 7 qubits are used in this demonstration, performing full state tomography is impractical and therefore the fidelities $\mathcal{F}$ were not calculated.
\item [$\text{b}$] See note b in Table~\ref{tab:VQE_data_SU3}.
\item [$\text{c}$] See discussion in Sec.~\ref{app: methods} regarding optimization methods.
\item [$\text{d}$] On account of the larger number of qubits and their fixed connectivity on IBM Quantum systems, we employed the noise adaptive layout method to map qubits onto hardware~\cite{Qiskit}. The mapping is chosen automatically by Qiskit's transpiler (based on current calibration data) and leads to {\update slight} variations in the number of $CX$ gates incurred by each circuit in a VQE run. Here, we reflect the full range of $CX$ numbers during self-mitigation by calculating $\kappa$ from the {\update physics} and mitigation circuits with the largest and smallest number of $CX$s respectively, over all measurement bases. In other words, we employ the maximum possible $\kappa$.
\end{tablenotes}
\end{threeparttable}
\end{center}
\label{tab:VQE_data_PC}
\end{table}

\begin{table}[htb]
\begin{center}
\begin{threeparttable}
\caption{Initial and final (optimized) variational parameters for each VQE demonstration.$^\text{a}$}
\begin{tabular}{|c|c|c|c|}
 \hline
 Model & Free Param. & $\vec{\theta}^{\hspace{0.05cm}(0)}$ & $\vec{\theta}^{\hspace{0.05cm} (\text{opt})}$ \\
 \hline\hline
${\update \mathbb{Z}_{2}}$
& \makecell[l]{$\lambda = 0.5$ \\ $\lambda = 0.63$ \\ $\lambda = 0.85$\\ $\lambda = 1.12$ \\ $\lambda = 1.52$ \\ $\lambda = 1.98$ \\ $\lambda = 2.33$ \\ $\lambda = 2.65$ \\ $\lambda = 2.88$ \\ $\lambda = 3.3$}  & \makecell{$[1.643, 5.56,  5.17,  2.6]$ \\ $[1.643,  5.56,  5.17,  2.6]$ \\ $[1.643,  5.56,  5.17,  2.6]$ \\ $[1.643,  5.56,  5.17,  2.6]$ \\ $[1.643,  5.56,  5.17,  2.6]$ \\ $[1.643,  5.56,  5.17,  2.6]$ \\ $[1.643,  5.56,  5.17,  2.6]$ \\ $[1.643,  5.56,  5.17,  2.6]$ \\ $[1.643,  5.56,  5.17,  2.6]$ \\ $[1.643,  5.56,  5.17,  2.6]$} & \makecell{$[1.438, 6.232, 4.260, 2.937]$ \\ $[1.038, 6.119, 3.337, 2.567]$ \\ $[1.593, 6.063, 5.170, 3.391]$ \\ $[1.533, 5.948, 5.131, 3.403]$ \\ $[1.502, 5.858, 3.573, 3.055]$ \\ $[1.261, 5.741, 3.629, 3.097]$ \\ $[1.502, 5.627, 3.517, 3.066]$ \\ $[1.709, 5.302, 3.555, 3.083]$ \\ $[1.410, 5.002, 3.493, 3.059]$ \\ $[1.718, 4.984, 3.557, 3.024]$} \\
\hline
${\update \text{QCD}}$
& \makecell[l]{$\tilde{m} = -1$ \\ $\tilde{m} = -0.5$ \\$\tilde{m} = -0.2$ \\ $\tilde{m}= -0.05$ \\ $\tilde{m} = 0.01$ \\ $\tilde{m} = 0.05$ \\ $\tilde{m} = 0.1$ \\ $\tilde{m} = 0.2$ \\ $\tilde{m} = 0.5$ \\ $\tilde{m} = 1$ } & \makecell{$[3\pi/2, \pi/2, \pi/2, \pi/2]$ \\ $[3\pi/2, \pi/2, \pi/2, \pi/2]$ \\ $[3\pi/2, \pi/2, \pi/2, \pi/2]$ \\ $[3\pi/2, \pi/2, \pi/2, \pi/2]$ \\ $[3\pi/2, \pi/2, \pi/2, \pi/2]$ \\ $[3\pi/2, \pi/2, \pi/2, \pi/2]$ \\ $[3\pi/2, \pi/2, \pi/2, \pi/2]$ \\ $[\pi, \pi/2, 0, \pi/2]$ \\ $[\pi, \pi/2, 0, \pi/2]$ \\$[\pi, \pi/2, 0, \pi/2]$} & \makecell{$[5.832, 1.551, 1.188, 1.728]$ \\ $ [5.713, 1.826, 1.001 , 1.829]$ \\ $[5.580, 1.682, 0.755, 1.887]$ \\ $[5.211, 1.571, 0.499, 1.848]$ \\ $[4.755, 1.502, 0.002, 1.795]$ \\ $[4.851, 1.722, 0.290, 1.551]$ \\ $[5.250, 2.131, 0.538, 1.571]$ \\ $[3.676, 1.581, 5.002, 2.185]$ \\ $[3.754, 1.671, 5.400 , 1.757]$ \\ $[2.356, 4.328, 3.740, 0.955]$}\\
\hline
PC
& \makecell[l]{$\xi = 0.01$ \\ \\ $\xi = 0.0316$ \\ \\ $\xi = 0.1$ \\ \\ $\xi = 0.316$ \\ \\ $\xi = 1.0$ \\ \\ $\xi = 1.5$ \\ \\ $\xi = 2.0$ \\ \\ $\xi = 3.0$ \\ \\ $\xi = 4.0$ \\ \\ $\xi = 6.0$ \\ \\ $\xi = 8.0$ \\ \\ $\xi = 10.0$ \\ }
& \makecell{$[\pi] \times 11$ \\ \\ $[\pi] \times 11$ \\ \\ $[\pi] \times 11$ \\ \\ $[\pi] \times 11$ \\ \\ $[\pi] \times 11$ \\ \\ $[\pi] \times 11$ \\ \\ $[\pi] \times 11$ \\ \\ $[\pi] \times 11$ \\ \\ $[\pi] \times 11$ \\ \\ $[\pi] \times 11$ \\ \\ $[\pi] \times 11$ \\ \\ $[\pi] \times 11$ \\}
& \makecell{$[5.278, 3.113, 3.118, 3.099, 3.103, 3.100,$ \\ $3.117, 1.047, 3.113, 3.115, 3.108]$ \\ $[1.051, 3.137, 3.127, 3.159, 3.043, 3.221,$ \\ $3.144, 6.272, 3.146, 3.147, 3.142]$ \\ $[1.047, 3.142, 3.159, 2.990, 3.142, 3.142,$ \\ $3.142, 6.021, 3.142, 3.142, 3.142]$ \\ $[1.852, 3.142, 3.142, 3.142, 3.142, 3.142,$ \\ $3.142, 5.236, 4.745, 3.142, 3.142]$ \\ $[1.032, 3.142, 3.094, 3.141, 3.130, 3.215,$ \\ $3.335, 6.283, 3.122, 3.120, 3.124]$ \\ $[1.046, 3.388, 2.844, 2.994, 3.146, 3.036,$ \\ $3.003, 6.282, 2.980, 3.142, 3.209]$ \\ $[1.651, 3.142, 1.964, 3.142, 3.227, 3.219,$ \\ $3.124, 6.160, 3.142, 3.142, 3.142]$ \\ $[1.679, 3.142, 1.711, 3.142, 3.142, 2.978,$ \\ $2.685, 5.775, 1.384, 6.283, 6.283]$ \\ $[1.832, 3.142, 1.832, 3.142, 3.166, 3.166,$ \\ $3.171, 6.018, 4.696, 4.718, 3.142]$ \\ $[1.702, 3.976, 1.369, 3.017, 2.552, 2.447,$ \\ $2.539, 6.254, 0.352, 5.622, 3.498]$ \\ $[1.703, 2.287, 2.094, 3.237, 3.136, 3.211,$ \\ $3.178, 6.283, 2.942, 5.658, 3.717]$ \\ $[1.499, 3.778, 1.360, 2.496, 2.355, 2.939,$ \\ $2.273, 6.078, 4.990, 5.584, 3.703]$ \\} \\
\hline
$\text{LiH}$
& \makecell[l]{${\update R} = 1.6$ \AA} & $[0.1] \times 9$ & \makecell{$[0.483,  1.713, 4.808, 4.600,  2.326, $ \\ $1.048, 6.157, 6.157, 6.157]$} \\ 
\hline
\end{tabular}
\begin{tablenotes}\footnotesize
\item [$\text{a}$] All $\vec{\theta}^{(0)}$ were chosen from trial-and-error runs on fake IBM backends that yielded satisfactory results, prior to runs on the corresponding real backends. 
\end{tablenotes}
\end{threeparttable}
\label{tab:VQE_params}
\end{center}
\end{table}
\clearpage
\newpage
\section{Simulation data for extrapolation} \label{app: extrapol}
{\update In this section, we extrapolate the results of our VQE simulations. To do so, we extend the models studied in the main text to larger systems, scale the depths of the ansatz circuits, and apply repeated instances of our techniques. For practical reasons, we perform these as classical simulations, either with noise completely absent or using a simplified noise model. {\update Nevertheless, our results and subsequent analyses provide a first indication of how our MB-infused circuit VQE techniques fare in more complex scenarios. \par
\textit{\ul{Scalability of ansatz modification}}. We first consider PC lattices with multiple plaquettes in both spatial directions. In addition to the $7$-qubit $(M, N) = (2, 1)$ system studied on a quantum computer, we also conduct larger-scale simulations for $(M, N) = (3, 1)$ and $(2, 2)$, which require lattices of $10$ and $12$ qubits respectively. The Hamiltonians for the three systems are
\begin{align}
\begin{split}
\hat{\mathcal{H}}^{(2, 1)} &= -\hat{X}_{1}\hat{X}_{2}\hat{X}_{3}\hat{X}_{4} - \hat{X}_{4}\hat{X}_{5}\hat{X}_{6}\hat{X}_{7} -\hat{Z}_{2}\hat{Z}_{4}\hat{Z}_{5} - \hat{Z}_{3}\hat{Z}_{4}\hat{Z}_{6} -  \hat{Z}_{1}\hat{Z}_{2} - \hat{Z}_{1}\hat{Z}_{3} - \hat{Z}_{5}\hat{Z}_{7} - 
\hat{Z}_{6}\hat{Z}_{7} + \xi\sum_{i=1}^{7}\hat{Z}_{i};
\label{eq:2x1PC}
\end{split} \\ 
\begin{split}
\hat{\mathcal{H}}^{(3, 1)} &= -\hat{X}_{1}\hat{X}_{2}\hat{X}_{3}\hat{X}_{4} -\hat{X}_{4}\hat{X}_{5}\hat{X}_{6}\hat{X}_{7} -
\hat{X}_{7}\hat{X}_{8}\hat{X}_{9}\hat{X}_{10} -
\hat{Z}_{2}\hat{Z}_{4}\hat{Z}_{5} - \hat{Z}_{3}\hat{Z}_{4}\hat{Z}_{6} -
\hat{Z}_{5}\hat{Z}_{7}\hat{Z}_{8} - 
\hat{Z}_{6}\hat{Z}_{7}\hat{Z}_{9} -
\hat{Z}_{1}\hat{Z}_{2} \\ & \dots- \hat{Z}_{1}\hat{Z}_{3} -\hat{Z}_{8}\hat{Z}_{10} - 
\hat{Z}_{9}\hat{Z}_{10} + \xi\sum_{i=1}^{10}\hat{Z}_{i}; 
\label{eq:3x1PC}
\end{split} \\
\begin{split}
\hat{\mathcal{H}}^{(2, 2)} &= \hspace{0.1cm} -\hat{X}_{1}\hat{X}_{3}\hat{X}_{4}\hat{X}_{6} - \hat{X}_{2}\hat{X}_{4}\hat{X}_{5}\hat{X}_{7} - \hat{X}_{6}\hat{X}_{8}\hat{X}_{9}\hat{X}_{11} - \hat{X}_{7}\hat{X}_{9}\hat{X}_{10}\hat{X}_{12} -  \hat{Z}_{4}\hat{Z}_{6}\hat{Z}_{7}\hat{Z}_{9} -\hat{Z}_{3}\hat{Z}_{6}\hat{Z}_{8} - \hat{Z}_{5}\hat{Z}_{7}\hat{Z}_{10}\\ &\dots- 
\hat{Z}_{1}\hat{Z}_{2}\hat{Z}_{4} - \hat{Z}_{9}\hat{Z}_{11}\hat{Z}_{12} - \hat{Z}_{1}\hat{Z}_{3} - \hat{Z}_{2}\hat{Z}_{5} - \hat{Z}_{8}\hat{Z}_{11} - \hat{Z}_{10}\hat{Z}_{12} + \xi\sum_{i=1}^{12}\hat{Z}_{i}, 
\label{eq:2x2PC}
\end{split}
\end{align}
which encompass all plaquette, star, and perturbation operators. From these Hamiltonians, one may derive the stabilizer graphs corresponding to their GS [Fig.~\ref{fig:pc_extr}(a)]. \par
To choose appropriate ansatz modifications for each $\text{CZ}$, we employ a strategic approach based on the entangling behaviour of each qubit in the graph. This may be quantified as the entanglement entropy, computed by partitioning qubit $i$ from the rest of the graph via the partial trace and then determining the von Neumann entropy
\begin{equation}
\mathcal{S}(\rho_{i}) = -\operatorname{Tr}_{j \neq i}{[\rho_{\mathrm{a}} \text{log}(\rho_{\mathrm{a}})],}
\end{equation}
where $\rho_{\mathrm{a}} = |\psi_{\mathrm{a}}\rangle\langle\psi_{\mathrm{a}}|$ and $i \in \{1,2,...,n_{\mathrm{q}}\}$. For a single qubit, $\mathcal{S}$ ranges between 0 (separable) and 1 (maximally entangled). As discussed below, this provides a method to characterize how such modifications “tune” the amount of entanglement during VQE runs. Based on the underlying model symmetry, we may deduce that for the GS ($\rho_{\mathrm{a}} = \rho_{0}$), qubits adjacent to corners of the lattice possess identical entropies, but are relatively lower than those of other (inner) qubits. The inner qubits are generally positioned near a majority of the edges (entanglement bonds), and therefore exert a greater influence on the entanglement structure via the area law. This influence-based categorization suggests the need to apply different ansatz modifications to $\text{CZ}$s spanning corner-adjacent or inner qubits. Here, we apply Mod.~1 (a minimal 2-qubit modification) to each $\text{CZ}$ between two corner-adjacent qubits and Mod.~2 (a 3-qubit modification) to any two symmetrically positioned $\text{CZ}$s involving at least one inner qubit. To curb the larger resource overhead of Mod.~2, we restrict the eligible connections to the most impactful ones, namely, a) $\text{CZ}$s spanning multiple plaquettes, and b) $\text{CZ}$s involving inner qubits only. After acting Mod.~1 on all remaining $\text{CZ}$s, we obtain the graphs shown in Fig.~\ref{fig:pc_extr}(a). \par 
When modifying a $\text{CZ}$ multiple times ($L \geq 2$), we work exclusively on the circuit level and regard it as a concatenation of $2L-1$ modification circuits (i.e. layers in the traditional sense).
Here, the odd number of concatenations is needed to avoid gate cancellation when $\theta \hspace{0.1cm} \text{mod} \hspace{0.1cm} 2\pi = 0$. Owing to the asymmetric nature of Mod.~1, we alternate placement of the $RY$ gate such that the modification acts as $CZ_{m, n}RY_{k}(\theta)$, where $k = n(m)$ for every odd (even) numbered repetition. This ensures that the state space is not restricted in situations where qubits are subjected to $\text{CZ}$s only (in these cases, zero amplitudes are incurred in the parameterized unitary matrix.)
\par
Since the perturbation acts uniformly on each qubit, the applied symmetry is preserved for all $\xi$ values at the GS (Sec.~\ref{app: pc_perturb}). This implies that qubits on \textit{equivalent} vertices of the graph [i.e. with similarly positioned neighbours and connections; see Fig.~\ref{fig:pc_extr}(a)] are described by the same $\theta$. We exploited this symmetry in the $(M, N, L) = (2, 1, 1)$ demonstration on IBM, which enabled a reduction from 22 ($2\times 7 + 8$)} to 11 ($2\times 4 + 3$) free parameters. Applying symmetry arguments to each layer separately, we find that with the chosen modifications, $(2, 1, L), (3, 1, L)$ and $(2, 2, L)$ require $6L+5$, $20L+2$ and $18L+3$ parameters respectively (reduced from $16L+6$, $36L+2$ and $38L+5$) [Fig.~\ref{fig:pc_extr}(a)]. \par
To obtain the full circuit ansatz, a temporal order for the modifications must be assigned. By convention, we perform all Mod.~2 $\text{CZ}$s first (if present) followed by all Mod.~1 $\text{CZ}$s, such that their corresponding edges proceed left-to-right, top-to-bottom in the graph. Since Mod.~2 influences a larger area of the graph, we may regard this variationally as performing coarse “knob” adjustments of the circuit ansatz, followed by finer adjustments from Mod.~1. Although such modifications may be performed in a nose-blind way (i.e. by directly modifying the initial stabilizer circuit), here we exploit qualitative entanglement properties of the graph so that our choice of modification for each $\text{CZ}$ is well-informed. Moreover, while it is possible that different arrangements of Mod.~1 and 2 $\text{CZ}$s may yield differences in performance and/or convergence (due to their noncommuting nature), we expect these differences to be minimal provided reasonable initial guesses, since the circuits are variationally equivalent up to permutations or multiples of parameters. Fig.~\ref{fig:pc_extr}(b) depicts the circuit ansatz designed for $(M, N, L) = (2, 2, 2)$, containing three modification circuits in series. As before, each circuit contains additional $H$ gates derived from the unperturbed graph, and local $RY$ and $RX$ layers appended at the beginning and end respectively. \par
With the circuit ansatzes in place, we now present the VQE simulation results for $(M, N, L) = (2, 1, 1), (2, 1, 2)$, $(3, 1, 1), (3, 1, 2), (2, 2, 1), (2, 2, 2)$ and $(2, 2, 3)$, with $ 10^{-2} \leq \xi \leq 10^{1}$. Each ansatz was optimized using the BFGS algorithm with a minimum gradient tolerance of $10^{-6}$. For thorough exploration of the state space, we perform the optimization multiple times (a quasi-basinhopping method) using:
\begin{enumerate}[label={(\arabic*)}]
\item{Fixed Clifford initial guesses: \\ $\vec{\theta}^{(0)} \in \{[\pi + 0.1], [2\pi + 0.1], [\pi + 0.01], [2\pi + 0.01]\} \times n_{\text{par}}$, where $n_{\text{par}}$ is the total number of variational parameters.}
\item{The parameters $\vec{\theta}_{opt}$ corresponding to the lowest energy found in (1) for: \\
a) $\xi = 10^{-2}$, and setting  $\vec{\theta}^{(0)} = \vec{\theta}^{\text{(opt)}}$  for the next largest $\xi$. \\
b) $\xi = 10^{1}$, and setting $\vec{\theta}^{(0)} = \vec{\theta}^{\text{(opt)}}$ for the next smallest $\xi$.}
\end{enumerate}
The small offsets added to the Clifford guesses are used to avoid eigenstates of $|G\rangle$, which behave as local minima when $\xi > 0$ (Sec.~\ref{app: pc_perturb}). Such guesses correspond to near-maximally entangled states with $0.9 \leq \mathcal{S}(\rho_{i}) < 1 \hspace{0.1cm} \forall \hspace{0.1cm} i \in \{1, 2,...,n_{\mathrm{q}}\}$. The result with the smallest relative error $\Delta E/E_{g}$ was then taken to yield the plot in Fig.~\ref{fig:pc_extr}(c). \par
\begin{figure}
     \centering\subfloat{\includegraphics[width=18.5cm]{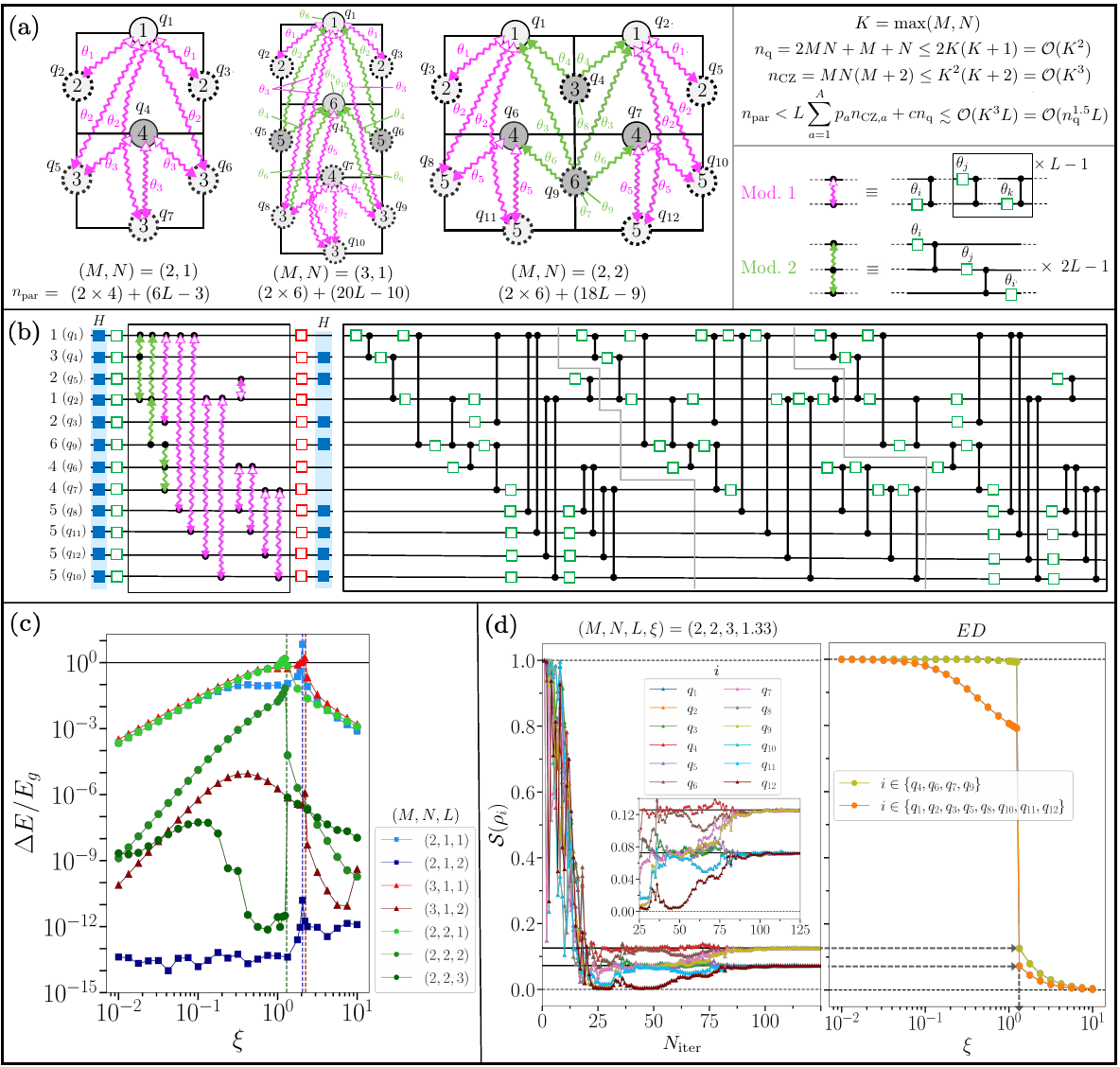}}
    \caption{\update{\textbf{Noiseless VQE simulation of ansatz modification, applied to the $M \times N$ planar code.} (a) The stabilizer graphs used to construct the gate-modified circuits for the VQE, employing the notation in Fig.~1. For ease of visualization, the modifications are denoted explicitly on the graphs themselves. The applied symmetry is represented as follows: qubits with the same number inside their circles are parameterized with the same angle (in the rotation layers), while each modified $\text{CZ}$-edge is parameterized with the angle(s) $\theta$ shown next to the modifications. The total number of parameters $n_{\text{par}}$ for each $(M, N)$ studied is given in terms of the number of modification layers $L$. A general scaling analysis is also shown, which relates $n_{\text{par}}$ to the number of entangling gates ($n_{\mathrm{CZ}}$), parameters per modification ($p$), layers ($L$), qubits ($n_{\mathrm{q}}$), and local rotation layers ($c$). Qubits (circles) are labelled as $q_{i}$, and grouped according to their entangling behaviour. The lightest shade of grey denotes corner-adjacent qubits while darker shades denote inner qubits with the same entanglement entropy ($\mathcal{S}$). (b) Schematic of the modified VQE circuit ansatz for $(M, N, L) = (2, 2, 2)$. The entire modification is enclosed with a dashed
    box, and the grey lines delimit each constituent circuit in the modification.  (c) VQE results corresponding to the modified ansatzes, showing plots of $\Delta E/E_{g}$ vs. $\xi$ for various $(M, N, L)$. The approximate crossover points for these systems are shown as vertical dotted lines. The optimizer used is BFGS. (d) Left panel: $\mathcal{S}$ corresponding to $\xi = 1.33$ for $(M, N, L) = (2, 2, 3)$, shown for each qubit $i$ as a progress plot (vs. $N_{\text{iter}}$). The dotted lines indicate entropy bounds and solid lines indicate  $\mathcal{S}_{i}$ corresponding to the exact GS energy ($E_{0}$) for $\xi = 1.33$. Right panel: $S_{i}$ (at $E_{0}$) vs. $\xi$, calculated via exact diagonalization (ED).}}
    \label{fig:pc_extr}
 \end{figure}
From Fig.~\ref{fig:pc_extr}(c), we observe that for $L=1$, the errors are small ($\approx 10^{-4}$) toward the limiting cases but tend to be large for intermediate $\xi$ ($10^{-1}$ to $6.8$, indicating optimized energies significantly above $E_{1}(\xi)$). This “crossover” regime is characterized by smaller $E_{g}$ (approaching near-degeneracy), and is where the GS switches from a highly entangled to a near-product state (see Sec.~\ref{app: pc_perturb}). The former leads to optimization difficulties in resolving $E_{g}$ while the latter leads to rapid changes followed by a discontinuity in the error's slope. This behaviour is similarly observed for $L = 2$, with  reduced errors stemming from the extra variational parameters  [$10^{-14}-10^{-11}$, $10^{-12}-10^{-5}$, $10^{-10}-10^{-1}$ for $(M, N) = (2, 1), (3, 1), (2, 2)$ respectively]. Of peculiarity is the $(M, N, L) = (2, 2, 3)$ result, where we observe smaller errors of $10^{-12}$ in the crossover regime but slightly larger errors than those of $(2, 2, 2)$ toward the limiting cases. This may be attributed to the search landscape being sufficiently high-dimensional, resulting in increased ansatz susceptibility to barren plateaus. Such plateaus tend to be prevalent for parameters near Clifford angles: since the limiting-case states $|G\rangle$ and $|1\rangle^{\otimes n_{\mathrm{q}}}$ require all parameters to be exact multiples of $\pi/2$, the optimizer efficacy is reduced in achieving these states. \par 
It is interesting to note the $\xi$-value where the crossover occurs depends on the PC dimensions and number of star operators (see Sec.~\ref{app: pc_perturb}). It occurs approximately when
\begin{flalign}
\begin{split}
 E_{0}(\xi) &= (+1)[2(M+N-2)] + (-1)[(M-1)(N-1)+4] - n_{\mathrm{q}}\xi \\
&= 3(M+N-3)-MN-n_{\mathrm{q}}\xi,
\end{split}
\end{flalign}
where the RHS is the value of $\langle 1^{\otimes n_{\mathrm{q}}}|\hat{\mathcal{H}}^{(M, N)}|1^{\otimes n_{\mathrm{q}}}\rangle$, the amplitude primarily affected by the perturbation. Numerically, we find these crossover points to be around $\xi = 2.059, 2.25, 1.3$ for $(M, N) = (2, 1), (3, 1), (2, 2)$ respectively. \par
As previously mentioned, we may also characterize the variational tunability of our modified ansatzes via the entanglement entropy. We illustrate this for $(M, N, L) = (2, 2, 3)$ at a point near the crossover ($\xi = 1.33$), as shown in the left panel of Fig.~\ref{fig:pc_extr}(d). Here, we compute $S(\rho_{i}) \hspace{0.1cm} \forall \hspace{0.1cm} i \in \{1, 2,...,12\}$ at each optimizer iteration to obtain a progress plot. In a display of the ansatz's full tunability, the entropies fluctuate between 0 and 1 during the initial stages of the VQE run, where the optimizer explores the full parameter space. As the cost function begins to converge and the optimization step size decreases, the entropies undergo more controlled adjustments. Several local minima are also encountered (most notably when $S_{12} = 0$) and subsequently overcome by the optimizer, as evidenced by extended plateau regions followed by step-like changes. At $N_{\text{iter}} \approx 90$, the entropies converge at two distinct values (0.0725 and 0.126), which are in agreement with the entropies of the exact GS for $\xi = 1.33$. Plotting these over all simulated $\xi$ [Fig.~\ref{fig:pc_extr}(d), right panel], we see that the entropies of all qubits are always classified into two such values. As expected, the smaller entropies (orange line) correspond to corner-adjacent qubits ($q_{1}, q_{2}, q_{3}, q_{5}, q_{8}, q_{10}, q_{11}, q_{12}$) while the larger entropies (dark yellow line) correspond to the inner qubits ($q_{4}, q_{6}, q_{7}, q_{9}$). In addition, the trend of the entropies closely mirrors the perturbative effects on the PC: all qubits have $S \approx 1$ at small $\xi$ (since $|G\rangle$ is maximally entangled), undergo abrupt decreases at the crossover point, then plateau towards $S = 0$ at larger $\xi$ (since $|1\rangle^{\otimes n_{\mathrm{q}}}$ is separable).\par
Our VQE results highlight the suitability of ansatz modification to the PC, with effective choices for the initial guesses and modifications. For each ($M, N$) studied, all errors fell below those of $E_{1}$ for $L \geq 2$, signifying that with moderate circuit depth, our modifications are capable of tuning the entanglement structure to any desired degree. Such advantages also extend to arbitrary modifications: since each modified $\text{CZ}$ introduces a constant number of parameters $p$ (i.e. independent of other variables), the scaling in $n_{\text{par}}$ (the total number) is dependent only on the graph. To illustrate, we consider a $K \times K$ square lattice to derive an upper bound where $K = \text{max}(M, N)$. The unperturbed graph state has $n_{\mathrm{q}} \leq 2K(K+1)$ vertices (qubits) and $n_{\mathrm{CZ}} \leq K^{2}(K+2)$ edges. If we add $L$ layers of $A$ different modifications, there will be (prior to applying symmetry) $p_{a}n_{\mathrm{CZ},{a}}L$ parameters per modification $a$, and $cn_{\mathrm{q}}$ parameters from $c$ local rotation layers. Therefore,
\begin{flalign}
\begin{split}
n_{\text{par}} &< L\sum_{a=1}^{A}p_{a}n_{\mathrm{CZ},{a}} + cn_{\mathrm{q}} \\
 &\leq \max_{a}(p_{a})LK^{2}(K+2) + 2cK(K+1) \\ & \sim \mathcal{O}(K^{3}L) = \mathcal{O}(n_{\mathrm{q}}^{1.5}L),
\end{split}
\end{flalign}
where $a \in \{1, 2, \dots, A\}$ and $\sum_{a}n_{\mathrm{CZ}, a} = n_{\mathrm{CZ}}$. This indicates sub-quadratic scaling with the system size, which is comparable to the quadratic scaling of UCCSD ansatzes (used in quantum chemistry) \cite{nirmal2024}, and linear scaling of heuristic ansatzes (e.g. efficient $SU(2)$) \cite{Qiskit}. We further remark that $n_{\text{par}}$ scales only quadratically at most ($\mathcal{O}(n_{\mathrm{q}}^{2}L)$), corresponding to graph states with all-to-all connectivity. Since their wavefunctions scale exponentially with $n_{\mathrm{q}}$, ansatz modification presents a feasible variational approach to models represented by arbitrarily complex graphs. \par}
{\update\textit{\ul{Scalability of Pauli gadgets: Rectangular $\mathrm{H}_4$}}.
    Next, we demonstrate a circuit ansatz containing multiple Pauli gadgets. As test system, we consider the rectangular $\mathrm{H}_4$ molecule which is well known for possessing strong correlations. This molecule is formed by two parallel $\mathrm{H}_2$ molecules [Fig.~\ref{fig:rect_h4}(a)] and have edge lengths $d$ and $R$; for concreteness we choose $d=1.5$\AA. We calculate the molecular Hamiltonian in the sto-3g basis using the OpenFermion~\cite{openfermion} and PySCF~\cite{pyscf} libraries, with the Jordan-Wigner transformation to map fermionic operators to qubits. The resulting 8-qubit Hamiltonian is shown in Table~\ref{table:H4ham} for $R=1.5$\AA\ and $R=2.5$\AA, with GS energies $E_{0}(1.5\text{\AA}) = -1.9551250116$ and $E_{0}(2.5\text{\AA}) = -1.9932657424$, and GS wave functions\par
\begin{align}
|\psi_{0}(d=1.5\text{\AA})\rangle = \hspace{0.2cm} & 0.6538|11110000\rangle -0.6538|11001100\rangle +0.1955|10010110\rangle +0.1955|01101001\rangle\nonumber \\
&+ 0.1232|00001111\rangle - 0.1232|00110011\rangle -0.0978|10011001\rangle -0.0978|01100110\rangle\nonumber \\
&-0.0978|10100101\rangle -0.0978|01011010\rangle;\nonumber \\
|\psi_{0}(d=2.5\text{\AA})\rangle = \hspace{0.2cm} & 0.9879|11110000\rangle -0.0663|11001100\rangle -0.0580|00111100\rangle -0.0580|11000011\rangle\nonumber \\
&+ 0.0578|10010110\rangle + 0.0578|01101001\rangle -0.0484|10011001\rangle -0.0484|01100110\rangle\nonumber \\
&-0.0367|00110011\rangle +0.0118|00001111\rangle -0.0094|01011010\rangle -0.0094|10100101\rangle.
\end{align} 
We build the circuit ansatz using single-qubit rotations and {\update $\mathcal{\hat{P}}^{(4)}$ gadgets, with $\mathcal{\hat{P}} = \hat{Y}\hat{X}\hat{X}\hat{Y}$} [Fig.~\ref{fig:rect_h4}(b)]. After the initial layer of single-qubit rotations, we apply a group of $g$ $\mathcal{\hat{P}}^{(4)}$ gadgets ($g=3, 5, 7$) followed by parameterized $RZ$ rotations on all qubits. The $g$ gadgets and the $RZ$ rotations are then repeated once. Each gadget is of the form $ \hat{Y}_i \hat{X}_j \hat{X}_k \hat{Y}_l$, where the qubit indices $(ijkl)$ are chosen as $\{(1278), (3456), (2367)\}$ for $g=3$, $\{(1278), (3456), (1256), (3478), (2367)\}$ for $g=5$, and $\{(1278), (3456), (1256), (3478), (1234), (5678), (2367)\}$ for $g=7$. Our choice of gadgets is motivated by the fact that all 4-qubit Pauli terms in the Hamiltonian act on these groups of qubits and contain two $X$ and two $Y$ (see Table~\ref{table:H4ham}). In addition, this choice entangles all qubits while (assuming all-to-all connectivity) reducing the circuit depth as much as possible: all but the last gadget $(2367)$ can be applied in parallel in pairs. While the initial single-qubit rotations may be arbitrarily chosen from $SU(2)$, our optimization results suggest that they can be chosen as $\hat{X}_1 \hat{X}_2 \hat{X}_3 \hat{X}_4$ (i.e. starting from the Hartree-Fock state) without reducing the performance. Thus, the total number of parameters to optimize is $2(g+8)$. We simulate VQE on a classical computer using state vectors, so that both statistical noises and device noises are absent.\par
The simulated VQE results are given in Fig.~\ref{fig:rect_h4}(c). In the left panel, we show the absolute error $\Delta E$ of the GS energy as functions of the intermolecular distance $R$ for $g=3, 5, 7$. $\Delta E$ is shown as a function of the number of {\update optimizer iterations} in the right panel for $R=2.5$\AA\ and $R=1.5$\AA. The exact GS energy and energy of the first excited state are plotted in the inset. The state infidelity $1-\mathcal{F}$ (not shown) behaves in a fashion qualitatively similar to $\Delta E$. For $g = 3, 5, 7$ respectively, we obtain $\mathcal{F} > 0.8718, 0.9977, 0.9999$ for $0.3 \leq R \leq 2.5$\AA. For the same $R$, with $g=5$, we achieve chemical accuracy ($\Delta E = 10^{-6}-10^{-3}$) except in a very small neighbourhood of $R=1.5$\AA; remarkably, the errors improve by orders of magnitude ($\Delta E = 10^{-11}-10^{-10}$) with only four additional gadgets ($g=7$). As expected, increasing the number of gadgets leads to slower convergence but higher accuracy. It is interesting to note that the ansatz with $g=5$ has a clear peak in $\Delta E$ at $R=1.5$\AA, corresponding to the strongly correlated square configuration, while neither $g=3$ nor $g=7$ exhibits strong features. \par Overall, our results showcase two key features of Pauli gadgets: 1) they may be sequenced in ways such that several gadgets can be performed in parallel (thereby reducing the scaling in depth), and 2) using an informed choice of gadgets based on specific Hamiltonian terms can greatly reduce the number $g$ needed to achieve a desired accuracy.} 

\begin{figure}
    \centering
    \subfloat{\includegraphics[width=18.5cm]{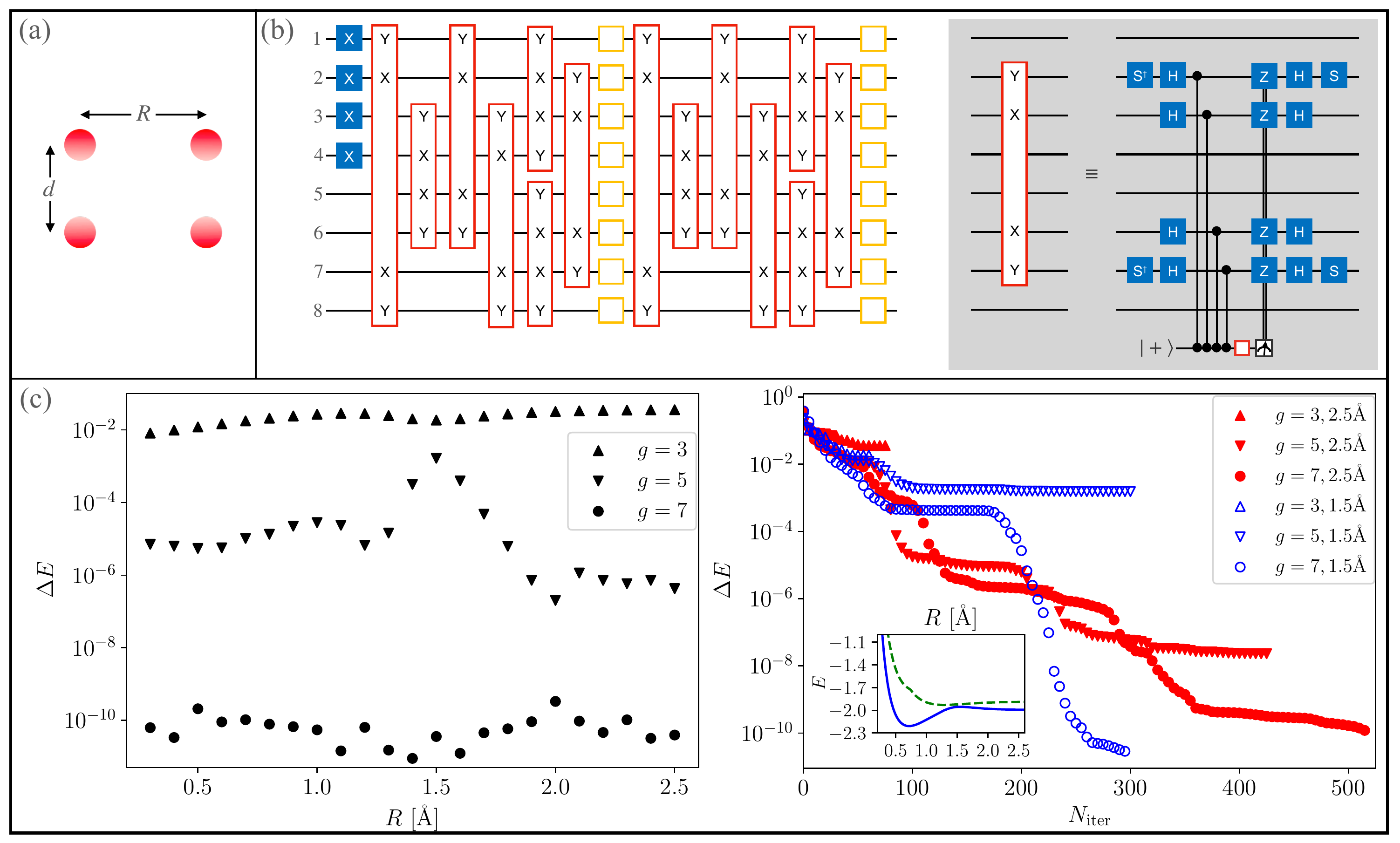}}
    \caption{{\update \textbf{VQE simulation of the rectangular $\mathrm{H}_4$ molecule using Pauli gadgets}. (a) Schematic of the molecule, with edge lengths of the rectangle $d$ and $R$. (b) Circuit ansatz with $2$ gadget layers and $2$ $RZ$ layers. Each gadget layer contains $g$ $\mathcal{\hat{P}}^{(4)}$ gadgets; here $\mathcal{\hat{P}} = \hat{Y}\hat{X}\hat{X}\hat{Y}$} and $g=7$. An example gadget is shown in the box, where the $Z$ gates are applied only if the ancilla measurement outcome is $1$ (see Fig.~\ref{z2_ansatz}). (c) Noiseless VQE simulation results with $d=1.5$\AA\ and various numbers of gadgets $g$ ($g = 3, 5, 7$). The absolute error in the GS energy $\Delta{E} = E_{\text{VQE}}-E_{0}$ is plotted in the left panel as a function of the intermolecular distance $R$, and in the right panel as a function of the number of optimizer iterations $N_{\text{iter}}$. The exact GS energy and the energy of the first excited state are shown as solid and dashed lines respectively in the inset. The optimizer used is BFGS.}
 \label{fig:rect_h4}
 \end{figure} 
 
\begin{table}[htb]
\caption{{\update 8-qubit Hamiltonian for the rectangular $\mathrm{H}_4$ molecule with edge lengths $(d, R)=(1.5, 1.5)$\AA\ and $(1.5, 2.5)$\AA.}}
\begin{center}
\begin{tabular}{|c|c|c|c|c|c|c|c|}
 
 \multicolumn{8}{c}{$(d, R)=(1.5, 1.5)$\AA}
 \\ \hline
\makecell{$IIIIIIII$ \\ $-0.8823$}  & \makecell{$ZIIIIIII$ \\ $0.1473$}  & \makecell{$IZIIIIII$ \\ $0.1473$}  & \makecell{$IIIIIIZZ$ \\ $0.1126$}  & \makecell{$IIZIIIIZ$ \\ $0.1093$}  & \makecell{$IIIZIIZI$ \\ $0.1093$}  & \makecell{$IIIIZIIZ$ \\ $0.1093$}  & \makecell{$IIIIIZZI$ \\ $0.1093$}  
 \\ \hline 
 \makecell{$IIIIZZII$ \\ $0.1080$}  & \makecell{$IIZZIIII$ \\ $0.1080$}  & \makecell{$ZIIIIIIZ$ \\ $0.1072$}  & \makecell{$IZIIIIZI$ \\ $0.1072$}  & \makecell{$ZZIIIIII$ \\ $0.1072$}  & \makecell{$IIIIIIIZ$ \\ $-0.1071$}  & \makecell{$IIIIIIZI$ \\ $-0.1071$}  & \makecell{$ZIIIIZII$ \\ $0.1062$}  
 \\ \hline 
 \makecell{$IZIIZIII$ \\ $0.1062$}  & \makecell{$ZIIZIIII$ \\ $0.1062$}  & \makecell{$IZZIIIII$ \\ $0.1062$}  & \makecell{$IIZIIZII$ \\ $0.1062$}  & \makecell{$IIIZZIII$ \\ $0.1062$}  & \makecell{$ZIIIIIZI$ \\ $0.0841$}  & \makecell{$IZIIIIIZ$ \\ $0.0841$}  & \makecell{$IIZIZIII$ \\ $0.0819$}  
 \\ \hline 
 \makecell{$IIIZIZII$ \\ $0.0819$}  & \makecell{$ZIZIIIII$ \\ $0.0721$}  & \makecell{$IZIZIIII$ \\ $0.0721$}  & \makecell{$ZIIIZIII$ \\ $0.0721$}  & \makecell{$IZIIIZII$ \\ $0.0721$}  & \makecell{$IIIIZIZI$ \\ $0.0710$}  & \makecell{$IIIIIZIZ$ \\ $0.0710$}  & \makecell{$IIZIIIZI$ \\ $0.0710$}  
 \\ \hline 
 \makecell{$IIIZIIIZ$ \\ $0.0710$}  & \makecell{$IIYXIIXY$ \\ $0.0383$}  & \makecell{$IIYYIIXX$ \\ $-0.0383$}  & \makecell{$IIXXIIYY$ \\ $-0.0383$}  & \makecell{$IIXYIIYX$ \\ $0.0383$}  & \makecell{$IIIIYXXY$ \\ $0.0383$}  & \makecell{$IIIIYYXX$ \\ $-0.0383$}  & \makecell{$IIIIXXYY$ \\ $-0.0383$}  
 \\ \hline 
 \makecell{$IIIIXYYX$ \\ $0.0383$}  & \makecell{$YZZXXZZY$ \\ $-0.0353$}  & \makecell{$YZZYXZZX$ \\ $0.0353$}  & \makecell{$XZZXYZZY$ \\ $0.0353$}  & \makecell{$XZZYYZZX$ \\ $-0.0353$}  & \makecell{$IYXIIXYI$ \\ $-0.0353$}  & \makecell{$IYYIIXXI$ \\ $0.0353$}  & \makecell{$IXXIIYYI$ \\ $0.0353$}  
 \\ \hline 
 \makecell{$IXYIIYXI$ \\ $-0.0353$}  & \makecell{$YZYIIYZY$ \\ $-0.0353$}  & \makecell{$YZYIIXZX$ \\ $-0.0353$}  & \makecell{$XZXIIYZY$ \\ $-0.0353$}  & \makecell{$XZXIIXZX$ \\ $-0.0353$}  & \makecell{$IYZYYZYI$ \\ $-0.0353$}  & \makecell{$IYZYXZXI$ \\ $-0.0353$}  & \makecell{$IXZXYZYI$ \\ $-0.0353$}  
 \\ \hline 
 \makecell{$IXZXXZXI$ \\ $-0.0353$}  & \makecell{$YXIIXYII$ \\ $0.0341$}  & \makecell{$YYIIXXII$ \\ $-0.0341$}  & \makecell{$XXIIYYII$ \\ $-0.0341$}  & \makecell{$XYIIYXII$ \\ $0.0341$}  & \makecell{$YXXYIIII$ \\ $0.0341$}  & \makecell{$YYXXIIII$ \\ $-0.0341$}  & \makecell{$XXYYIIII$ \\ $-0.0341$}  
 \\ \hline 
 \makecell{$XYYXIIII$ \\ $0.0341$}  & \makecell{$IIYXXYII$ \\ $0.0243$}  & \makecell{$IIYYXXII$ \\ $-0.0243$}  & \makecell{$IIXXYYII$ \\ $-0.0243$}  & \makecell{$IIXYYXII$ \\ $0.0243$}  & \makecell{$YZZXIXYI$ \\ $0.0237$}  & \makecell{$YZZYIYYI$ \\ $0.0237$}  & \makecell{$XZZXIXXI$ \\ $0.0237$}  
 \\ \hline 
 \makecell{$XZZYIYXI$ \\ $0.0237$}  & \makecell{$IYXIXZZY$ \\ $0.0237$}  & \makecell{$IYYIYZZY$ \\ $0.0237$}  & \makecell{$IXXIXZZX$ \\ $0.0237$}  & \makecell{$IXYIYZZX$ \\ $0.0237$}  & \makecell{$YXIIIIXY$ \\ $0.0231$}  & \makecell{$YYIIIIXX$ \\ $-0.0231$}  & \makecell{$XXIIIIYY$ \\ $-0.0231$}  
 \\ \hline 
 \makecell{$XYIIIIYX$ \\ $0.0231$}  & \makecell{$IIZIIIII$ \\ $0.0207$}  & \makecell{$IIIZIIII$ \\ $0.0207$}  & \makecell{$IIIIIZII$ \\ $0.0207$}  & \makecell{$IIIIZIII$ \\ $0.0207$}  & \makecell{$YZXIXZYI$ \\ $-0.0116$}  & \makecell{$XZYIYZXI$ \\ $-0.0116$}  & \makecell{$IYZXIXZY$ \\ $-0.0116$}  
 \\ \hline 
 \makecell{$IXZYIYZX$ \\ $-0.0116$}  & \makecell{$YZYIYZYI$ \\ $-0.0116$}  & \makecell{$XZXIXZXI$ \\ $-0.0116$}  & \makecell{$IYZYIYZY$ \\ $-0.0116$}  & \makecell{$IXZXIXZX$ \\ $-0.0116$}  
\\ \hline 
 \multicolumn{8}{c}{$(d, R)=(1.5, 2.5)$\AA}
\\ \hline
\makecell{$IIIIIIII$ \\ $-0.9795$}  & \makecell{$ZIIIIIII$ \\ $0.1071$}  & \makecell{$IZIIIIII$ \\ $0.1071$}  & \makecell{$IIIIIIZZ$ \\ $0.0982$}  & \makecell{$IIIIZIIZ$ \\ $0.0971$}  & \makecell{$IIIIIZZI$ \\ $0.0971$}  & \makecell{$IIIIZZII$ \\ $0.0961$}  & \makecell{$IIZIIIIZ$ \\ $0.0955$}  
 \\ \hline 
 \makecell{$IIIZIIZI$ \\ $0.0955$}  & \makecell{$IIZZIIII$ \\ $0.0949$}  & \makecell{$IIZIIZII$ \\ $0.0945$}  & \makecell{$IIIZZIII$ \\ $0.0945$}  & \makecell{$ZIIIIIIZ$ \\ $0.0943$}  & \makecell{$IZIIIIZI$ \\ $0.0943$}  & \makecell{$ZIIZIIII$ \\ $0.0939$}  & \makecell{$IZZIIIII$ \\ $0.0939$}  
 \\ \hline 
 \makecell{$ZIIIIZII$ \\ $0.0935$}  & \makecell{$IZIIZIII$ \\ $0.0935$}  & \makecell{$ZZIIIIII$ \\ $0.0932$}  & \makecell{$IIIZIIII$ \\ $0.0794$}  & \makecell{$IIZIIIII$ \\ $0.0794$}  & \makecell{$ZIIIIIZI$ \\ $0.0680$}  & \makecell{$IZIIIIIZ$ \\ $0.0680$}  & \makecell{$IIZIZIII$ \\ $0.0674$}  
 \\ \hline 
 \makecell{$IIIZIZII$ \\ $0.0674$}  & \makecell{$ZIIIZIII$ \\ $0.0637$}  & \makecell{$IZIIIZII$ \\ $0.0637$}  & \makecell{$IIZIIIZI$ \\ $0.0637$}  & \makecell{$IIIZIIIZ$ \\ $0.0637$}  & \makecell{$ZIZIIIII$ \\ $0.0495$}  & \makecell{$IZIZIIII$ \\ $0.0495$}  & \makecell{$IIIIYXXY$ \\ $0.0488$}  
 \\ \hline 
 \makecell{$IIIIYYXX$ \\ $-0.0488$}  & \makecell{$IIIIXXYY$ \\ $-0.0488$}  & \makecell{$IIIIXYYX$ \\ $0.0488$}  & \makecell{$IIIIZIZI$ \\ $0.0483$}  & \makecell{$IIIIIZIZ$ \\ $0.0483$}  & \makecell{$IIIIIIIZ$ \\ $-0.0475$}  & \makecell{$IIIIIIZI$ \\ $-0.0475$}  & \makecell{$YZYIIYZY$ \\ $0.0456$}  
 \\ \hline 
 \makecell{$YZYIIXZX$ \\ $0.0456$}  & \makecell{$XZXIIYZY$ \\ $0.0456$}  & \makecell{$XZXIIXZX$ \\ $0.0456$}  & \makecell{$IYZYYZYI$ \\ $0.0456$}  & \makecell{$IYZYXZXI$ \\ $0.0456$}  & \makecell{$IXZXYZYI$ \\ $0.0456$}  & \makecell{$IXZXXZXI$ \\ $0.0456$}  & \makecell{$YXXYIIII$ \\ $0.0444$}  
 \\ \hline 
 \makecell{$YYXXIIII$ \\ $-0.0444$}  & \makecell{$XXYYIIII$ \\ $-0.0444$}  & \makecell{$XYYXIIII$ \\ $0.0444$}  & \makecell{$IIYXIIXY$ \\ $0.0318$}  & \makecell{$IIYYIIXX$ \\ $-0.0318$}  & \makecell{$IIXXIIYY$ \\ $-0.0318$}  & \makecell{$IIXYIIYX$ \\ $0.0318$}  & \makecell{$YZZXXZZY$ \\ $0.0307$}  
 \\ \hline 
 \makecell{$YZZYXZZX$ \\ $-0.0307$}  & \makecell{$XZZXYZZY$ \\ $-0.0307$}  & \makecell{$XZZYYZZX$ \\ $0.0307$}  & \makecell{$IYXIIXYI$ \\ $0.0307$}  & \makecell{$IYYIIXXI$ \\ $-0.0307$}  & \makecell{$IXXIIYYI$ \\ $-0.0307$}  & \makecell{$IXYIIYXI$ \\ $0.0307$}  & \makecell{$YXIIXYII$ \\ $0.0298$}  
 \\ \hline 
 \makecell{$YYIIXXII$ \\ $-0.0298$}  & \makecell{$XXIIYYII$ \\ $-0.0298$}  & \makecell{$XYIIYXII$ \\ $0.0298$}  & \makecell{$IIYXXYII$ \\ $0.0271$}  & \makecell{$IIYYXXII$ \\ $-0.0271$}  & \makecell{$IIXXYYII$ \\ $-0.0271$}  & \makecell{$IIXYYXII$ \\ $0.0271$}  & \makecell{$YZZXIXYI$ \\ $-0.0267$}  
 \\ \hline 
 \makecell{$YZZYIYYI$ \\ $-0.0267$}  & \makecell{$XZZXIXXI$ \\ $-0.0267$}  & \makecell{$XZZYIYXI$ \\ $-0.0267$}  & \makecell{$IYXIXZZY$ \\ $-0.0267$}  & \makecell{$IYYIYZZY$ \\ $-0.0267$}  & \makecell{$IXXIXZZX$ \\ $-0.0267$}  & \makecell{$IXYIYZZX$ \\ $-0.0267$}  & \makecell{$YXIIIIXY$ \\ $0.0264$}  
 \\ \hline 
 \makecell{$YYIIIIXX$ \\ $-0.0264$}  & \makecell{$XXIIIIYY$ \\ $-0.0264$}  & \makecell{$XYIIIIYX$ \\ $0.0264$}  & \makecell{$IIIIZIII$ \\ $-0.0239$}  & \makecell{$IIIIIZII$ \\ $-0.0239$}  & \makecell{$YZYIYZYI$ \\ $0.0189$}  & \makecell{$XZXIXZXI$ \\ $0.0189$}  & \makecell{$IYZYIYZY$ \\ $0.0189$}  
 \\ \hline 
 \makecell{$IXZXIXZX$ \\ $0.0189$}  & \makecell{$YZYIXZXI$ \\ $0.0149$}  & \makecell{$XZXIYZYI$ \\ $0.0149$}  & \makecell{$IYZYIXZX$ \\ $0.0149$}  & \makecell{$IXZXIYZY$ \\ $0.0149$}  & \makecell{$YZXIXZYI$ \\ $0.0040$}  & \makecell{$XZYIYZXI$ \\ $0.0040$}  & \makecell{$IYZXIXZY$ \\ $0.0040$}  
 \\ \hline 
 \makecell{$IXZYIYZX$ \\ $0.0040$}  &  & & & & & &\\
 \hline
\end{tabular}
\end{center}
\label{table:H4ham}
\end{table}
\clearpage
\section{Perturbative analysis of the \texorpdfstring{$M \times N$}{M*N} planar code} 
\label{app: pc_perturb}
In this section, we analyze the $M \times N$ planar code on a more physical level by studying how the local perturbations act on its qubits. To achieve this, we employ time-independent perturbation theory to approximate the GS and associated energy as a function of the perturbation strength $\xi$. Since the free part of the Hamiltonian [Eq.~(1)] is $\hat{\mathcal{H}}_{0} = \hat{\mathcal{H}}_{\square} + \hat{\mathcal{H}}_{+}$, its corresponding GS energy is given by the contributions from all its corresponding plaquette and star operators:
\begin{equation}
E_{0}^{(0)} = -\underbrace{MN}_{\text{\# plaquettes}}-\underbrace{(M-1)(N-1)}_{\text{\# 4-body star}}-\underbrace{2(M+N-2)}_{\text{\# 3-body star}}-\underbrace{4}_{\text{\# 2-body star}} = \sum_{i=1}^{n_{\mathrm{q}}+1}(-1), \\
\end{equation} \\
where each operator contributes a minimum eigenvalue of \text{--}1 and the superscript $(0)$ indicates the unperturbed (uncorrected) case. On the other hand, the associated GS {\update{$|\psi_{0}^{(0)}\rangle$}} arises from the contribution of all self- and multi-plaquette interactions:
\begin{equation}
|\psi_{0}^{(0)}\rangle = \frac{1}{\sqrt{2^{MN}}}\bigg[1 + \sum_{p_1=1}^{MN} \prod_{p_{1}}\square_{p_{1}} + \sum_{\substack{p_1, p_2 = 1 \\ p_1 < p_2}}^{MN}\prod_{p_{1}, p_{2}}\square_{p_{1}}\square_{p_{2}} + ... + 
\square_{p_{1}}\square_{p_{2}}...\square_{p_{MN}}\bigg]|0\rangle^{\otimes n_{\mathrm{q}}}.
\label{EE2}
\end{equation}
Here, {\update $n_{\mathrm{q}} = 2MN + M + N$} is the number of qubits and each $\square_{p}$ is a 4-body plaquette operator $\hat{X}_{A}\hat{X}_{B}\hat{X}_{C}\hat{X}_{D}$, where $A,B,C,D$ labels the four qubits forming the edges of plaquette $p$. Since the $M \times N$ lattice possesses a graph state representation {\update $|G\rangle$}, we may cast Eq.~\eqref{EE2} in a simpler form involving its stabilizers. Denoting $Q$ as the set of all qubits \{1, 2,...,$n_{\mathrm{q}}$\}, $E_{\text{graph}}$ as the set of all $\text{CZ}$ edges (acting on qubits $a$ and $b$), and $LC \subset Q$ as the subset of qubits requiring local Clifford ($H$) operations, we have
\begin{equation}
{\update |G\rangle = \hspace{0.1cm}} |\psi_{0}^{(0)}\rangle = \bigg(\bigotimes_{q}H_{q}\bigg)\bigg(\prod_{E_{\text{graph}}}CZ_{a, b}\bigg)|+\rangle^{\otimes{n_{\mathrm{q}}}}; \hspace{0.5cm} q \in LC. 
\label{EE3}
\end{equation}
 From this expression, one may systematically generate all excited states by applying combinations of $\hat{X}$ and $\hat{Z}$ operators, which serve to flip the constituent eigenvalues of $|\psi_{0}\rangle$ from --1 to +1. These states form a complete, mutually orthonormal basis, and are given as
\begin{flalign}
|\psi_{n}^{(0)}\rangle = \bigg(\bigotimes_{j=1}^{n_{\mathrm{q}}}(\hat{X}_{j}^{\vec{x}_{j}})^{k}(\hat{Z}_{j}^{\vec{x}_{j}})^{1-k}\bigg)|\psi_{0}^{(0)}\rangle \hspace{0.3cm}  (n=1,2...),
\label{EE4} 
\end{flalign}
where $\vec{x} \in \{0,1\}^{n_q}: 
      \vec{x} \neq 0^{n_{\mathrm{q}}}$, and 
      $k = 0 \hspace{0.05cm} (1)$ for $j \notin LC \hspace{0.05cm} (j \in LC)$.
Based on Eq.~\eqref{EE4}, we can derive the action of the free Hamiltonian on an excited state, and obtain an exact form for the excited state energies (each energy corresponding to a specific $\vec{x}$):
\begin{flalign}
\begin{split}
\hat{\mathcal{H}}_{0}|\psi_{n}^{(0)}\rangle &= \sum_{i=1}^{n_{\mathrm{q}}+1}\hat{\mathcal{H}}_{0, i}\bigg(\bigotimes_{j=1}^{n_{\mathrm{q}}}(\hat{X}_{j}^{\vec{x}_{j}})^{k}(\hat{Z}_{j}^{\vec{x}_{j}})^{1-k}\bigg)|\psi_{0}^{(0)}\rangle
\end{split} \\
\begin{split}
&=  \sum_{i=1}^{n_{\mathrm{q}}+1}\bigotimes_{j=1}^{n_{\mathrm{q}}}\hat{A}_{j}^{i}|\psi_{0}^{(0)}\rangle; \hspace{0.2cm} \hat{A}_{j}^{i} \in \{\hat{I}_{j}, \hat{X}_{j}, \hat{Z}_{j}, \hat{Z}_{j}\hat{X}_{j}, \hat{X}_{j}\hat{Z}_{j}\}
\end{split} \\
\begin{split}
\implies E_{n}^{(0)} &= \sum_{i=1}^{n_{\mathrm{q}}+1}\prod_{j=1}^{n_q}a_{j}^{i}
, \hspace{0.2cm} \text{where} \hspace{0.2cm}
a_{i}^{j} = \begin{cases} 
             -1, & \hat{A}_{j}^{i} \in \{(\hat{X}_{j}\hat{Z}_{j} \hspace{0.1cm} | \hspace{0.1cm} j \notin LC), (\hat{Z}_{j}\hat{X}_{j} \hspace{0.1cm} | \hspace{0.1cm} j \in LC)\}; \\
            +1, & \hat{A}_{j}^{i} \in \{\hat{I}_{j}, (\hat{Z}_{j} \hspace{0.1cm} | \hspace{0.1cm} j \notin LC), (\hat{X}_{j} \hspace{0.1cm} | \hspace{0.1cm} j \in LC)\},
             \end{cases}
\end{split} 
\end{flalign}
\vspace{-0.035cm}and $\hat{\mathcal{H}}_{0, i}$ is the $i^{\mathrm{th}}$ term in $\hat{\mathcal{H}}_{0}$. We can now proceed with the perturbative analysis, where we determine the first and second order corrections in $E_{0}$ and $|\psi_{0}\rangle$ respectively. The first-order energy correction vanishes since
\begin{flalign}
\begin{split}
E_{0}^{(1)}
&= \langle\psi_{0}^{(0)}|\hat{\mathcal{H}}_{\triangle}|\psi_{0}^{(0)}\rangle  \\
&= \xi\langle+|^{\otimes{n_{\mathrm{q}}}}\bigg(\bigotimes_{j=1}^{n_{\mathrm{q}}}{\hat{Z}_{j}}\bigg)|+\rangle^{\otimes{n_{\mathrm{q}}}} = \xi\langle+|-\rangle^{\otimes{n_{\mathrm{q}}}} = 0,
\end{split}
\end{flalign}
and is indicative of a non-polarized (singlet) state in which all the up and down spins are paired. For the second-order correction, we have
\begin{flalign}
\begin{split}
E_{0}^{(2)} &= \sum_{n}\frac{|\langle\psi_{n}^{(0)}|\hat{\mathcal{H}}_{\triangle}|\psi_{0}^{(0)}\rangle|^{2}}{E_{0}^{(0)}-E_{n}^{(0)}} \\ 
&= \xi^{2}\sum_{\vec{x}}\frac{|\sum_{j'}\langle+|^{\otimes n_{\mathrm{q}}}(\prod_{E_{\text{graph}}}CZ_{a,b})\hat{X}_{j'}^{k}(\prod_{E_{\text{graph}}}CZ_{a,b})(\hat{Z}_{j'}^{1-k}\bigotimes_{j}\hat{Z}_{j}^{\vec{x}_{j}})|+\rangle^{\otimes n_{\mathrm{q}}}|^{2}}{\sum_{i}(-1+\prod_{j}a_{j}^{i})} \\
&= \xi^{2}\sum_{\vec{x}}\frac{|\sum_{j'}\langle+|^{\otimes n_{\mathrm{q}}}(\bigotimes_{j''\in{V_{j'' \rightarrow j'}}}\hat{Z}_{j''}^{k})(\hat{X}_{j'}^{k}\hat{Z}_{j'}^{\vec{x}_{j'}+1-k})(\bigotimes_{j \neq j'}\hat{Z}_{j}^{\vec{x}_{j}})|+\rangle^{\otimes n_{\mathrm{q}}}|^{2}}{\sum_{i}(-1+\prod_{j}a_{j}^{i})} \\
&=  \xi^{2}\sum_{\vec{x}}\frac{|\sum_{j'}\langle+|^{\otimes n_{\mathrm{q}}}\hat{X}_{j'}^{k}(\hat{Z}_{j'}^{\vec{x}_{j'}+1-k}\bigotimes_{j \neq j'}\hat{Z}_{j}^{\vec{x}_{j}+ck})|+\rangle^{\otimes n_{\mathrm{q}}}|^{2}}{\sum_{i}(-1+\prod_{j}a_{j}^{i})},
\label{EE10}
\end{split} 
\end{flalign}
where $V \subset Q$ is the set of all qubits $j''$ connected via $\text{CZ}$-edges to the qubit $j'$, and $c = 0 \hspace{0.05cm} (1)$ for $j \notin V \hspace{0.05cm} (j \in V)$.  
The non-zero contributions in $E_{0}^{(2)}$ occur when each qubit is acted on by an even number of perturbative $\hat{Z}$ operators, which prevents orthogonality between $\hat{\mathcal{H}}_{\triangle}|\psi_{0}^{(0)}\rangle$ and $|\psi_{n}^{(0)}\rangle$. Since the perturbation is odd under parity transformation, its action above imposes interactions between antisymmetric states which is a characteristic of spin-1/2 systems. From Eq.~\eqref{EE10}, we may now determine the first order correction to $|\psi_{0}\rangle$:
\begin{flalign}
|\psi_{0}^{(1)}\rangle &= \sum_{n}\frac{\langle\psi_{n}^{(0)}|\hat{\mathcal{H}}_{\triangle}|\psi_{0}^{(0)}\rangle}{E_{0}^{(0)}-E_{n}^{(0)}}|\psi_n^{(0)}\rangle \\
&= \bigg[\xi\sum_{\vec{x}}\frac{\sum_{j'}\langle+|^{\otimes n_{\mathrm{q}}}\hat{X}_{j'}^{k}(\hat{Z}_{j'}^{\vec{x}_{j'}+1-k}\bigotimes_{j \neq j'}\hat{Z}_{j}^{\vec{x}_{j}+ck})|+\rangle^{\otimes n_{\mathrm{q}}}}{\sum_{i}(-1+\prod_{j}a_{j}^{i})} \bigg(\bigotimes_{j}(\hat{X}_{j}^{\vec{x}_{j}})^{k}(\hat{Z}_{j}^{\vec{x}_{j}})^{1-k}\bigg)\bigg]|\psi_{0}^{(0)}\rangle.
\label{EE12}
\end{flalign}
{\update Applying Eqs.~\eqref{EE10} and \eqref{EE12} to the $(M, N)$ = ($2, 1$) graph, $E_{0}$ and $|\psi_{0}\rangle$ to their respective orders are
\begin{align}
E_{0}(\xi) &\approx -8 - \frac{37}{4}\xi^{2}; \\ 
|\psi_{0}(\xi)\rangle &\approx \bigg(\frac{1}{2} - \frac{13}{8}\xi\bigg)|0\rangle^{\otimes 7} + \bigg(\frac{1}{2} + \frac{1}{8}\xi\bigg)(|1111000\rangle + |0001111\rangle) + \bigg(\frac{1}{2} + \frac{11}{8}\xi\bigg)|1110111\rangle.
\end{align}
\vspace{-0.03cm}
For the ($2, 2$) graph, we have
\begin{flalign}
E_{0}(\xi) &\approx -13 - 9\xi^{2}; \\ 
\begin{split}
|\psi_{0}(\xi)\rangle &\approx \bigg(\frac{1}{4} - \frac{5}{4}\xi\bigg)|0\rangle^{\otimes 12} + \bigg(\frac{1}{4} - \frac{1}{2}\xi\bigg)(|101101000000\rangle + |010110100000\rangle + |000001011010\rangle + |000000101101\rangle) \\ &+ \bigg(\frac{1}{4} + \frac{1}{2}\xi\bigg)(|111010111010\rangle + |111011001101\rangle + |010111010111\rangle + |101100110111\rangle) \\ &+ \bigg(\frac{1}{4} + \frac{1}{4}\xi\bigg)(|010111111010\rangle + |101101101101\rangle) + \bigg(\frac{1}{4} + \frac{3}{4}\xi\bigg)|111010010111\rangle \\ &+ \frac{1}{4}(  
|000001110111\rangle + |111011100000\rangle + |101100011010\rangle + |010110001101\rangle).
\end{split}
\end{flalign}}{\update From these equations, we observe that increasing the perturbation $\xi$ shifts $E_{0}$ downwards, while $|\psi_{0}\rangle$  shifts towards a superposition of basis states containing a majority of qubits in the  $|1\rangle$ state.} This is expected since the dominance of the terms in $\hat{\mathcal{H}}_{\triangle}$ (each containing a single $\hat{Z}$ operator) acting on $|1\rangle$ contributes eigenvalues $\propto -\xi$ for large $\xi$. With these expressions valid for small-to-intermediate $\xi$, these $\hat{Z}$-operator terms guarantee a crossover to $|1\rangle^{\otimes n_{\mathrm{q}}}$ as $\xi \rightarrow \infty$. It is also worth mentioning the equal positive and negative contributions by the perturbation on $|\psi_{0}\rangle$, which is due to its uniform nature. This preserves the underlying symmetry of the problem, and 
thus enables redundant parameters to be eliminated in the VQE implementation. We remark that if different perturbation strengths were applied to each qubit, then a symmetry breaking is expected and all free parameters must be optimized separately.
\newpage
\section{VQE demonstrations --- error analysis} \label{app: err}
In this section, we elaborate on the method used to compute observable errors with statistical correlations and after self-mitigation is applied (Sec.~\ref{app: methods}). For a given observable $\hat{O} \in \hat{\mathcal{H}}$ measured with $N$ shots, the variance is given by
\begin{equation}
\mathrm{Var}(\hat{O}) = \frac{1}{N}\sum_{i}N^{(i)}({\hat{O}^{(i)}-\langle\hat{O}\rangle})^{2},
\end{equation}
where $O^{(i)}$ is the $i^{\mathrm{th}}$ unbiased estimator (i.e. a computational basis measurement outcome) of $\hat{O}$ with frequency $N^{(i)}$, $\sum_i N^{(i)}=N$, and $\langle\hat{O}\rangle = \frac{1}{N}\sum_i N^{(i)} O^{(i)}$ is the average over all estimators. The uncorrelated error is then calculated as the standard deviation $[\mathrm{Var}(\hat{O})]^{1/2}$. \par
To lower resource requirements, we exploit the property that any set of commuting observables in $\hat{\mathcal{H}}$ can be measured \textit{simultaneously} (i.e. with a single circuit). 
Specifically, if $\hat{O}$ is an $n$-qubit Pauli string $\hat{P}_{1}\hat{P}_{2}...\hat{P}_{n} \in \{\hat{I}, \hat{X}, \hat{Y}, \hat{Z}\}^{\otimes{n}}$, then any $n$-qubit Pauli $\hat{Q}_{1}\hat{Q}_{2}...\hat{Q}_{n}$ where $\hat{Q}_{i} \in \{\hat{P}_{i}, \hat{I}\}$ will commute with $\hat{O}$. These observables are grouped into a set $G$ from which we designate a single circuit capable of measuring each of them during readout. We note that this is guaranteed since each $\hat{O} \in G$ must act either $\hat{I}$ or the same $\hat{X}, \hat{Y}$ or $\hat{Z}$ on a given qubit. Thus, one may set each qubit basis to the corresponding non-identity Pauli (if present in at least one $\hat{O}$, else it remains unmeasured) and this constitutes a valid $n$-qubit measurement basis for the circuit. \par
The grouping requires us to account for correlations that exist between observables in $G$. In the context of error analysis, these manifest as covariances since each observable is influenced by the statistics of the same counts distribution. Partitioning $\hat{\mathcal{H}}$ into commuting groups $G_{1}, G_{2},\dots, G_{K}$, we obtain
\begin{equation}
\hat{\mathcal{H}} = \sum_{j_{1}=1}^{|G_{1}|}c_{j_{1}}
\hat{O}_{j_{1}}+ \sum_{j_{2}=1}^{|G_{2}|}c_{j_{2}}
\hat{O}_{j_{2}}+...+ \sum_{j_{K}=1}^{|G_{K}|}c_{j_{K}}\hat{O}_{j_{K}},
\end{equation}
with coefficients $c_{1},c_{2},...,c_{|G_{k}|}$ and measurement basis $\hat{O}_{j_{k}}$ corresponding to the group $G_{k}$, where $k \in S$ and $S = \{1,2,...,K\}$. Note that a specific $\hat{O}_{j_{k}}$ may belong to multiple $G_{k}$; its expectation value is then computed from the combined counts of multiple circuits, which contributes to increased estimation accuracy at no added cost~\cite{adaptive_estimation}. Let $\hat{O}'$ denote such an observable (with coefficient $c$) belonging to $M \leq K$ groups. Then its contribution to each $G_{m}$ is $cR_{m}\hat{O'}$, where $m \in S' \subseteq S$, $R_{m} = N_{m}/(\sum_{j \in S'}N_{j})$, and $M = |S'|$.
To compute the covariance, we consider all bitwise products between $\hat{O}_{a}$ and $\hat{O}_{b}$ (cross-terms) within each $G$. We then have the set $G^{\times}$, which we define as $G$ augmented with the cross-terms:
\begin{equation}
G^{\times} = G \cup \{\hat{O}_{a}\hat{O}_{b} \hspace{0.1cm} \forall \hspace{0.1cm} a, b \leq |G|: \hspace{0.1cm} a\neq b\}.
\end{equation}
Therefore, the variance of $\hat{\mathcal{H}}$ with correlations included (denoted with a tilde) is
\begin{equation}
\widetilde{\mathrm{Var}}(\hat{\mathcal{H}}) = \sum_{k=1}^{K}\frac{\widetilde{\mathrm{Var}}{(\hat{O}_{k})}}{N_{k}},
\end{equation} where
\begin{equation}
\widetilde{\mathrm{Var}}{(\hat{O}_{k})} = c_{k}^{2}\hspace{0.05cm}\mathrm{Var}{(\hat{O}_{k})} + \sum_{\substack{\ell=1 \\ {\hat{O}_{k} \neq \hat{O}_{\ell}}}}^{|G_{k}^{\times}|}\bigg[c_{\ell}^{2}R_{\ell}^{2}\mathrm{Var}(\hat{O}_{\ell})+ 2c_{k}c_{\ell}R_{k}\big(\underbrace{\langle\hat{O}_{k}\hat{O}_{\ell}\rangle - \langle\hat{O}_{k}\rangle\langle\hat{O}_{\ell}\rangle\big)}_{\text \ \mathrm{Cov}(\hat{O}_{k}, \hat{O}_{\ell})}\bigg],
\end{equation}
and $\mathrm{Cov}(\hat{O}_{k}, \hat{O}_{\ell})$ denotes the covariance between $\hat{O}_{k}$ and $\hat{O}_{\ell}$. \par 
Now we proceed to calculate the errors after self-mitigation. Since $\langle\hat{O}_{k}\rangle_{\text{phys, true}}$ is a function of $\langle\hat{O}_{k}\rangle_{\text{phys, meas}}$ and $\langle\hat{O}_{k}\rangle_{ \text{mitig, meas}}$ [Eq.~\eqref{EB9}], we must propagate their respective errors. Applying the variance formula to Eq.~\eqref{EB9} with $\langle\hat{O}_{k}\rangle_{\text{mitig, true}}$ set to 1, we obtain \\
\vspace{-\baselineskip}
\begin{flalign}
\begin{split}
\label{eq:propgate_var}
\widetilde{\mathrm{Var}}(\hat{O}_{k, \text{phys, true}}) &= \bigg(\frac{\partial \langle\hat{O}_{k}\rangle_{\text{phys, true}}}{\partial\langle\hat{O}_{k}\rangle_{\text{phys, meas}}}\bigg)^{2}\widetilde{\mathrm{Var}}(\hat{O}_{k, \text{phys, meas}}) + \bigg(\frac{\partial \langle\hat{O}_{k}\rangle_{\text{phys, true}}}{\partial\langle\hat{O}_{k}\rangle_{\text{mitig, meas}}}\bigg)^{2}\widetilde{\mathrm{Var}}(\hat{O}_{k, \text{mitig, meas}}) \\
 &= \bigg(\frac{1}{\langle\hat{O}_{k}\rangle_\text{mitig, meas}}\bigg)^{2\kappa}\widetilde{\mathrm{Var}}(\hat{O}_{k, \text{phys, meas}}) + \kappa^{2} \bigg(\frac{\langle\hat{O}_{k}\rangle_\text{phys, meas}}{(\langle\hat{O}_{k}\rangle_ \text{mitig, meas})^{\kappa+1}}\bigg)^{2}\widetilde{\mathrm{Var}}(\hat{O}_{k, \text{mitig, meas}}).
 \end{split}
\end{flalign}
 The errors in the VQE data tables (Sec.~\ref{app: suppdat}) are then reported as the standard deviation $[\widetilde{\mathrm{Var}}(\hat{O}_{k, \text{phys, true}})]^{1/2}$. We remark that the variance formula in Eq.~\eqref{eq:propgate_var} linearly approximates $\langle\hat{O}_{k}\rangle_{\text{phys, true}}$, and therefore assumes the errors from the {\update physics} and mitigation runs to be uncorrelated and relatively small. This is a fair assumption since their respective observables are calculated from separate circuits and retain the same counts distribution after correcting the expectation values. However, if self-mitigation is instead used to correct each frequency in the counts distribution, then a reshaping {\update of the distribution} can occur and potentially amplify the errors. These situations would require alternate techniques such as Quantum Monte Carlo that analyze how the observables' error distributions transform under propagation, owing to their nonlinear relations with the function.
\bibliography{Bibliography}